\documentclass[letterpaper]{article}

\usepackage{authblk}
\usepackage{amssymb}
\usepackage{amsmath}
\usepackage{amsthm}
\usepackage{amsfonts}

\usepackage{indentfirst}
\usepackage{subfig}
\usepackage{mathtools}
\usepackage{bm}
\usepackage{mathrsfs}
\usepackage{fancyhdr}
\usepackage{cite}
\usepackage{enumerate}
\usepackage{booktabs}
\usepackage{colortbl}
\usepackage{multicol}
\usepackage{multirow}
\usepackage{cancel,soul}
\setstcolor{red}
\usepackage{xcolor}
\usepackage{nth}

\usepackage{stfloats}
\usepackage[font=small,skip=0pt]{caption}
\usepackage[boxed,linesnumbered]{algorithm2e}
\usepackage{listings}
\newcommand{\argmin}{\operatornamewithlimits{argmin}}
\newcommand{\argmax}{\operatornamewithlimits{argmax}}
\title{Sparse-promoting Full Waveform Inversion based on Online Orthonormal Dictionary Learning}

\author{Lingchen Zhu}
\author{Entao Liu}
\author{James H. McClellan}
\affil{Center for Energy and Geo Processing (CeGP) at Georgia Tech and KFUPM\\75 Fifth Street NW, Atlanta, GA 30308}

\begin{document}
\maketitle

\renewcommand{\thefootnote}{\fnsymbol{footnote}}

\begin{abstract}
Full waveform inversion (FWI) delivers high-resolution images of the subsurface by minimizing iteratively the misfit between the recorded and calculated seismic data. It has been attacked successfully with the Gauss-Newton method and sparsity promoting regularization based on fixed multiscale transforms that permit significant subsampling of the seismic data when the model perturbation at each FWI data-fitting iteration can be represented with sparse coefficients. Rather than using analytical transforms with predefined dictionaries to achieve sparse representation, we introduce an adaptive transform called the Sparse Orthonormal Transform (SOT) whose dictionary is learned from many small training patches taken from the model perturbations in previous iterations. The patch-based dictionary is constrained to be orthonormal and trained with an online approach to provide the best sparse representation of the complex features and variations of the entire model perturbation. The complexity of the training method is proportional to the cube of the number of samples in one small patch. By incorporating both compressive subsampling and the adaptive SOT-based representation into the Gauss-Newton least-squares problem for each FWI iteration, the model perturbation can be recovered after an $\ell_1$-norm sparsity constraint is applied on the SOT coefficients. Numerical experiments on synthetic models demonstrate that the SOT-based sparsity promoting regularization can provide robust FWI results with reduced computation.
\end{abstract}

\section{Introduction}
Seismic imaging reveals properties of the Earth's subsurface by recording and processing seismic waves. One way to achieve this goal is by full waveform inversion (FWI), which is a data-fitting procedure that minimizes the misfit between recorded and calculated seismic data to create high-resolution models. A conventional FWI method is carried out iteratively. Each iteration consists of solving wave equations with the current model parameters to generate seismic data, calculating the value as well as the gradient of the misfit function, and updating the model parameters with an optimization method \cite{Tarantola:1984aa,Tarantola:1986aa,Virieux:2009aa}. The efficiency of these three components determines the industrial applicability of FWI in the real world. By recording the response of sequential sources on the surface or in the water, a wide-aperture seismic survey typically covers a large area of interest.
Because the discretized sizes of seismic datasets and models could both be huge, computation of forward modeling, misfit calculation and model updating in FWI could be very intensive. In this paper, we discretize the wave equations in the frequency domain with a finite-difference method, which yields compact expressions and numerical advantages \cite{Pratt:1999aa}. Since each source, or each frequency component, parametrizes an individual wave equation to solve, the computational cost is proportional to the number of frequency components (for frequency-domain forward modeling) and/or sources (for iterative solvers). Such a computational burden is usually termed the ``curse of dimensionality'' \cite{Herrmann:2012aa} and has to be relieved by reducing the problem dimensionality.

\subsection{Related Work}
Reducing the computational cost of FWI has been an active research area for many years. When a frequency-domain FWI is carried out, one can divide the frequency range of interest into several bands \cite{Bunks:1995aa}, and invert only a few frequencies per band, sequentially from the low to high frequency bands, to help reduce the cost \cite{Sirgue:2004aa}. Another well-known method for cost reduction is to generate simultaneous shots by linearly combining one or several different sequential shots at different source positions with random weights \cite{Krebs:2009aa,Moghaddam:2010aa,Ben-Hadj-Ali:2011aa,Castellanos:2015aa}, or randomly choosing a few sequential shots at each FWI iteration \cite{Li:2012ab,Warner:2013aa}. Though these methods employ different data acquisition strategies, they essentially reduce the amount of data used in the FWI misfit function in the frequency and spatial domains.

In recent years, compressive sensing (CS) has attracted considerable attention by proving that it is possible to recover sparse information vectors from a subsampled dataset with high robustness in the presence of noise and artifacts.
CS has been widely applied in seismic data processing tasks such as recovery \cite{Herrmann:2008ab,Yu:2015aa} and denoising \cite{Hennenfent:2006aa}. The basic idea is that seismic data can be represented with a sparse set of coefficients over a carefully chosen domain, which then implies that we no longer need to maintain high temporal or spatial sampling rates. Instead, we can acquire data at a much reduced sampling rate and process the smaller data sets in the compressed domain to reduce complexity. A typical CS problem involves three steps: (1) data randomization, (2) subsampling and (3) sparsity promotion. Randomization of the data plays an important role due to its ability to suppress coherent subsampling-related interferences such as aliasing and crosstalk into relatively harmless Gaussian noise \cite{Donoho:2010aa}. It is then safe to extract the underlying sparsity through the solution of an optimization problem coupled with an $\ell_p$-norm ($0 \leq p \leq 1$) constraint, even when the signal sampling rate is well below Nyquist.

Researchers have extended the above idea from seismic data processing to more sophisticated seismic imaging and inversion problems by assuming sparsity on the velocity models.
For example, \cite{Loris:2007aa} regularized the model in the wavelet domain with $\ell_1$-norm constraints, allowing sharp velocity discontinuities (model perturbations) to be superimposed on a smooth background model. Celebrated works reported by \cite{Herrmann:2011aa}, \cite{Li:2012aa} and \cite{Li:2016aa} discover the $\ell_1$-norm sparsity of model perturbations in the curvelet domain \cite{Candes:2005aa,Candes:2006aa} so that they can be efficiently updated via a random subset of the full data. \cite{Ma:2012aa} introduced the image-guided gradient to FWI which leads to a significant reduction in the number of model parameters to be inverted. \cite{Xue:2015aa} imposed seislet-domain $\ell_1$-norm sparsity regularization on the model to improve the robustness and efficiency of FWI. All these methods reduce the computational complexity of the FWI problem without degrading the quality of the result by regularizing the sparsity of the model (or its perturbation) over a predefined transform domain.

\subsection{Main Contribution}
In this paper, we propose a new framework for the FWI Gauss-Newton method where the model perturbation is compressed by adaptive data-driven dictionaries \cite{Sezer:2012aa,Cai:2014aa,Sezer:2015aa,Zhu:2016aa} so that it has a very sparse representation in the resulting adaptive transform.
Each data-driven dictionary is learned from a set of small patches taken from a model perturbation. Based on the resulting sparsity promoting transform, the dimensionality of FWI can then be greatly reduced via randomized phase encoding and subsampling. Compared to the traditional model-based transforms such as wavelet, curvelet and seislet with predefined dictionaries, dictionary learning methods are better able to adapt to nonintuitive signal regularities beyond piecewise smoothness. Therefore, they have become a promising technique for sparse signal representation and approximation. In fact, the complex subsurface model parameters may be difficult to capture with any concise analytical model, leaving dictionary learning as the only alternative. However, prevailing dictionary learning algorithms such as K-Singular Value Decomposition (K-SVD) \cite{Aharon:2006aa}, as well as its variants \cite{Rubinstein:2010aa,Rubinstein:2013aa} are too expensive, so we prefer to learn orthonormal dictionaries that have simple implementations for sparse representation in FWI.

The contributions in this paper are the following:
\begin{itemize}
  \item We propose an online approach for orthonormal dictionary learning by minimizing the expectation of the cost function when new training patches join.
  \item We design a fast block-wise orthonormal transform called the sparse orthonormal transform (SOT) based on the learned dictionaries in order to exploit the sparsity of model perturbations.
  \item We implement a CS framework in the FWI problems by admitting sparse representation of model perturbations over learned dictionaries and reducing the problem dimensionality by selecting random sequential shots.
  \item Tests on several synthetic velocity models demonstrate that our method can significantly reduce the amount of data needed for FWI and therefore decrease the computation time without introducing visible artifacts.
\end{itemize}

\subsection{Notation}
Although our work is exemplified on 2D models in this paper, an extension to 3D case would not be difficult. Before we describe our methodology in detail, we give a summary of the notation that is going to be used throughout this paper in Table \ref{tb:notations}.
\begin{table}[htbp]
	\renewcommand{\arraystretch}{1.2} 
	\begin{center}
	\begin{tabular}{ c | l }
	\toprule[2pt]
	$\mathbf{x}$ & 2D coordinate $\mathbf{x} \triangleq (z, x)$, where $z$ is vertical, $x$ lateral \\ \hline
	$\mathcal{S}$ & shots and receivers domain on the surface \\ \hline
	$\mathcal{U}$ & scattering domain in the subsurface \\ \hline

	$\omega$ & angular frequency \\ \hline
	$\Omega$ & angular frequency set \\ \hline
	$\nabla^2$ & Laplacian operator $\nabla^2 \triangleq \partial^2/\partial z^2 + \partial^2/\partial x^2$ \\ \hline
	$v(\mathbf{x}), \mathbf{v}$ & acoustic wave velocity model \\ \hline
	{$m(\mathbf{x}), \mathbf{m}$} & squared slowness $m(\mathbf{x}) \triangleq 1/v^2(\mathbf{x})$, which is the actual model used in FWI 
	\\ \hline
	{$p(\mathbf{x}; \omega, \mathbf{x}_s)$} & 2D acoustic wavefield at frequency $\omega$ is response to a point source $f(\omega)\delta(\mathbf{x} - \mathbf{x}_s)$\\ \hline
	{$G(\mathbf{x}; \omega, \mathbf{x}_s)$} & 2D Green's function at frequency $\omega$ is response to an impulse source $\delta(\mathbf{x} - \mathbf{x}_s)$ \\ \hline
	$\mathbf{M}$ & matrices are written with bold capital letters \\ \hline
	{$\mathbf{w}$} & vectors are written in bold lower-case; $\mathbf{w}_i$ refers to the $i$-th column of $\mathbf{W}$ \\ \hline
	$\text{diag}(\mathbf{w})$ & a diagonal matrix with diagonal elements $\mathbf{w}$ \\ \hline
	{$c$} & scalars are written in lower-case;
	 $c_{ij}$ refers to the $(i,j)$ element of $\mathbf{C}$ \\ \hline
	$\mathbf{A} \otimes \mathbf{B}$ & Kronecker product of $\mathbf{A}$ and $\mathbf{B}$  \\ \hline
	$\overline{z}$ & complex conjugate of complex number $z$ \\ \hline
	$\Re\{z\}$ & real part of complex number $z$ \\ \hline
	{$\left(\cdot\right)^{\dag}$} & adjoint of the operator $\left(\cdot\right)$\\
	\toprule[2pt]
	\end{tabular}
	\end{center}
	\caption{Notation used throughout the paper}
	\label{tb:notations}
\end{table}

\subsection{Outline}
The rest of the paper is organized as follows: In the second section, we review the FWI problem in the frequency domain based on updating the model perturbation via the Gauss-Newton method. 
The following section describes the orthonormal dictionary learning method and introduces an online learning approach designed for FWI. The next section combines the above two topics together and formulates an FWI framework based on CS that comes with randomized phase encoding, subsampling and sparsity regularization. A practical method to solve this problem is also presented in this section. Numerical results from synthetic models are given in the next section, followed by discussion and conclusions.

\section{Full Waveform Inversion}
The purpose of FWI \cite{Tarantola:1984aa} is to recover velocity models by fitting the forward modeling data to the recorded data. The recorded data are acquired from an array of seismic receivers and denoted as $\mathbf{d}_{\text{obs}}$. The forward modeling data are calculated by simulating full wave equations in a model $\mathbf{m}$ with finite-difference methods and then sampling the wavefields at the receiver positions. Note that the model $\mathbf{m} \triangleq [m(\mathbf{x}_1), \dots, m(\mathbf{x}_{N_zN_x})]^T$ is a vector of parameters of length $N_zN_x$ where $N_z$ and $N_x$ are the number of grid points in the vertical and lateral directions, respectively, i.e., the size of model can be regarded as $N_z \times N_x$. In the model, $m(\mathbf{x}_i)$ indicates the squared-slowness value at the 2D coordinate $\mathbf{x}_i$, $\forall i = 1, \dots, N_zN_x$. Such a calculation process from model to data can be denoted as $\mathbf{d}_{\text{cal}} \triangleq \bm{\mathcal{F}}(\mathbf{m})$ where $\bm{\mathcal{F}}(\cdot)$ is the nonlinear forward modeling operator. Based on these definitions, FWI seeks a model $\mathbf{m}$ that minimizes the following nonlinear least-squares misfit function
\begin{equation}
  \label{eq:fwi-nl-misfit}
  E(\mathbf{m}) \triangleq \frac{1}{2} \| \mathbf{d}_{\text{obs}} - \mathbf{d}_{\text{cal}} \|_2^2 = \frac{1}{2} \| \mathbf{d}_{\text{obs}} - \bm{\mathcal{F}}(\mathbf{m}) \|_2^2.
\end{equation}

\begin{figure}[htbp]
	\centering
	\begin{minipage}{\linewidth}
	\centering
	\includegraphics[width=\textwidth]{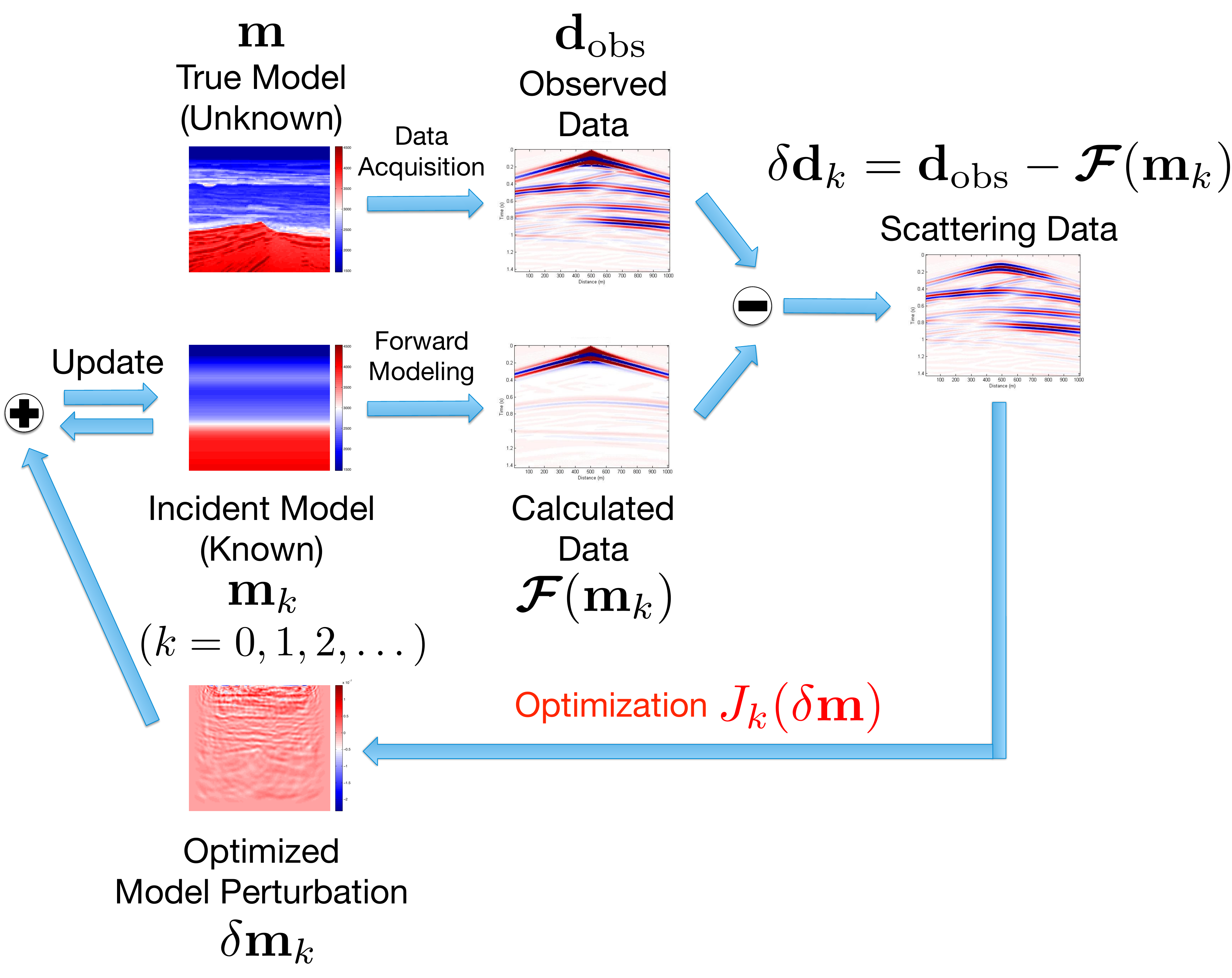}
	\end{minipage}
	\caption{Schematic FWI Workflow based on the Gauss-Newton method}
\label{fig:fwiWorkflow}
\end{figure}

In practice, the minimum of $E(\mathbf{m})$ needs to be searched in an iterative manner $\mathbf{m}_{k+1} = \mathbf{m}_k + \delta \mathbf{m}_k$, $k = 0, 1, 2, \dots$ where $\delta \mathbf{m}_k$ is the optimal model perturbation that minimizes $E(\mathbf{m})$ in the vicinity of the current model $\mathbf{m}_k$. Hence, we expand $E(\mathbf{m})$ in a small vicinity $\delta \mathbf{m}$ of $\mathbf{m}_k$ with a Taylor polynomial of degree two
\begin{equation}
  \label{eq:fwi-nl-misfit-taylor}
  E(\mathbf{m}) = E(\mathbf{m}_k) + \delta \mathbf{m}^T \mathbf{g}_k + \frac{1}{2} \delta \mathbf{m}^T \mathbf{H}_k \delta \mathbf{m} + o(\|\delta \mathbf{m}\|^3),
\end{equation}
where $\mathbf{g}_k \triangleq \dfrac{\partial E(\mathbf{m}_k)}{\partial \mathbf{m}}$ denotes the gradient of the misfit function $E(\mathbf{m})$ evaluated at $\mathbf{m}_k$ and $\mathbf{H}_k \triangleq \dfrac{\partial^2 E(\mathbf{m}_k)}{\partial \mathbf{m}^2}$ denotes the Hessian matrix whose elements are the second-order partial derivatives of $E(\mathbf{m})$ at $\mathbf{m}_k$. In each iteration, FWI seeks a model perturbation $\delta \mathbf{m}$ such that $E(\mathbf{m}_k + \delta \mathbf{m})$ is a minimum. Setting the gradient of \eqref{eq:fwi-nl-misfit-taylor} with respect to $\delta \mathbf{m}$ equal to zero yields the solution $\delta \mathbf{m}_k$ that satisfies
\begin{equation}
  \label{eq:fwi-nl-newton}
  \mathbf{H}_k\delta \mathbf{m}_k = -\mathbf{g}_k.
\end{equation}
Specifically, by taking the gradient of the misfit function in \eqref{eq:fwi-nl-misfit}, $\mathbf{g}_k$ can be evaluated as
\begin{equation}
  \label{eq:fwi-nl-misfit-grad}
  \mathbf{g}_k \triangleq \frac{\partial E(\mathbf{m}_k)}{\partial \mathbf{m}} = -\Re\left\{ \left[ \frac{\partial \bm{\mathcal{F}}(\mathbf{m}_k)}{\partial \mathbf{m}} \right]^{\dag} (\mathbf{d}_{\text{obs}} - \bm{\mathcal{F}}(\mathbf{m}_k)) \right\} = -\Re\left\{ \mathbf{J}_k^{\dag}\delta \mathbf{d}_k \right\},
\end{equation}
where $\delta \mathbf{d}_k \triangleq \mathbf{d}_{\text{obs}} - \bm{\mathcal{F}}(\mathbf{m}_k)$, and $\mathbf{J}_k \triangleq \dfrac{\partial \bm{\mathcal{F}}(\mathbf{m}_k)}{\partial \mathbf{m}}$ is the Jacobian matrix of $\bm{\mathcal{F}}(\cdot)$ which indicates the sensitivity of the forward modeling data with respect to the model perturbation. By taking another derivative, the Hessian matrix $\mathbf{H}_k$ can be expressed as
\begin{equation}
  \label{eq:fwi-nl-misfit-hessian}
  \mathbf{H}_k \triangleq \frac{\partial^2 E(\mathbf{m}_k)}{\partial \mathbf{m}^2} = \Re\left\{ \mathbf{J}_k^{\dag}\mathbf{J}_k \right\} - \Re\left\{ \left[ \left( \frac{\partial \mathbf{J}_k^{\dag}}{\partial m_1} \right) \delta \mathbf{d}_k, \cdots, \left( \frac{\partial \mathbf{J}_k^{\dag}}{\partial m_N} \right) \delta \mathbf{d}_k\right] \right\}.
\end{equation}
The second term of $\mathbf{H}_k$ is often dropped off due to its complexity \cite{Tarantola:1987aa,Pratt:1998aa}. When $\mathbf{H}_k$ in \eqref{eq:fwi-nl-misfit-hessian} is approximated with its first term $\mathbf{H}_k = \Re\left\{ \mathbf{J}_k^{\dag}\mathbf{J}_k \right\}$, we have the Gauss-Newton method in which the solution $\delta \mathbf{m}_k$ satisfies the normal equation
\begin{equation}
  \label{eq:fwi-gauss-newton}
  \left[ \Re\left\{ \mathbf{J}_k^{\dag}\mathbf{J}_k \right\} \right]\delta \mathbf{m}_k = \Re\left\{ \mathbf{J}_k^{\dag}\delta \mathbf{d}_k \right\}.
\end{equation}
When $\mathbf{J}_k$ is full-rank, \eqref{eq:fwi-gauss-newton} has a unique solution $\delta \mathbf{m}_k$ that actually minimizes the following linear least-squares objective function
\begin{equation}
  \label{eq:fwi-gauss-newton-linear}
  J_k(\delta \mathbf{m}) \triangleq \frac{1}{2} \|\delta \mathbf{d}_k - \mathbf{J}_k\delta \mathbf{m}\|_2^2.
\end{equation}
From now on, throughout the paper, we work with the Gauss-Newton method and update the model $\mathbf{m}$ iteratively by minimizing the objective function \eqref{eq:fwi-gauss-newton-linear} to solve the FWI problem. The schematic workflow is shown in Figure \ref{fig:fwiWorkflow}.

\subsection{Computation of the gradient and Hessian}
The forward modeling operator $\bm{\mathcal{F}}(\mathbf{m})$ used in this paper is based on the frequency-domain constant-density acoustic wave equation
\begin{equation}
  \label{eq:acoustic-wave-equation-freq}
  \left( -m(\mathbf{x})\omega^2 - \nabla^2 \right) p(\mathbf{x}; \omega, \mathbf{x}_s) = f(\omega)\delta(\mathbf{x} - \mathbf{x}_s),
\end{equation}
where $p(\mathbf{x}; \omega, \mathbf{x}_s)$ is the acoustic pressure wavefield generated by a point source term $f(\omega)\delta(\mathbf{x} - \mathbf{x}_s)$ at source position $\mathbf{x}_s$, parametrized by a given angular frequency $\omega$. We assume the source shot signature $f(\omega)$ is known and fixed. Given the Green's function $G(\mathbf{x}; \omega, \mathbf{x}_s)$ of the model $m(\mathbf{x})$ defined by
\begin{equation}
  \label{eq:greens-function}
  \left( -m(\mathbf{x})\omega^2 - \nabla^2 \right) G(\mathbf{x}; \omega, \mathbf{x}_s) = \delta(\mathbf{x} - \mathbf{x}_s),
\end{equation}
the solution of \eqref{eq:acoustic-wave-equation-freq} can be expressed as
\begin{equation}
  \label{eq:solution-p}
  p(\mathbf{x}; \omega, \mathbf{x}_s) = f(\omega)G(\mathbf{x}; \omega, \mathbf{x}_s).
\end{equation}
The calculated data $\mathbf{d}_{\text{cal}} \triangleq \{p(\mathbf{x}_r; \omega, \mathbf{x}_s)\}$ is a set of wavefield samples collected at all receiver locations $\mathbf{x}_r$ for all source positions $\mathbf{x}_s$ and all frequencies $\omega$.

The Jacobian matrix $\mathbf{J}_k \triangleq \dfrac{\partial \bm{\mathcal{F}}(\mathbf{m}_k)}{\partial \mathbf{m}}$ can be computed under the framework of the Born approximation \cite{Gubernatis:1977aa,Wu:1985aa}. Assuming a small perturbation $\delta m(\mathbf{x})$ of the model $m_k(\mathbf{x})$, the resulting wavefield perturbation $\delta p(\mathbf{x}; \omega, \mathbf{x}_s)$ satisfies the following wave equation
\begin{equation}
  \label{eq:born-approximation}
  \left( -m_k(\mathbf{x})\omega^2 - \nabla^2 \right) \delta p(\mathbf{x}; \omega, \mathbf{x}_s) = \omega^2 \delta m(\mathbf{x}) p_k(\mathbf{x}; \omega, \mathbf{x}_s),
\end{equation}
and the solution collected at $\mathbf{x} = \mathbf{x}_r$ is
\begin{equation}
  \label{eq:scattering-wavefield-solution}
  \delta p(\mathbf{x}_r; \omega, \mathbf{x}_s) = \omega^2 f(\omega) \sum\limits_{\mathbf{x} \in \mathcal{U}} \delta m(\mathbf{x}) G_k(\mathbf{x}; \omega, \mathbf{x}_r) G_k(\mathbf{x}; \omega, \mathbf{x}_s).
\end{equation}
where the sum is taken over the $N_zN_x$ points in the 2-D subsurface scattering domain $\mathcal{U}$.
Therefore, for one specified source $\mathbf{x}_s$ and frequency $\omega$, the $(i, j)$-th element of the Jacobian matrix $\mathbf{J}_k(\omega, \mathbf{x}_s)$ is given by
\begin{equation}
  \label{eq:jacobian-matrix}	
  \left[\mathbf{J}_k(\omega, \mathbf{x}_s)\right]_{i, j} \triangleq \lim\limits_{\delta m_j \rightarrow 0} \dfrac{\delta p(\mathbf{x}_{r_i}; \omega, \mathbf{x}_s)}{\delta m_j} = \omega^2 f(\omega) G_k(\mathbf{x}_j; \omega, \mathbf{x}_{r_i}) G_k(\mathbf{x}_j; \omega, \mathbf{x}_s).
\end{equation}
The size of $\mathbf{J}_k(\omega, \mathbf{x}_s)$ is $N_r \times N_zN_x$ where $N_r$ is the number of receivers. To obtain the entire Jacobian $\mathbf{J}_k$, \eqref{eq:jacobian-matrix} would be used to determine $\mathbf{J}_k(\omega, \mathbf{x}_s)$ for all sources and frequencies of interest, e.g., $N_s$ sources $\mathbf{x}_s \in \mathcal{S}$ and $N_{\omega}$ frequencies $\omega \in \Omega$. Finally, all of these different $\mathbf{J}_k(\omega, \mathbf{x}_s)$ are vertically concatenated to form a huge matrix $\mathbf{J}_k$ of size $N_{\omega}N_sN_r \times N_zN_x$ that can be used in the objective function \eqref{eq:fwi-gauss-newton-linear}. Inserting $\mathbf{J}_k$ back into the normal equation \eqref{eq:fwi-gauss-newton}, we have the gradient $\mathbf{g}_k$ and the approximate Hessian matrix $\mathbf{H}_k$ as follows
\begin{equation}
  \label{eq:gradient-dm-closedform}
  \mathbf{g}_k(\mathbf{x}) = - \Re \left\{ \sum\limits_{\omega \in \Omega} \omega^2 f(\omega) \sum\limits_{\mathbf{x}_s \in \mathcal{S}} \sum\limits_{\mathbf{x}_r \in \mathcal{S}} G_k(\mathbf{x}; \omega, \mathbf{x}_r) G_k(\mathbf{x}; \omega, \mathbf{x}_s)
    \left( \overline{\delta d_k(\mathbf{x}_r; \omega, \mathbf{x}_s)} \right) \right\},
\end{equation}
\begin{equation}
  \label{eq:hessian-dm-closedform}
  	\mathbf{H}_k(\mathbf{x}, \mathbf{y}) = \Re \left\{ \sum\limits_{\omega \in \Omega} \omega^4 |f(\omega)|^2
  	\sum\limits_{\mathbf{x}_s \in \mathcal{S}} G_k(\mathbf{x}; \omega, \mathbf{x}_s) \overline{G_k(\mathbf{y}; \omega, \mathbf{x}_s)}
  	\sum\limits_{\mathbf{x}_r \in \mathcal{S}} G_k(\mathbf{x}; \omega, \mathbf{x}_r)
  	\overline{G_k(\mathbf{y}; \omega, \mathbf{x}_r)} \right\}.
\end{equation}

\subsection{Dimensionality Reduction by Compressive Sensing}
From Equations \eqref{eq:gradient-dm-closedform} and \eqref{eq:hessian-dm-closedform}, we can see that the complexity of the Gauss-Newton method comes primarily from the computation and inversion of the Hessian matrix $\mathbf{H}_k$. Unfortunately, due to the large number of $N_{\omega}$, $N_s$, $N_r$ and $N_z$, $N_x$, it is prohibitive to compute $\mathbf{H}_k^{-1}$ directly with all data in industrial-scale FWI problems. In order to reduce the computational complexity of FWI, it has been widely reported that for the cases with large acquisition aperture and wide frequency bandwidth, $\mathbf{H}_k$ is almost diagonally dominant and $\mathbf{H}_k^{-1}$ can be approximated with a diagonal matrix \cite{Beylkin:1985aa,Shin:2001aa,Plessix:2004aa,Jang:2009aa,Ren:2013aa,Pan:2015aa}. The computational complexity can also be reduced by approximating $\mathbf{H}_k$ with quasi-Newton methods such as the limited-memory Broyden-Fletcher-Goldfarb-Shanno (l-BFGS) algorithm \cite{Nocedal:1980aa,Nocedal:2006aa}.

The development of CS theories provides another perspective to lower the complexity of Gauss-Newton FWI by reducing the problem dimensionality rather than simplifying the Hessian matrix, when sparsity of the model can be exploited. This suggests minimizing the linear least-squares objective function in \eqref{eq:fwi-gauss-newton-linear} for each Gauss-Newton iteration can be replaced by the following optimization problem
\begin{equation}
  \label{eq:fwi-gauss-newton-random}
  \min\limits_{\bm{\alpha}} \left\{ J_k^{(\text{W})}(\bm{\alpha}) \triangleq \frac{1}{2} \|\mathbf{W}_k\delta \mathbf{d}_k - \mathbf{W}_k\mathbf{J}_k\bm{\mathcal{D}}(\bm{\alpha})\|_2^2 \right\} \quad \text{s.t.} \quad \|\bm{\alpha}\|_1 \leq \tau_k
\end{equation}
where $\mathbf{W}_k$ is a compressive sampling matrix for dimensionality reduction which can be different for each iteration $k$ for better performance \cite{Krebs:2009aa,Herrmann:2011aa,Warner:2013aa,Li:2012aa,Li:2016aa}; $\bm{\mathcal{D}}$ is a transform such that the model perturbation can be represented as $\delta \mathbf{m} = \bm{\mathcal{D}}(\bm{\alpha})$ with the coefficient vector $\bm{\alpha}$ being sparse. Each element of $\bm{\alpha}$ corresponds to a scalar weighting factor of a unique function called atom and a collection of these atoms is called a dictionary $\mathbf{D}$. An atom is used interchangeably as a column vector of $\mathbf{D}$ if it is an explicit matrix.

Leaving the design of $\mathbf{W}_k$ aside for a while, a fundamental consideration in employing this representation of the model perturbation is the choice of the transform $\bm{\mathcal{D}}$. It is usually appealing to choose multiscale transforms such as wavelets, curvelets, seislets, etc. These fixed transforms have proven their analytical optimality on sparse representation of multidimensional signals with assumed features such as smooth lines or curves, and hence their success in applications relies on how suitable the signals in question fit the assumptions. In most cases, these multiscale transforms have efficient algorithmic implementations in the spatial-frequency domain and, as a result, their representations as dictionaries $\mathbf{D}$ are implicit. In the last several years, many authors \cite{Loris:2007aa,Herrmann:2011aa,Li:2012aa,Xue:2015aa,Li:2016aa} have developed methods that exploit the sparsity of $\delta \mathbf{m}$ by using various multiscale transforms to solve FWI problems efficiently.

In this paper, we investigate how to exploit the sparsity of $\delta \mathbf{m}$ with a novel transform based on explicit adaptive dictionaries rather than implicit fixed dictionaries for some assumed feature characteristic. Particularly, in the FWI iteration \eqref{eq:fwi-gauss-newton-random}, we leave a place for an adaptive transform that changes at each FWI iteration. The key to this approach is to infer explicit dictionary matrices $\mathbf{D}_k$ from a set of training examples and construct a transform $\bm{\mathcal{D}}_k$ based on these dictionaries that synthesizes $\mathbf{\alpha}$ to $\delta \mathbf{m}$. The similarity among different model perturbations suggests that small patches of previously optimized model perturbations $\{\delta \mathbf{m}_i\}_{i=0}^{k-1}$ could be an appropriate choice for a training set. In the next section, we will discuss dictionary learning algorithms that derive an adaptive dictionary from a set of training examples for sparse representation as well as the way to construct a transform operator based on this dictionary.

\section{Sparse Orthonormal Transform}
The CS technique can help to reduce the problem dimensionality of each Gauss-Newton problem in FWI, as long as the model perturbation $\delta \mathbf{m}$ is sparse with respect to some transform. Rather than using fixed transforms based on off-the-shelf dictionaries such as the wavelets, curvelets, seislets, etc., we design transforms based on adaptive dictionaries that discover the inherent sparsity of $\delta \mathbf{m}$ at each FWI iteration. These dictionaries are learned via maximum likelihood (ML) estimation using training data from previously obtained model perturbations $\{\delta \mathbf{m}_i\}$ in an online manner.

The sparse approximation model represents an arbitrary vector $\mathbf{y}$ as
\begin{equation}
  \label{eq:sparsity-model}
  \mathbf{y} = \mathbf{Dx} + \mathbf{n} \approx \mathbf{Dx}
\end{equation}
where $\mathbf{x}$ is a sparse vector and the residual $\mathbf{n}$ is Gaussian noise. Given a matrix of $R$ training examples $\mathbf{Y} \triangleq [\mathbf{y}_1, \mathbf{y}_2, \dots, \mathbf{y}_R] \in \mathbb{R}^{N \times R}$, the dictionary learning method seeks the dictionary matrix $\mathbf{D}$ that maximizes the likelihood function $P(\mathbf{Y}|\mathbf{D})$ based on the \textit{a priori} constraint that each column vector of the coefficient matrix $\mathbf{X} \triangleq [\mathbf{x}_1, \mathbf{x}_2, \dots, \mathbf{x}_R]$ be sparse. Using a probabilistic framework based on ML estimation \cite{Olshausen:1996aa,Lewicki:1999aa}, the dictionary learning method minimizes the following empirical cost function
\begin{equation}
  \label{eq:empirical-cost-function}
  \begin{aligned}
  e_R(\mathbf{Y}, \mathbf{D}) &\triangleq \frac{1}{R} \sum\limits_{i=1}^R \left( \| \mathbf{y}_i - \mathbf{D}\mathbf{x}_i\|_2^2 + \lambda\|\mathbf{x}_i\|_0 \right) \\
  &= \frac{1}{R} \left( \| \mathbf{Y} - \mathbf{DX} \|_F^2 + \lambda\|\mathbf{X}\|_0 \right),
  \end{aligned}
\end{equation}
where $\lambda$ is a Lagrange multiplier, $\|\cdot\|_F$ is the Frobenius norm, and $\|\cdot\|_0$ is the $\ell_0$-norm that promotes sparsity by counting the nonzero entries of a vector or a matrix.

In this work, we pick $R$ training examples $\mathbf{y}_i$, $i = 1, \dots, R$, to form the matrix $\mathbf{Y}$ from patches of the optimized model perturbation $\delta \mathbf{m}_{k-1}$ obtained from the previous $(k-1)$-th FWI iteration. These patches cover all of $\delta \mathbf{m}_{k-1}$ and can be overlapping so that the matrix $\mathbf{D}$ will be a generative dictionary that provides sparse representations for all patches of $\delta \mathbf{m}$ in the following $k$-th FWI iteration. This updating strategy, which is called online learning, plays a critical role for the iterative problems such as FWI.
\begin{figure}[htbp]
  \centering
  \subfloat[]
  {
    \begin{minipage}{0.55\linewidth}
      \centering
      \includegraphics[width=\textwidth]{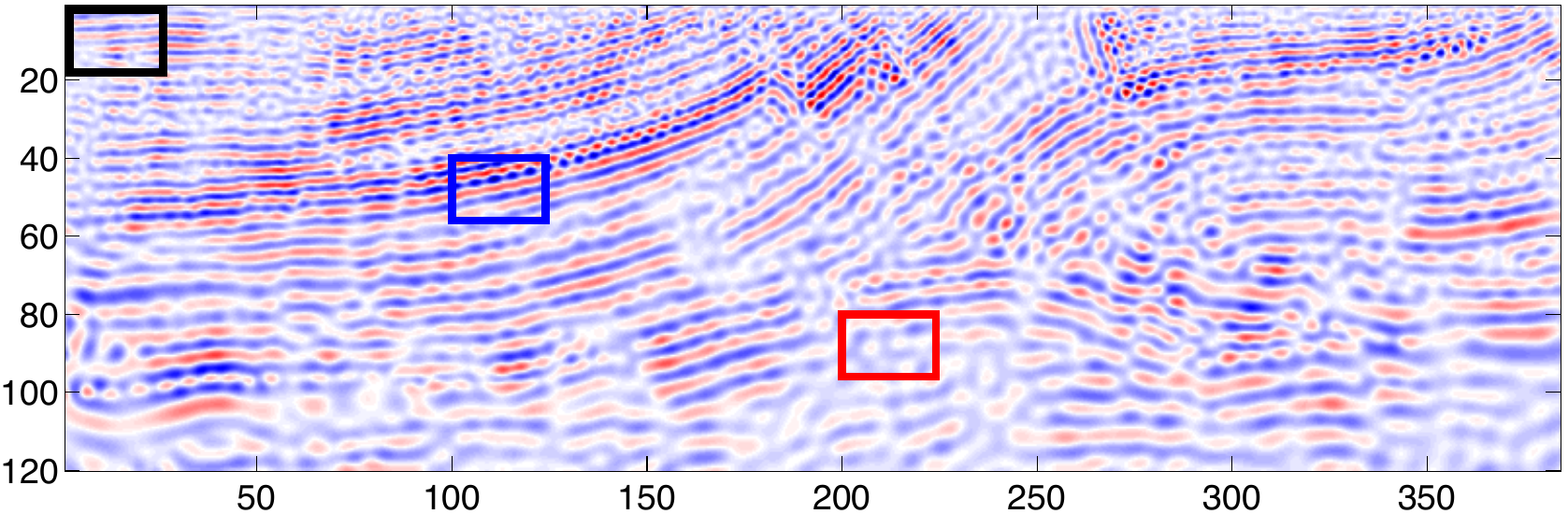}
    \end{minipage}
    \label{fig:dm_marmousi_patches}
  }
  \subfloat[]
  {
    \begin{minipage}{0.4\linewidth}
      \centering
      \includegraphics[width=\textwidth]{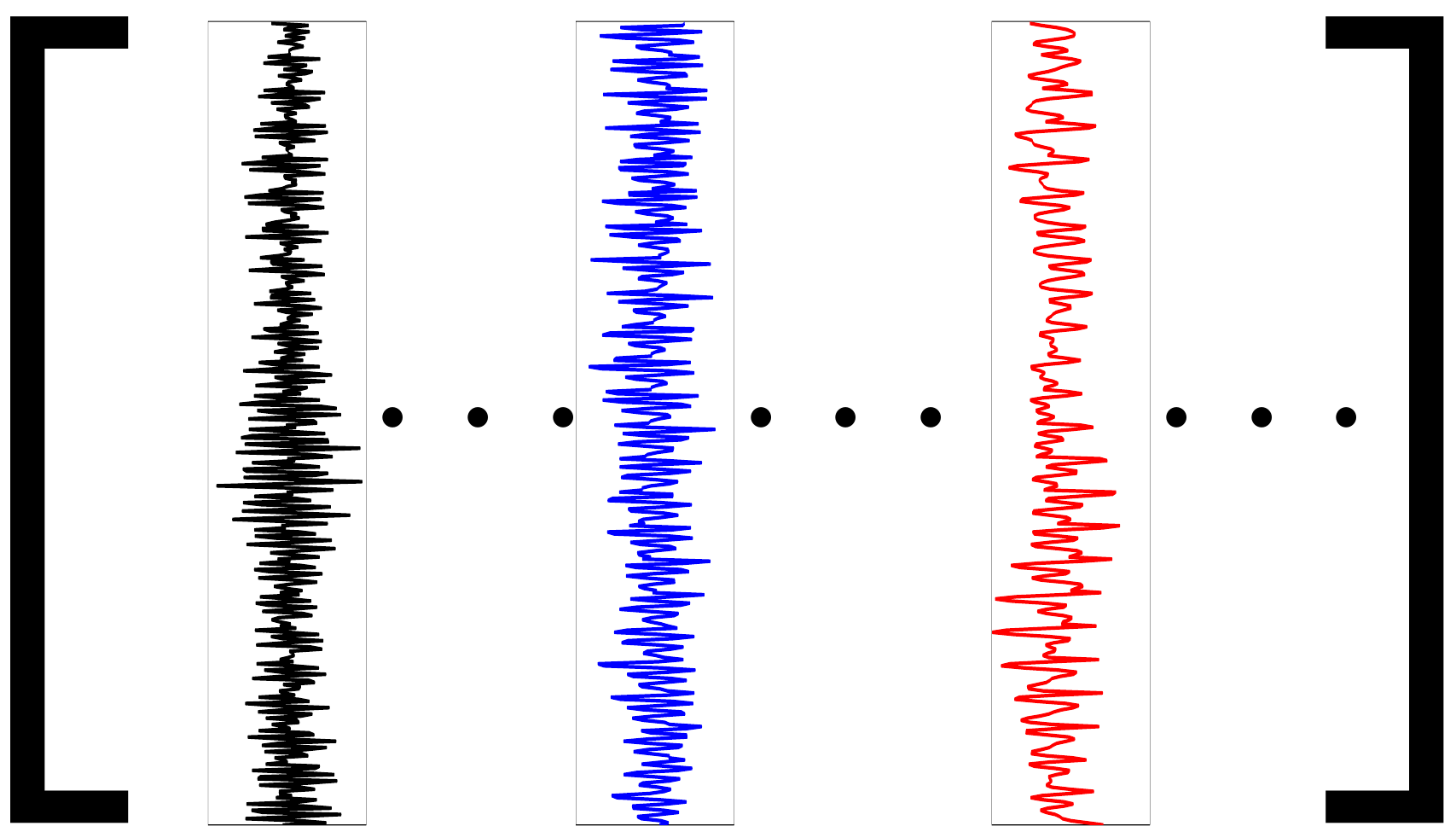}
    \end{minipage}
    \label{fig:dm_marmousi_Y}
  }
  \caption{Example of training patches vectorized into columns. (a) Three training patches of size $n_z \times n_x$ extracted from a model perturbation are (b) reshaped into three column vectors of length $N = n_zn_x$ in the training set matrix $\mathbf{Y}$.}
  \label{fig:dm_marmousi_patches_Y}
\end{figure}
\begin{figure}[htbp]
  \centering
  \subfloat[]
  {
    \begin{minipage}{0.4\linewidth}
      \centering
      \includegraphics[width=\textwidth]{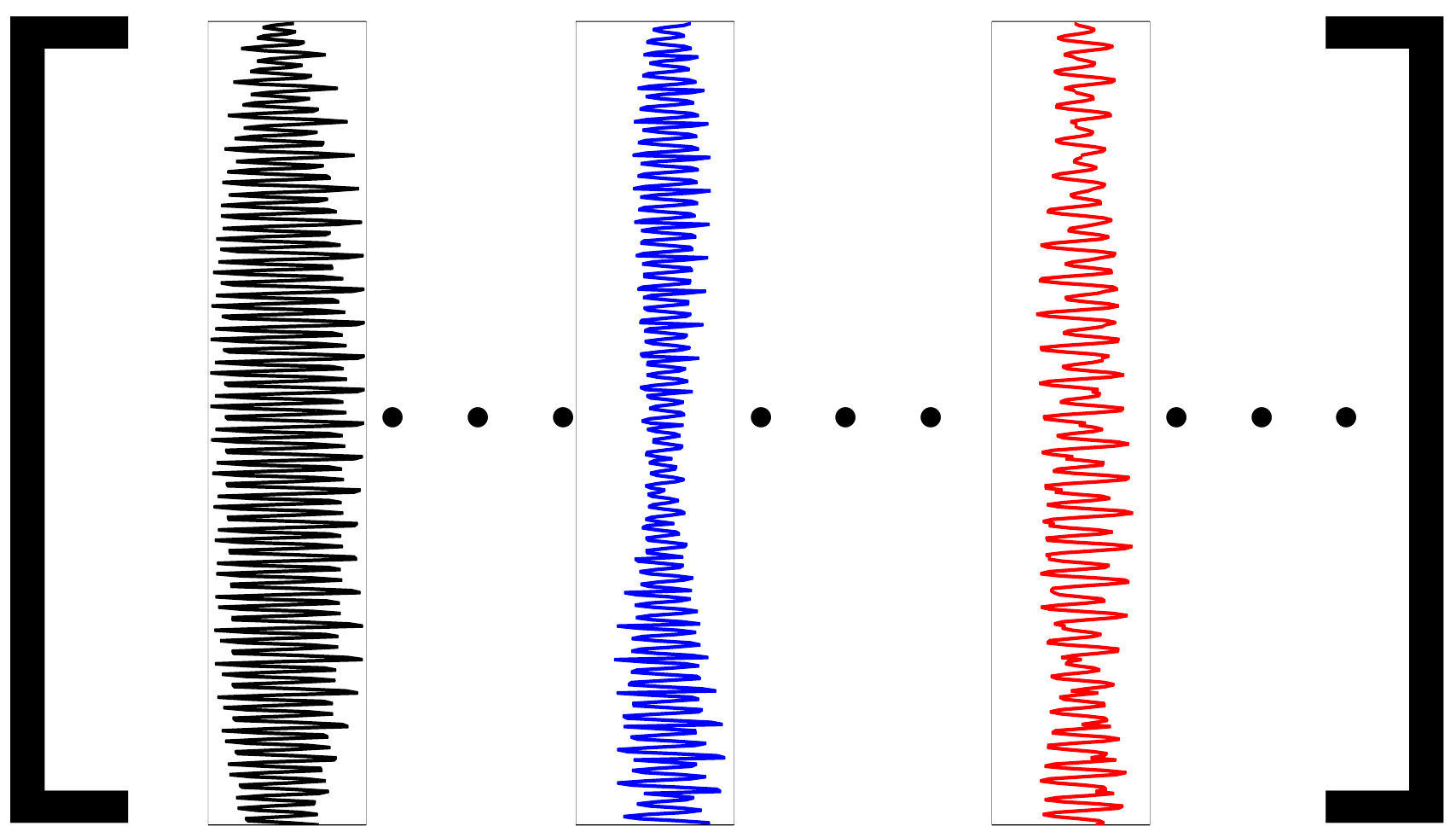}
    \end{minipage}
    \label{fig:D_illustration}
  }
  \subfloat[]
  {
    \begin{minipage}{0.57\linewidth}
      \centering
      \includegraphics[width=\textwidth]{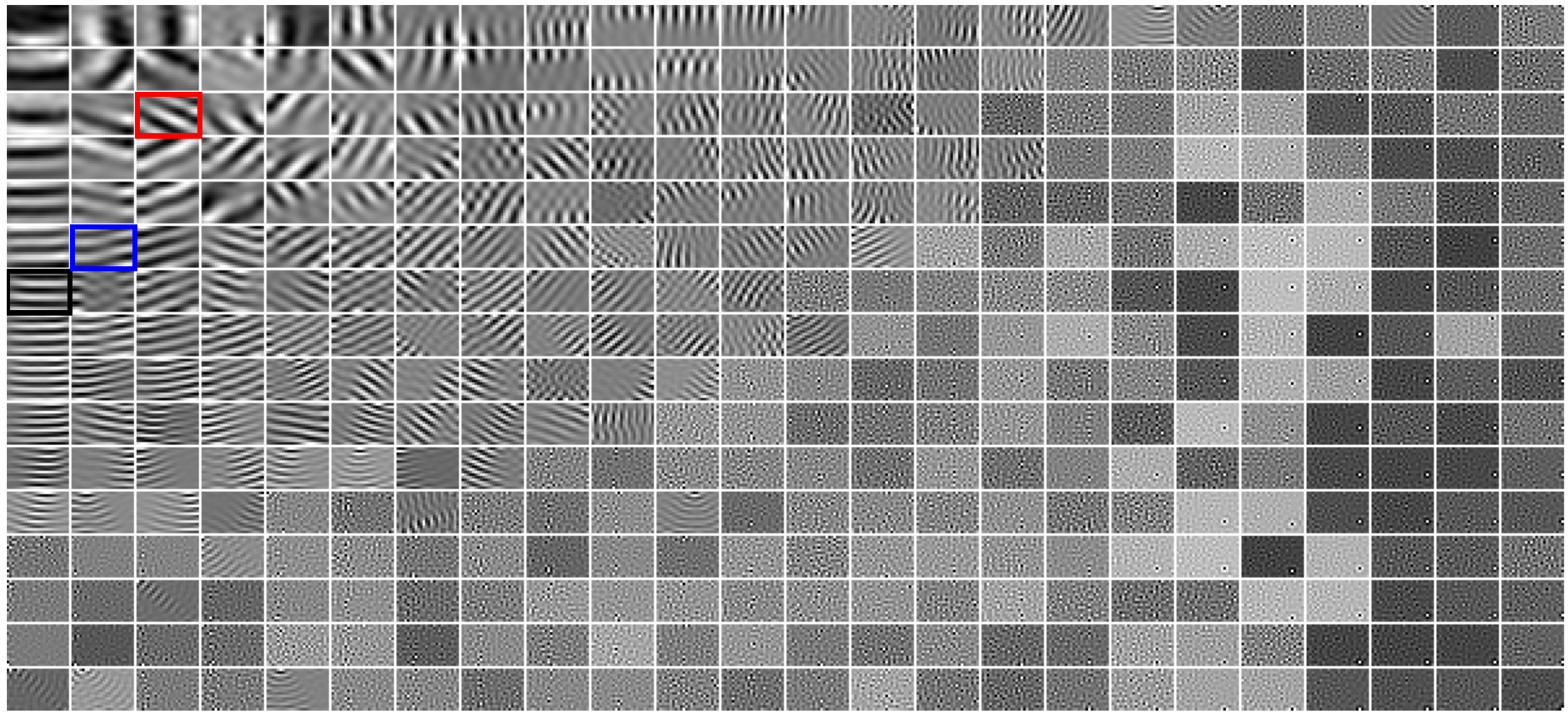}
    \end{minipage}
    \label{fig:D_example}
  }
  \caption{Dictionary atoms are (a) the columns of $\mathbf{D}$ with length $N = n_zn_x$, which can be reshaped into blocks of size $n_z \times n_x$ as in (b) for better visualization of $\mathbf{D}$.}
  \label{fig:D_illustration_example}
\end{figure}
\begin{figure}[htbp]
	\centering
	\begin{minipage}{0.67\linewidth}
	\centering
	\includegraphics[width=\textwidth]{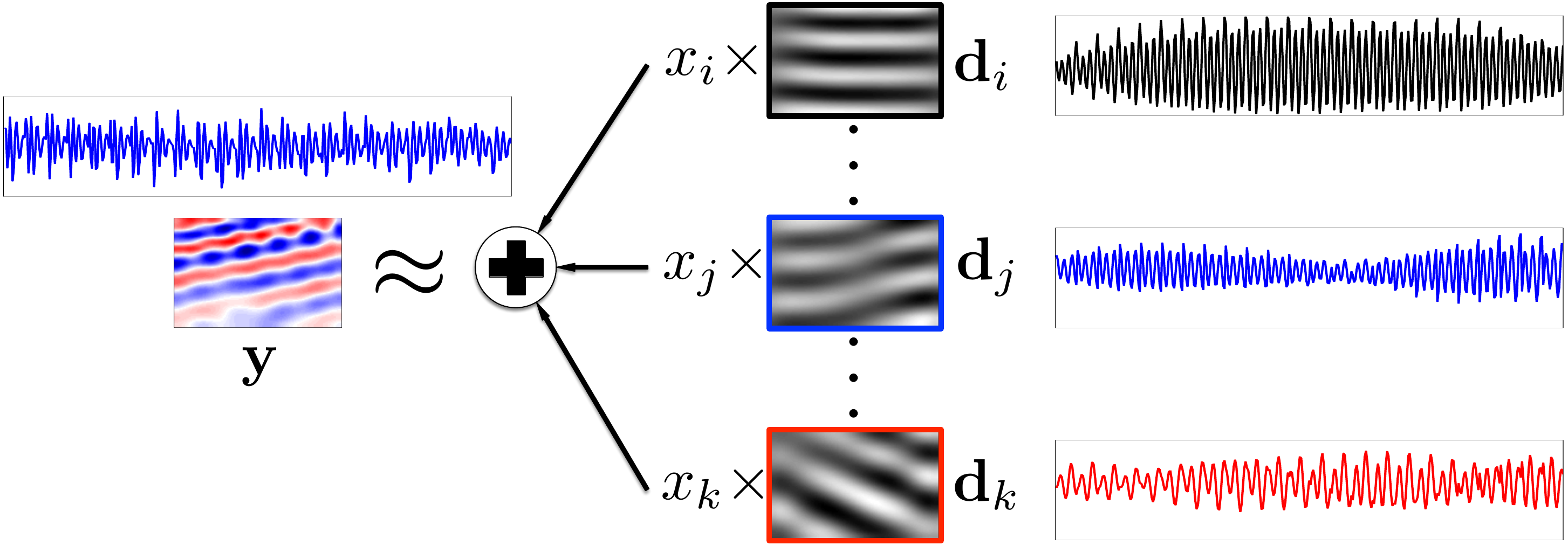}
	\end{minipage}
	\caption{Approximation of the patch $\mathbf{y}$ by a few dictionary atoms is written as a matrix-vector product $\mathbf{y}\approx \mathbf{Dx} = \sum\limits_i x_i\mathbf{d}_i$, but is equivalent to summing a few blocks (atoms) from the dictionary visualized in Figure \ref{fig:D_illustration_example}.}
\label{fig:patch_representation}
\end{figure}

A patch from the model perturbation is originally a matrix of size $n_z \times n_x$ and it is reshaped to be a column of length $N = n_zn_x$ in the training patch matrix $\mathbf{Y}$, as shown in Figure \ref{fig:dm_marmousi_patches_Y}, so we use the term patch and its reshaped vector interchangeably.

For patch-based dictionary learning, it is also a convention to present $\mathbf{D}$ by reshaping each dictionary atom of length $N = n_zn_x$ back to a matrix of size $n_z \times n_x$ for better visualization, as shown in Figure \ref{fig:D_illustration_example}. Hence one can illustrate atoms of $\mathbf{D}$ as blocks. The sparse approximation $\mathbf{y} \approx \mathbf{Dx}$ can be illustrated in Figure \ref{fig:patch_representation} with vectors or patch blocks interchangeably.

\subsection{Orthonormal Dictionary Learning}
Imposing orthonormality on $\mathbf{D}$ provides a key property to solve the $\ell_0$-norm regularized minimization problem so that the computational complexity of dictionary learning is greatly reduced. An efficient implementation of orthonormal dictionary learning has been successfully applied in natural image compression \cite{Sezer:2012aa,Sezer:2015aa} and seismic data denoising \cite{Cai:2014aa,Yu:2015aa}. By viewing $\mathbf{\delta m}$ as an image, we use the previous results to build a transform named the Sparse Orthonormal Transform (SOT) based on orthonormal dictionary learning to find a sparse representation of $\delta \mathbf{m}$ without introducing significant extra computational complexity to FWI.

Orthonormal dictionary learning seeks the square dictionary matrix $\mathbf{D} \in \mathbb{R}^{N \times N}$ that minimizes the empirical cost function $e_R(\mathbf{Y}, \mathbf{D})$ defined in \eqref{eq:empirical-cost-function} with the orthonormality constraint $\mathbf{D}^T\mathbf{D} = \mathbf{I}$
\begin{equation}
  \label{eq:ortho-dl-l0-lagrangian-optimization}
  \min\limits_{\mathbf{D}, \mathbf{X}} \frac{1}{R} \left( \| \mathbf{Y} - \mathbf{DX} \|_F^2 + \lambda\|\mathbf{X}\|_0 \right) \quad \text{s.t.} \quad \mathbf{D}^T\mathbf{D} = \mathbf{I}.
\end{equation}
Since there are two unknowns, an alternating optimization approach is used to solve \eqref{eq:ortho-dl-l0-lagrangian-optimization}. The first step is to find the sparsest representations of all columns of $\mathbf{Y} \in \mathbb{R}^{N \times R}$ over a fixed orthonormal dictionary $\mathbf{D}$, which is
\begin{equation}
  \label{eq:ortho-dl-sparse-coding}
  \hat{\mathbf{X}} = \argmin\limits_{\mathbf{X}}\left( \| \mathbf{Y} - \mathbf{DX} \|_F^2 + \lambda \|\mathbf{X}\|_0 \right).
\end{equation}
Since $\mathbf{D}$ is orthonormal, $\| \mathbf{Y} - \mathbf{DX} \|_F^2 = \| \mathbf{D}^T\mathbf{Y} - \mathbf{X} \|_F^2$, the solution to \eqref{eq:ortho-dl-sparse-coding} is straightforward by hard-thresholding the entries of $\mathbf{C} = \mathbf{D}^T\mathbf{Y}$ with the threshold $\sqrt{\lambda}$ \cite{Blumensath:2009aa,Sezer:2015aa}
\begin{equation}
  \label{eq:hard-thresholding}
  \hat{x}_{ij} = \left\{
  \begin{aligned}
  	c_{ij}, \quad &|c_{ij}| \geq \sqrt{\lambda}\\
  	0, \quad &|c_{ij}| < \sqrt{\lambda}.
  \end{aligned}
  \right.
\end{equation}
The second step in solving \eqref{eq:ortho-dl-l0-lagrangian-optimization} is to optimize the orthonormal dictionary $\mathbf{D}$ that minimizes the reconstruction error for the sparse coefficients $\hat{\mathbf{X}} = [\hat{\mathbf{x}}_1, \hat{\mathbf{x}}_2, \dots, \hat{\mathbf{x}}_R] \in \mathbb{R}^{N \times R}$, i.e.,
\begin{equation}
  \label{eq:orthogonal-procrustes}
  \hat{\mathbf{D}} = \argmin\limits_{\mathbf{D}} \|\mathbf{Y} - \mathbf{D}\hat{\mathbf{X}}\|_F^2 \quad \text{s.t.} \quad \mathbf{D}^T\mathbf{D} = \mathbf{I}.
\end{equation}
Such a problem is called the ``orthogonal Procrustes problem'' and it is proved by \cite{Schonemann:1966aa} and \cite{Sezer:2015aa} that if we define a matrix $\mathbf{P} = \hat{\mathbf{X}}\mathbf{Y}^T \in \mathbb{R}^{N \times N}$ and let $\mathbf{P} = \mathbf{U}\bm{\Sigma}\mathbf{V}^T$ denote its singular value decomposition (SVD), then the orthonormal matrix $\hat{\mathbf{D}} = \mathbf{V}\mathbf{U}^T$ solves \eqref{eq:orthogonal-procrustes}. The orthonormal dictionary $\mathbf{D}$ can thus be learned by alternating between the above two steps iteratively until the cost function $e_R(\mathbf{Y}, \mathbf{D})$ reaches a limit.

For each learning iteration, orthonormal dictionary learning needs three matrix multiplications that cost $\mathcal{O}(2RN^2+N^3)$ and one SVD operation that costs $\mathcal{O}(N^3)$ to obtain both the sparse coding and the updated dictionary.
Specifically, the model size is $N_z \times N_x$ and the training patch size is $n_z \times n_x$, where $n_z \ll N_z$, $n_x \ll N_x$, and $N = n_zn_x$. If all possible overlapping patches are used for training, then the number of training patches $R = (N_z-n_z+1)(N_x-n_x+1)$, and each training iteration costs $\mathcal{O}((n_zn_x)^2(N_z-n_z+1)(N_x-n_x+1)+(n_zn_x)^3)$. The number of training iterations does not depend on these sizes, hence the overall complexity does not change in the sense of the Big-$\mathcal{O}$ notation. Such analysis motivates the fact that the patch size should be small for dictionary learning algorithms; otherwise, the complexity would grow dramatically if large $n_z$ or $n_x$ were used.

Compared to the overcomplete dictionary learning method K-SVD \cite{Aharon:2006aa,Elad:2006aa}, the computational complexity of orthonormal dictionary learning is much less because it does not involve complex iterative algorithms such as Basis Pursuit (BP) \cite{Chen:1998aa}, Least Absolute Shrinkage and Selection Operator (LASSO) \cite{Tibshirani:1994aa} or Matching Pursuit (MP) \cite{Mallat:1993aa,Pati:1993aa,Tropp:2007aa} that have been widely used in K-SVD.

\subsection{Dictionary-based Block-wise Transform}
Dictionary learning methods use patches to form dictionaries, and therefore the learned dictionary can only be applied on the patches (see Figure \ref{fig:patch_representation}) rather than the whole image. Generally, dictionary learning is used in ``nearly-local'' problems such as signal denoising \cite{Beckouche:2014aa,Zhu:2015aa} or inpainting \cite{Yu:2015aa} where patches are processed one by one. However, the FWI problem needs to recover the entire model perturbation $\delta \mathbf{m}$ from compressive measurements, so an invertible transform that can be applied to the whole $\delta \mathbf{m}$ is required. In this section we show how to convert the local dictionaries $\mathbf{D}$ into a global transform $\bm{\mathcal{D}}$ that can be applied on the whole domain of $\delta \mathbf{m}$, and such a transform is named the Sparse Orthonormal Transform (SOT).

It is always true that the whole model perturbation $\delta \mathbf{m}$ can be exactly represented as
\begin{equation}
  \label{eq:dm-block-expression}
  \delta \mathbf{m} = \bm{\mathcal{T}}^{-1} \sum\limits_{(i, j)} \bm{\mathcal{R}}_{ij}^{\dag} \left( \bm{\mathcal{R}}_{ij} \left( \delta \mathbf{m} \right) \right)
\end{equation}
where the operator $\bm{\mathcal{R}}_{ij}$ extracts the $(i, j)$-th block of size $N = n_zn_x$ from $\delta \mathbf{m}$, and its adjoint $\bm{\mathcal{R}}_{ij}^{\dag}$ tiles the $(i, j)$-th block of size $N = n_zn_x$ back to $\delta \mathbf{m}$. The operator $\bm{\mathcal{T}} \triangleq \sum\limits_{(i, j)} \bm{\mathcal{R}}_{ij}^{\dag} \bm{\mathcal{R}}_{ij}$ is an invertible diagonal matrix if the set of all $(i,j)$ patches fully covers $\delta \mathbf{m}$ in which case $\bm{\mathcal{T}}^{-1}$ is a grid-by-grid operation. Every block $\bm{\mathcal{R}}_{ij} \left( \delta \mathbf{m} \right) \in \mathbb{R}^N$ has a sparse representation $\bm{\alpha}_{ij} \in \mathbb{R}^N$ over a learned orthonormal dictionary $\mathbf{D} \in \mathbb{R}^{N \times N}$, i.e., $\bm{\mathcal{R}}_{ij} \left( \delta \mathbf{m} \right) = \mathbf{D}\bm{\alpha}_{ij}$, so the above representation of $\delta \mathbf{m}$ can be written as

\begin{equation}
  \label{eq:dm-block-expression-dl}
  \delta \mathbf{m} = \bm{\mathcal{T}}^{-1} \sum\limits_{(i, j)} \bm{\mathcal{R}}_{ij}^{\dag}\left(\mathbf{D}\bm{\alpha}_{ij}\right).
\end{equation}
Since $\bm{\alpha}_{ij}$ has the same length as $\bm{\mathcal{R}}_{ij} \left( \delta \mathbf{m} \right)$, $\bm{\alpha}_{ij}$ fits into the same $(i, j)$-th block of a global SOT coefficient $\bm{\alpha}$ by $\bm{\alpha}_{ij} = \bm{\mathcal{R}}_{ij}(\bm{\alpha})$. Therefore, the invertible SOT can be expressed as
\begin{equation}
  \label{eq:dm-block-expression-dl-globalx}
  \begin{aligned}
  \delta \mathbf{m} &= \bm{\mathcal{T}}^{-1} \sum\limits_{(i, j)} \bm{\mathcal{R}}_{ij}^{\dag} \left(\mathbf{D} \bm{\mathcal{R}}_{ij}(\bm{\alpha}) \right) = \left[ \bm{\mathcal{T}}^{-1} \sum\limits_{(i, j)} \bm{\mathcal{R}}_{ij}^{\dag} \mathbf{D} \bm{\mathcal{R}}_{ij} \right](\bm{\alpha}) = \bm{\mathcal{D}}(\bm{\alpha})\\
  \bm{\alpha} &= \bm{\mathcal{T}}^{-1} \sum\limits_{(i, j)} \bm{\mathcal{R}}_{ij}^{\dag} \left( \mathbf{D}^T\bm{\mathcal{R}}_{ij}(\delta \mathbf{m}) \right) = \left[ \bm{\mathcal{T}}^{-1} \sum\limits_{(i, j)} \bm{\mathcal{R}}_{ij}^{\dag}\mathbf{D}^T\bm{\mathcal{R}}_{ij} \right](\delta \mathbf{m}) = \bm{\mathcal{D}}^{\dag}(\delta \mathbf{m})
  \end{aligned}
\end{equation}
where $\bm{\mathcal{D}} \triangleq \bm{\mathcal{T}}^{-1} \sum\limits_{(i, j)} \bm{\mathcal{R}}_{ij}^{\dag} \mathbf{D} \bm{\mathcal{R}}_{ij}$ is the global SOT synthesis (i.e., inverse transform) operator. 
The operator $\bm{\mathcal{D}}$ decomposes the global coefficients $\bm{\alpha}$ into blocks, reconstructs all blocks into model patches with $\mathbf{D}$, and tiles the patches back to $\delta \mathbf{m}$ at correct positions.
Its adjoint operator $\bm{\mathcal{D}}^{\dag} \triangleq \bm{\mathcal{T}}^{-1} \sum\limits_{(i, j)} \bm{\mathcal{R}}_{ij}^{\dag}\mathbf{D}^T\bm{\mathcal{R}}_{ij}$ is the global SOT analysis (i.e., transform) operator that decomposes the whole model perturbation $\delta \mathbf{m}$ into patches, converts all patches into coefficient blocks with $\mathbf{D}^T$, and concatenates them into the global coefficient vector $\bm{\alpha}$.

Choosing non-overlapping blocks from $\delta \mathbf{m}$ is preferred so that the length of the global SOT coefficient $\bm{\alpha}$ is not longer than that of $\delta \mathbf{m}$, and more importantly, $\bm{\mathcal{D}}\bm{\mathcal{D}}^{\dag} = \bm{\mathcal{D}}^{\dag}\bm{\mathcal{D}} = \mathbf{I}$; otherwise, for the overlapping patch case, $\bm{\mathcal{D}}$ is overcomplete and nonorthogonal. Similar to the computational complexity analysis for orthonormal dictionary learning, with the model size being $N_z \times N_x$ and the patch size being $n_z \times n_x$, where $N_z$ is divisible by $n_z$ and $N_x$ is divisible by $n_x$, then the computational complexity of applying SOT or inverse SOT is $\mathcal{O}(n_zn_xN_zN_x)$ since each patch transform costs $\mathcal{O}((n_zn_x)^2)$ and there are $(N_zN_x)/(n_zn_x)$ non-overlapping model patches.

\subsection{Choice of Lagrange Parameter $\lambda$}
\begin{figure}[htbp]
  \centering
  \subfloat[]
  {
    \begin{minipage}{0.5\linewidth}
      \centering
      \includegraphics[width=\textwidth]{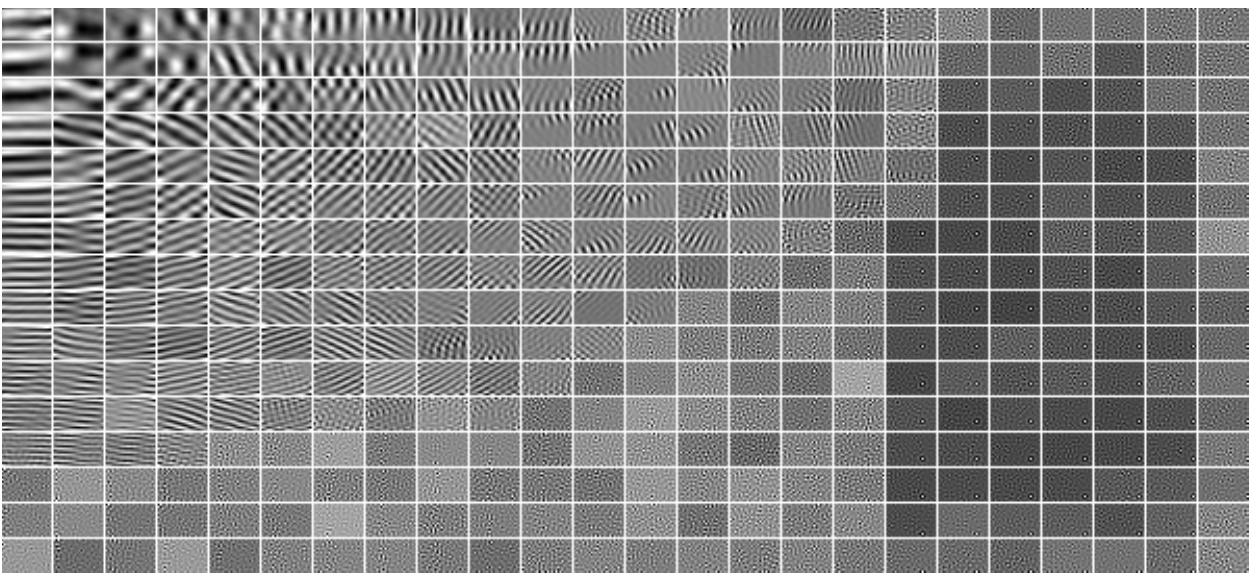}
    \end{minipage}
    \label{fig:D_sqrtlambda01}
  }
  \subfloat[]
  {
    \begin{minipage}{0.5\linewidth}
      \centering
      \includegraphics[width=\textwidth]{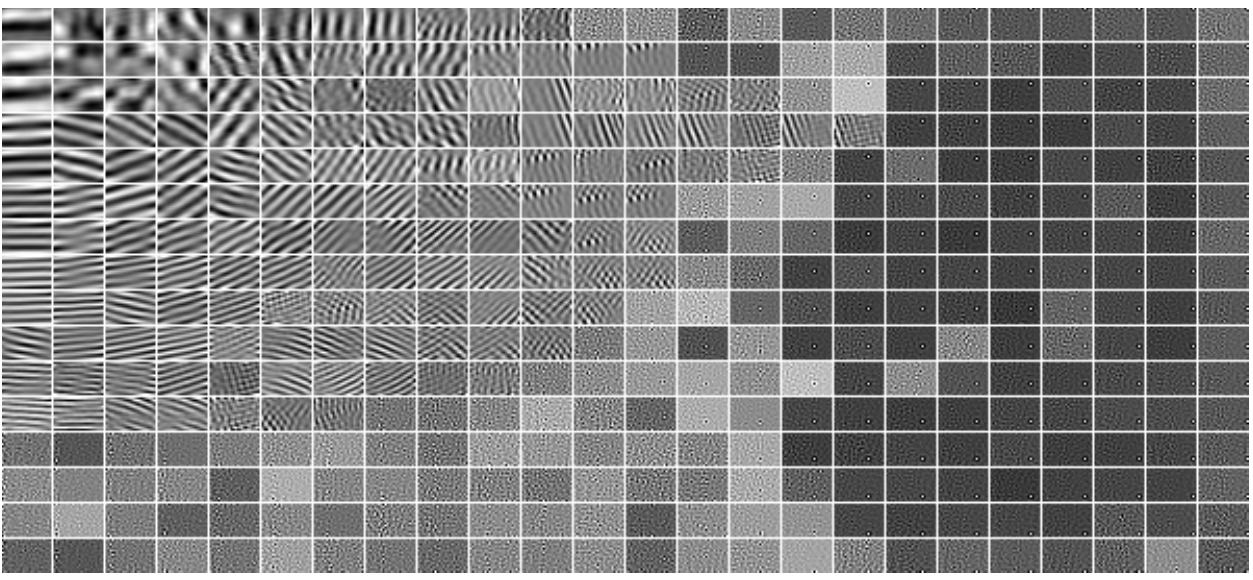}
    \end{minipage}
    \label{fig:D_sqrtlambda02}
  }\\
  \subfloat[]
  {
    \begin{minipage}{0.5\linewidth}
      \centering
      \includegraphics[width=\textwidth]{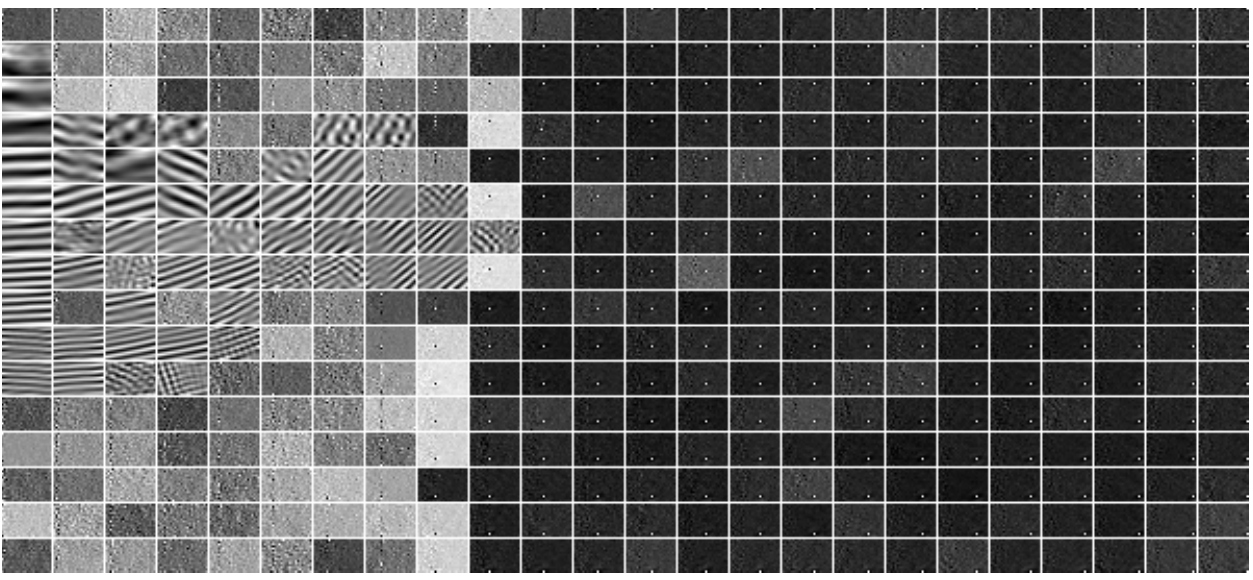}
    \end{minipage}
    \label{fig:D_sqrtlambda05}
  }
  \subfloat[]
  {
    \begin{minipage}{0.5\linewidth}
      \centering
      \includegraphics[width=\textwidth]{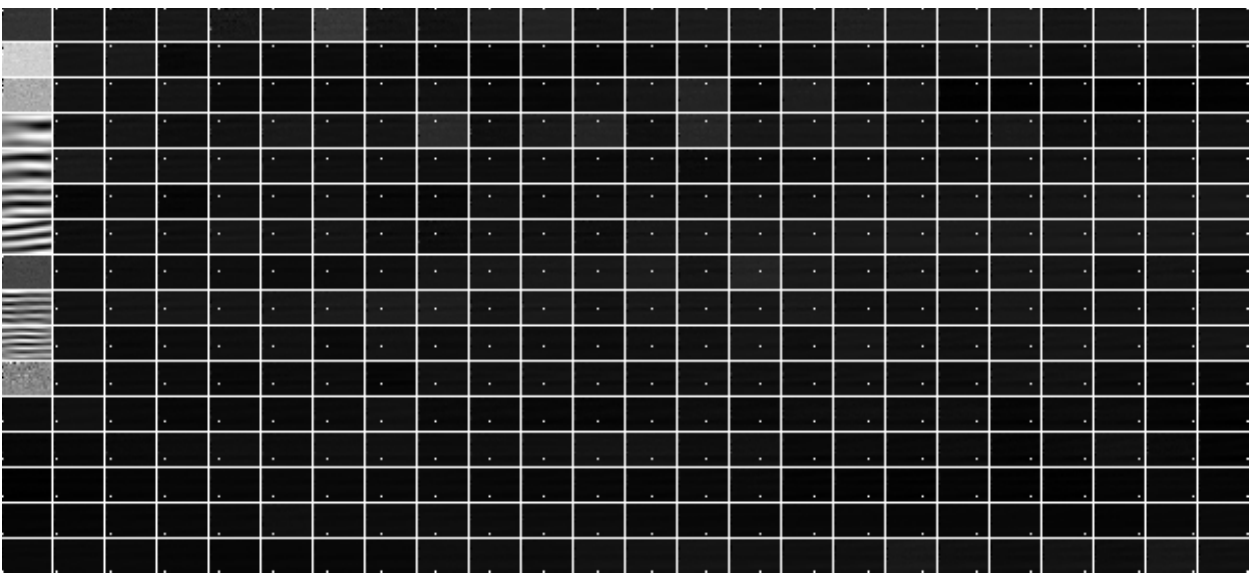}
    \end{minipage}
    \label{fig:D_sqrtlambda08}
  }
  \caption{Dictionaries $\mathbf{D} \in \mathbb{R}^{384 \times 384}$ trained with different values of $\lambda$. (a) $\lambda = 0.1$, (b) $\lambda = 0.2$, (c) $\lambda = 0.5$, (d) $\lambda = 0.8$.}
  \label{fig:D-with-different-lambdas}
\end{figure}

The value of Lagrange parameter $\lambda$ in orthonormal dictionary learning \eqref{eq:ortho-dl-l0-lagrangian-optimization} controls sparsity because it determines the design of dictionaries for a particular sparsity level and shapes the atoms of $\mathbf{D}$, as transform coefficients with absolute values smaller than $\sqrt{\lambda}$ are hard-thresholded to zero. A small $\lambda$ would yield marginal change of $\mathbf{D}$ after each iteration since most elements in $\mathbf{C} = \mathbf{D}^T\mathbf{Y}$ would remain unchanged for $\hat{\mathbf{X}}$.
The extreme case is when $\lambda = 0$, then $\hat{\mathbf{X}} = \mathbf{C} = \mathbf{D}^T\mathbf{Y}$, and it is trivial to solve \eqref{eq:orthogonal-procrustes} to obtain $\hat{\mathbf{D}} = \mathbf{D}$ which does not change at all. On the contrary, if $\lambda$ were large, then most elements in $\mathbf{C} = \mathbf{D}^T\mathbf{Y}$ would be hard-thresholded to zeros for $\hat{\mathbf{X}}$, and $\mathbf{P} = \hat{\mathbf{X}}\mathbf{Y}^T = \mathbf{U}\bm{\Sigma}\mathbf{V}^T$ would be a low-rank matrix, resulting in many atoms in $\hat{\mathbf{D}} = \mathbf{V}\mathbf{U}^T$ resembling the trivial standard basis.
The extreme case is when $\lambda = 1$, which gives $\text{rank}(\mathbf{P}) = 0$, and $\hat{\mathbf{D}}$ degrades to $\mathbf{I}$. Some examples of $\mathbf{D} \in \mathbb{R}^{384 \times 384}$ learned with different values of $\lambda$ are shown in Figure \ref{fig:D-with-different-lambdas}, where each atom of $\mathbf{D}$ is reshaped into a 2D block of size $16 \times 24$ for visualization.

Nonlinear approximation (NLA) can be used to verify the sparse representation capability of the learned orthonormal dictionary $\mathbf{D}$ (and the global SOT synthesis operator $\bm{\mathcal{D}}$ as well) for a $\delta \mathbf{m}$. It keeps the $l$ largest-magnitude coefficients from $\bm{\alpha}$ as $\widetilde{\bm{\alpha}}$, and then evaluates the normalized mean square error (NMSE) of the reconstruction
\begin{equation}
  \label{eq:nla-nmse}
  \text{NMSE}(\delta \mathbf{m}; \bm{\mathcal{D}}, l) = 1 - \left\|\frac{ \delta \mathbf{m} - \bm{\mathcal{D}}(\widetilde{\bm{\alpha}}) }{\delta \mathbf{m} - \text{mean}(\delta \mathbf{m}) }\right\|_2^2,
\end{equation}
which varies between $-\infty$ (bad fit) to 1 (perfect fit).
\begin{figure}[htbp]
  \centering
  \subfloat[]
  {
    \begin{minipage}{0.9\linewidth}
      \centering
      \includegraphics[width=\textwidth]{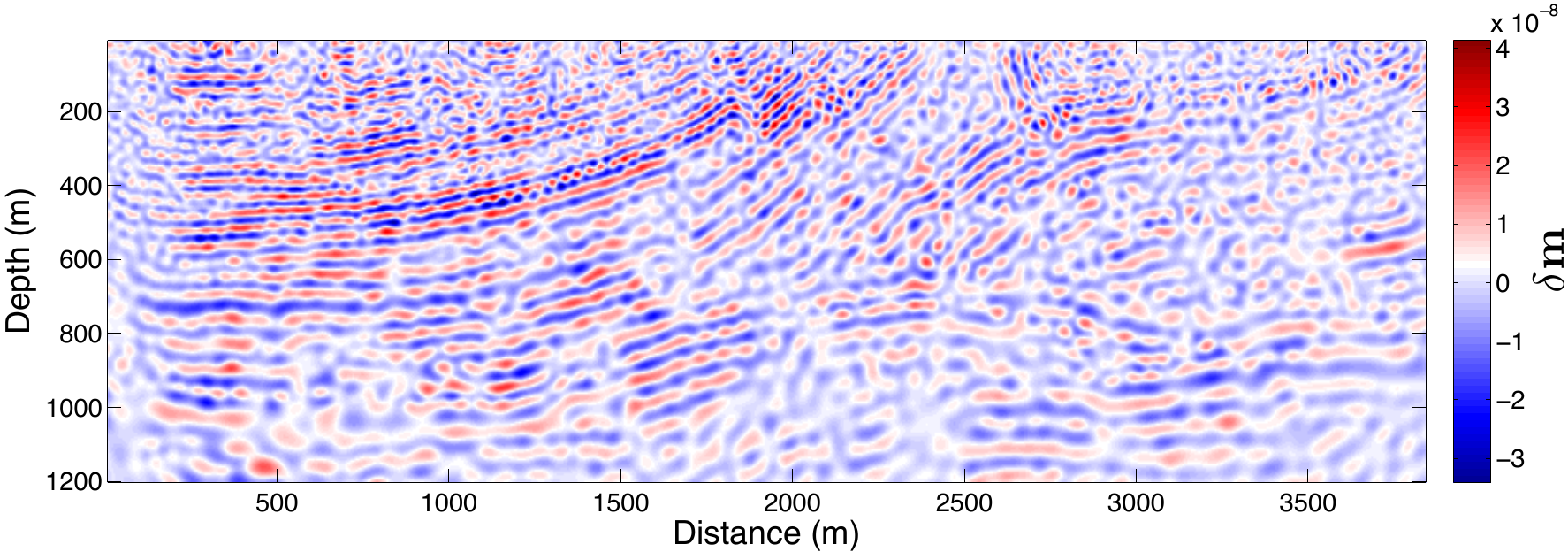}
    \end{minipage}
    \label{fig:dm_marmousi_train}
  }\\
  \subfloat[]
  {
    \begin{minipage}{0.9\linewidth}
      \centering
      \includegraphics[width=\textwidth]{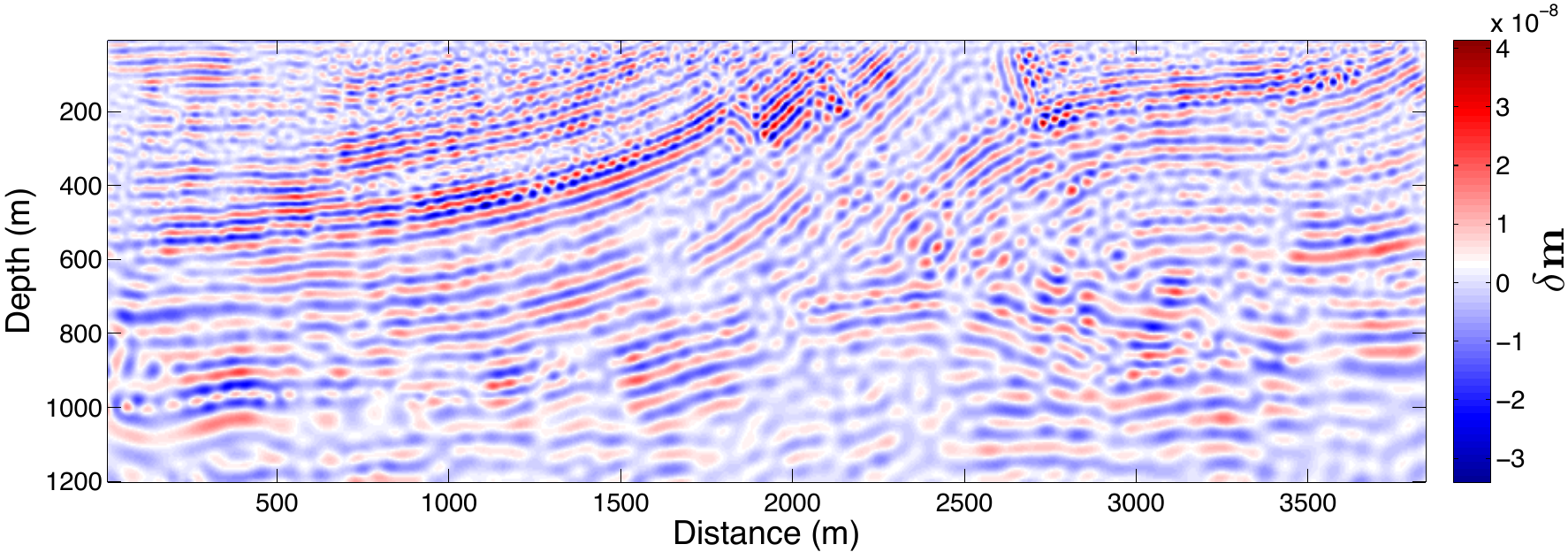}
    \end{minipage}
    \label{fig:dm_marmousi_test}
  }
  \caption{Two model perturbations $\delta \mathbf{m}$ extracted from consecutive FWI iterations and used for the nonlinear approximation test, (a) for training, (b) for testing.}
  \label{fig:nla-test-training}
\end{figure}
\begin{figure}[htbp]
  \centering
  \subfloat[]
  {
    \begin{minipage}{0.5\linewidth}
      \centering
      \includegraphics[width=\textwidth]{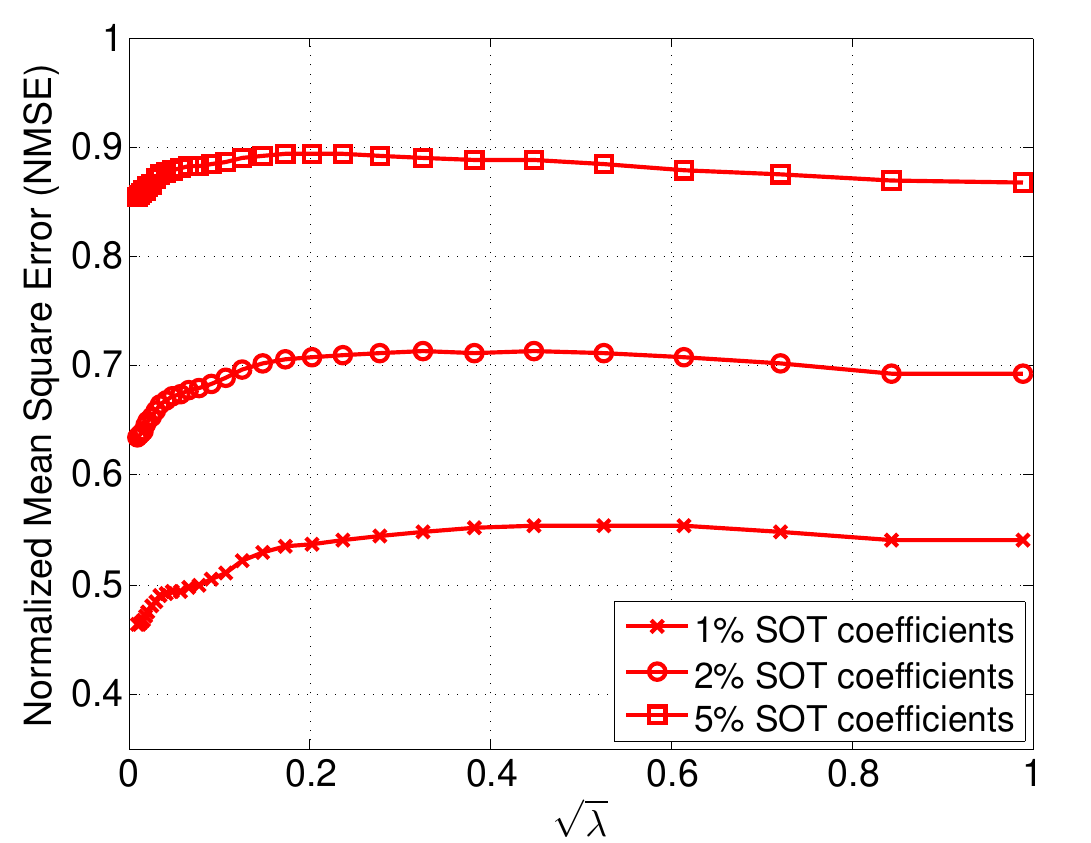}
    \end{minipage}
    \label{fig:NLA_marmousi_dm_10x12}
  }\\
  \subfloat[]
  {
    \begin{minipage}{0.5\linewidth}
      \centering
      \includegraphics[width=\textwidth]{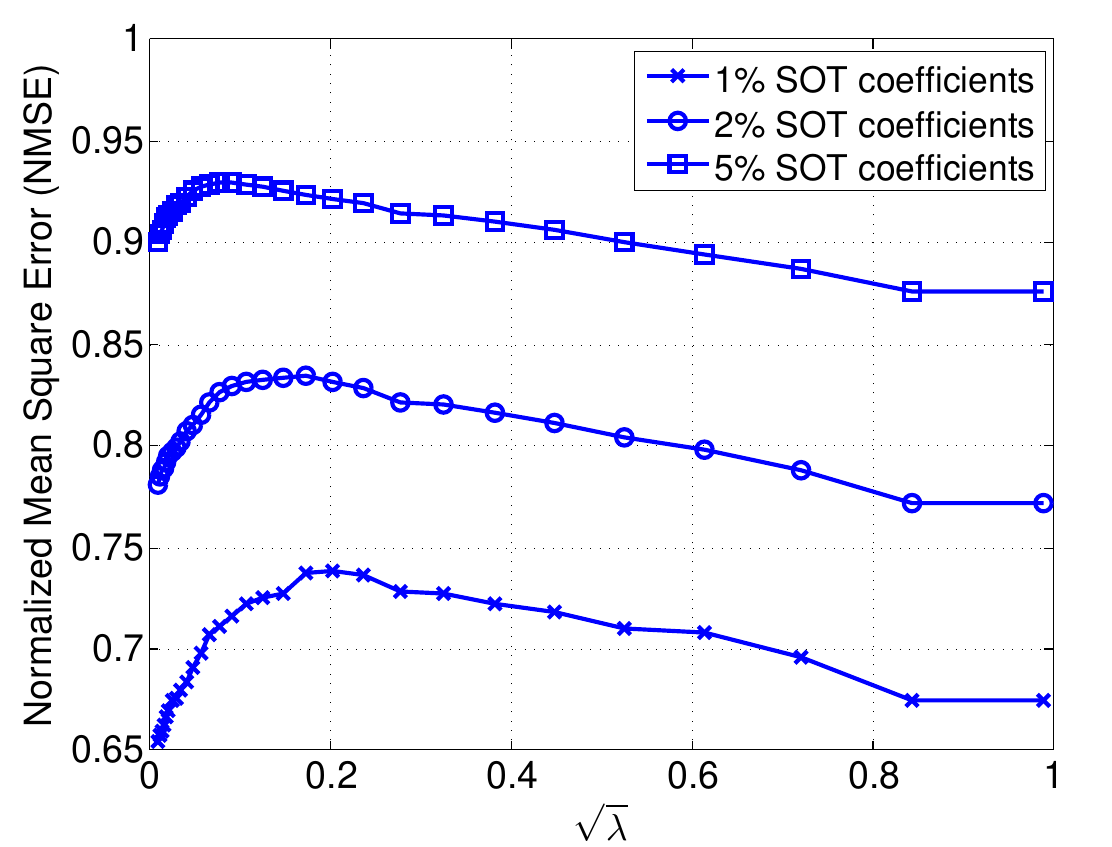}
    \end{minipage}
    \label{fig:NLA_marmousi_dm_20x24}
  }\\
  \subfloat[]
  {
    \begin{minipage}{0.5\linewidth}
      \centering
      \includegraphics[width=\textwidth]{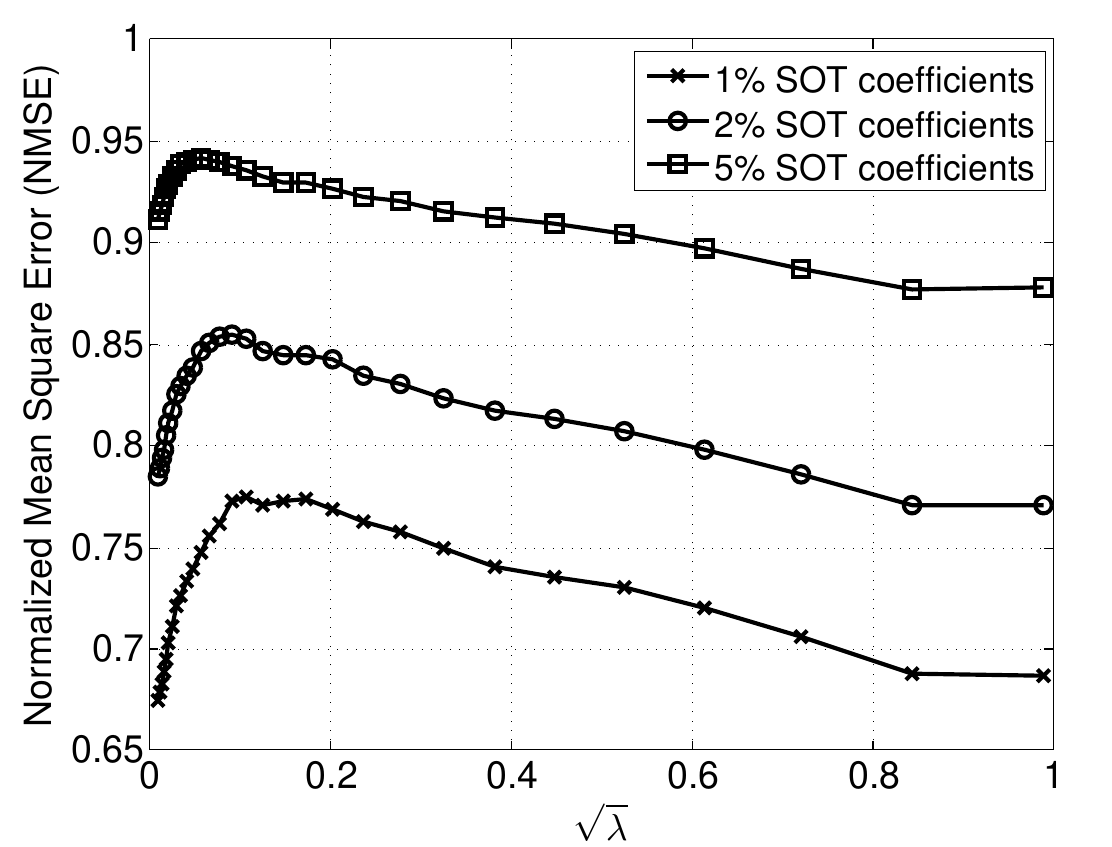}
    \end{minipage}
    \label{fig:NLA_marmousi_dm_30x32}
  }
  \caption{NLA performance curves of keeping 1\%, 2\% and 5\% largest-magnitude coefficients using (a) dictionary $\mathbf{D} \in \mathbb{R}^{120 \times 120}$ with patch size $10 \times 12$, (b) dictionary $\mathbf{D} \in \mathbb{R}^{480 \times 480}$ with patch size $20 \times 24$, (c) dictionary $\mathbf{D} \in \mathbb{R}^{960 \times 960}$ with patch size $30 \times 32$.}
  \label{fig:nla-results}
\end{figure}

The Lagrange parameter $\lambda$ is usually related to an approximate noise level if dictionary learning is applied in the denoising problems \cite{Elad:2006aa}. However, its selection in such a CS-based sparsity recovery problem still remains an open problem \cite{Starck:2015aa} so we have to choose $\lambda$ empirically. A simplified experiment can be conducted to compare the NLA performance of learned orthonormal dictionaries $\mathbf{D}$ trained with different values of $\lambda$. In addition, different dictionaries $\mathbf{D}$ are learned from training patches of different sizes extracted from a training model perturbation shown in Figure \subref*{fig:dm_marmousi_train} and tested on a testing model perturbation shown in Figure \subref*{fig:dm_marmousi_test}. The NLA performance curves that indicate the relationship between NMSE and $\lambda$ for different sparsity levels $l$ (1\%, 2\% and 5\% of largest-magnitude coefficients) are shown in Figure \ref{fig:nla-results}. In Figures \subref*{fig:NLA_marmousi_dm_10x12} to \subref*{fig:NLA_marmousi_dm_30x32}, we see that the optimal $\lambda$ that yields the highest NMSE depends on both the sparsity level $l$ and the patch size $n_z \times n_x$, and the optimal $\lambda$ tends to decrease if either the sparsity level $l$ or the patch size $n_z \times n_x$ increases. These results indicate that $\lambda \in (0.15^2, 0.25^2)$ is expected to deliver good reconstructions for reasonable sparsity levels and patch sizes.

\subsection{Online Orthonormal Dictionary Learning}
The above orthonormal dictionary learning algorithm takes the training patch set as a whole so that a dictionary $\mathbf{D}$ could be learned offline and would remain static as a sparse representation. Generally speaking, such an offline approach cannot effectively handle very large training sets, or dynamic training sets that vary over time. In practice, FWI is an iterative problem where the optimized $\delta \mathbf{m}_k$ that offers training patches is changing over iterations. Therefore, to exploit the availability of new training patches from $\delta \mathbf{m}_k$, we propose an online approach for orthonormal dictionary learning that minimizes the following expected cost function
\begin{equation}
  \label{eq:expected-cost-function}
  \begin{aligned}
  e(\mathbf{D}) &\triangleq \mathbb{E}_{\mathbf{y}}\left[ \| \mathbf{y} - \mathbf{D}\mathbf{x}\|_2^2 + \lambda\|\mathbf{x}\|_0 \right] 
  = \lim\limits_{R \rightarrow \infty} e_R(\mathbf{Y}, \mathbf{D}) \quad \text{almost surely}.
  \end{aligned}
\end{equation}
Rather than spending too much effort on accurately minimizing the empirical cost function $e_R(\mathbf{Y}, \mathbf{D})$ in \eqref{eq:empirical-cost-function}, \cite{Bousquet:2007aa} suggest minimizing $e(\mathbf{D})$ since $e_R(\mathbf{Y}, \mathbf{D})$ is merely an approximation of $e(\mathbf{D})$. Minimizing $e(\mathbf{D})$ does not rely on the number of patches $R$, but instead on the (unknown) stochastic characteristics of the training patches. The online approach learns a new dictionary $\mathbf{D}_k$ every time a new $\delta \mathbf{m}_{k-1}$ is ready, and the sequence of learned dictionaries can adapt to the variations of patches in later iterations. 

Algorithm \ref{alg:online-ortho-dl} summarizes a general version of the online orthonormal dictionary learning method in which the training examples $\mathbf{y}$ are drawn from a data stream source. In particular, at the end of the $(k-1)$-th FWI iteration, we first draw a batch of $R$ training patches from $\delta \mathbf{m}_{k-1}$, each of size $n_z \times n_x$, to form the matrix $\mathbf{Y}_{k-1} \in \mathbb{R}^{N \times R}$ and normalize each column into the range of $[0, 1]$. Then we use the previous dictionary $\mathbf{D}_{k-1} \in \mathbb{R}^{N \times N}$ as a warm start to represent $\mathbf{Y}_{k-1}$ with sparse coefficients $\mathbf{X}_{k-1} \in \mathbb{R}^{N \times R}$ by hard thresholding with $\sqrt{\lambda}$, and obtain the updated dictionary $\mathbf{D}_k$ for the following $k$-th FWI iteration with the orthonormal matrices of singular vectors of $\mathbf{P}_k \in \mathbb{R}^{N \times N}$ that accumulates $\mathbf{X}_i\mathbf{Y}_i^T$ for $i = 0, 1, \ldots, k-1$. Essentially, the above two alternating steps for learning $\mathbf{D}_k$ keep reducing the value of the function
\begin{equation}
  \label{eq:surrogate-expected-cost-function}
  \hat{e}_k(\mathbf{D}) \triangleq \frac{1}{kR} \sum\limits_{i=0}^{k-1} \left( \| \mathbf{Y}_i - \mathbf{D}\mathbf{X}_i\|_F^2 + \lambda\|\mathbf{X}_i\|_0 \right)
\end{equation}
which, in effect, takes training patches of all previously optimized model perturbations $\{\delta \mathbf{m}_i\}_{i=0}^{k-1}$ into account. It is proved in \cite{Bousquet:2007aa} that $\hat{e}_k(\mathbf{D})$ converges to $e(\mathbf{D})$ with probability one if $k$ is sufficiently large and, therefore, the online orthonormal dictionary learning converges to a stationary point.
\begin{algorithm}
  \caption{Online Orthonormal Dictionary Learning}
  \label{alg:online-ortho-dl}
  \SetAlgoLined
  \SetAlFnt{\small}
  \SetKwInOut{KwInit}{Initialization}
  \AlFnt
  \KwIn{a data source from which input data $\mathbf{y} \in \mathbb{R}^{N}$ are drawn, initial orthonormal dictionary $\mathbf{D}_0 \in \mathbb{R}^{N \times N}$, Lagrange multiplier $\lambda$, number of update iterations $T$, mini-batch size $R$, Cauchy's convergence error bound $\epsilon$}
  \KwOut{learned orthonormal dictionary $\mathbf{D}_K$, sparse representation matrix $\mathbf{X}_K$}
  \KwInit{$\mathbf{P}_0 = \mathbf{0}$}
  \For{$k = 1$ \KwTo $K$}
  {
  	Draw a mini-batch of data $\mathbf{Y}_{k-1} \triangleq [\mathbf{y}^{(k-1)}_1, \mathbf{y}^{(k-1)}_2, \dots, \mathbf{y}^{(k-1)}_R]$ from a data source\;
  	Normalization: $\mathbf{y}^{(k-1)}_i \leftarrow \mathbf{y}^{(k-1)}_i / \|\mathbf{y}^{(k-1)}_i\|_2$, $\forall i = 1, \dots, R$\;
  	$\mathbf{D} = \mathbf{D}_{k-1}$\;
  	\While{$\| \mathbf{Y}_{k-1} - \mathbf{D}\mathbf{X}_{k-1} \|_F^2 + \lambda\|\mathbf{X}_{k-1}\|_0$ not converged with error bound $\epsilon$}
  	{
	  	$\mathbf{C}_{k-1} = \mathbf{D}^T\mathbf{Y}_{k-1}$\;
	    $x^{(k-1)}_{ij} = \left\{
	  		\begin{aligned}
	  			c^{(k-1)}_{ij}, \quad &|c^{(k-1)}_{ij}| \geq \sqrt{\lambda}\\
	  			0, \quad &|c^{(k-1)}_{ij}| < \sqrt{\lambda}
	  		\end{aligned}
	  	\right.$\;
	  	$\mathbf{P}_k = \mathbf{P}_{k-1} + \mathbf{X}_{k-1}\mathbf{Y}_{k-1}^T$\;
	  	$\mathbf{U}\bm{\Sigma}\mathbf{V}^T = \mathbf{P}_k$ \tcp*{Compute SVD}
	  	$\mathbf{D} = \mathbf{V}\mathbf{U}^T$\;
  	}
  	$\mathbf{D}_k = \mathbf{D}$\;
  }
\end{algorithm}

\section{Full Waveform Inversion with Dictionary-based Sparsity Regularization}
Recall the LASSO optimization problem of each Gauss-Newton iteration $k = 0, 1, 2, \dots$ under the CS framework:
\begin{equation}
  \tag{\ref{eq:fwi-gauss-newton-random} revisited}
  \min\limits_{\bm{\alpha}} \left\{ J_k^{(\text{W})}(\bm{\alpha}) \triangleq \frac{1}{2} \|\mathbf{W}_k\delta \mathbf{d}_k - \mathbf{W}_k\mathbf{J}_k\bm{\mathcal{D}}_k(\bm{\alpha})\|_2^2 \right\} \quad \text{s.t.} \quad \|\bm{\alpha}\|_1 \leq \tau_k
\end{equation}
where right now $\bm{\mathcal{D}}_k$ is the SOT synthesis operator based on the orthonormal dictionary $\mathbf{D}_k$ trained for the $k$-th FWI iteration and $\bm{\alpha}$ is the SOT coefficient vector. If $\bm{\alpha}_k$ minimizes the objective function $J_k^{(\text{W})}(\bm{\alpha})$ in \eqref{eq:fwi-gauss-newton-random}, then its inverse SOT recovers the optimal model perturbation $\delta \mathbf{m}_k$ via
\begin{equation}
  \label{eq:dm-Dx}
  \delta \mathbf{m}_k = \bm{\mathcal{D}}_k(\bm{\alpha}_k).
\end{equation}

In \eqref{eq:fwi-gauss-newton-random} the subsampling matrix $\mathbf{W}_k$ must be designed. The construction of $\mathbf{W}_k$ can take advantage of the linearity property of wave equations. Because the computational cost of an FWI iteration is proportional to the number of seismic wave modeling processes with different source functions, a random source encoding method has been proposed to combine a large number of sequential sources with random weights into only a few simultaneous shots. These simultaneous shots are named \textit{supershots} in the literature \cite{Romero:2000aa,Krebs:2009aa,Ben-Hadj-Ali:2011aa}. Due to the linear relationship between a seismic wavefield and its source function, the random weights can be incorporated into $\mathbf{W}_k$ and become the key to subsampling. Summing many individual sources into a few simultaneous shots introduces crosstalk artifacts, but crosstalk can be mitigated during the inversion process by enforcing a sparsity constraint in the SOT domain.

\subsection{Random Source Encoding -- The Supershot Method}
Modeling supershots and their corresponding wavefields in the frequency domain is a common practice in recent research \cite{Ben-Hadj-Ali:2011aa}. Consider a conventional seismic survey with $N_s$ sequential shots, in which each shot produces a 2D acoustic seismic wavefield $p_j(\mathbf{x}, t; \mathbf{x}_{s_j})$ modeled by a frequency-domain PDE
\begin{equation}
  \label{eq:acoustic-wave-equation-freq-1shot}
  \left( -m(\mathbf{x})\omega^2 - \nabla^2 \right) p_j(\mathbf{x}; \omega, \mathbf{x}_{s_j}) = f(\omega)\delta(\mathbf{x} - \mathbf{x}_{s_j}), \quad \forall j = 1, \dots, N_s
\end{equation}
where $f(\omega)\delta(\mathbf{x} - \mathbf{x}_{s_j})$ is the point source function excited at location $\mathbf{x}_{s_j}$.

A supershot can be simulated as a linear combination of simultaneous shots with random coefficients, i.e.,
\begin{equation}
  \label{eq:supershot-source-freq}
  f_i^{(\text{s})}(\mathbf{x}; \omega) \triangleq \sum\limits_{j=1}^{N_s} w_{ij}(\omega)f(\omega)\delta(\mathbf{x} - \mathbf{x}_{s_j}), \quad \forall i = 1, \dots, N_s'
\end{equation}
where $i = 1, \dots, N_s'$ is the index of the supershot ($N_s' \ll N_s$) and $w_{ij}(\omega)$ is a frequency-dependent random Gaussian variable that encodes $f(\omega)\delta(\mathbf{x} - \mathbf{x}_{s_j})$ for the $i$-th supershot. By replacing the right-hand side of the PDE \eqref{eq:acoustic-wave-equation-freq-1shot} with a supershot,
\begin{equation}
  \label{eq:acoustic-wave-equation-supershot}
  \left( -m(\mathbf{x})\omega^2 - \nabla^2 \right) p_i^{(\text{s})}(\mathbf{x}; \omega) = f_i^{(\text{s})}(\mathbf{x}; \omega),
\end{equation}
the excited wavefield becomes
\begin{equation}
  \label{eq:supershot-wavefield-freq}
  p_i^{(\text{s})}(\mathbf{x}; \omega) = \sum\limits_{j=1}^{N_s} w_{ij}(\omega) p_j(\mathbf{x}; \omega, \mathbf{x}_{s_j}), \quad \forall i = 1, \dots, N_s'
\end{equation}
due to the linearity of the wave equation.

\begin{figure}[htbp]
  \centering
  \subfloat[]
  {
    \begin{minipage}{0.9\linewidth}
      \centering
      \includegraphics[width=\textwidth]{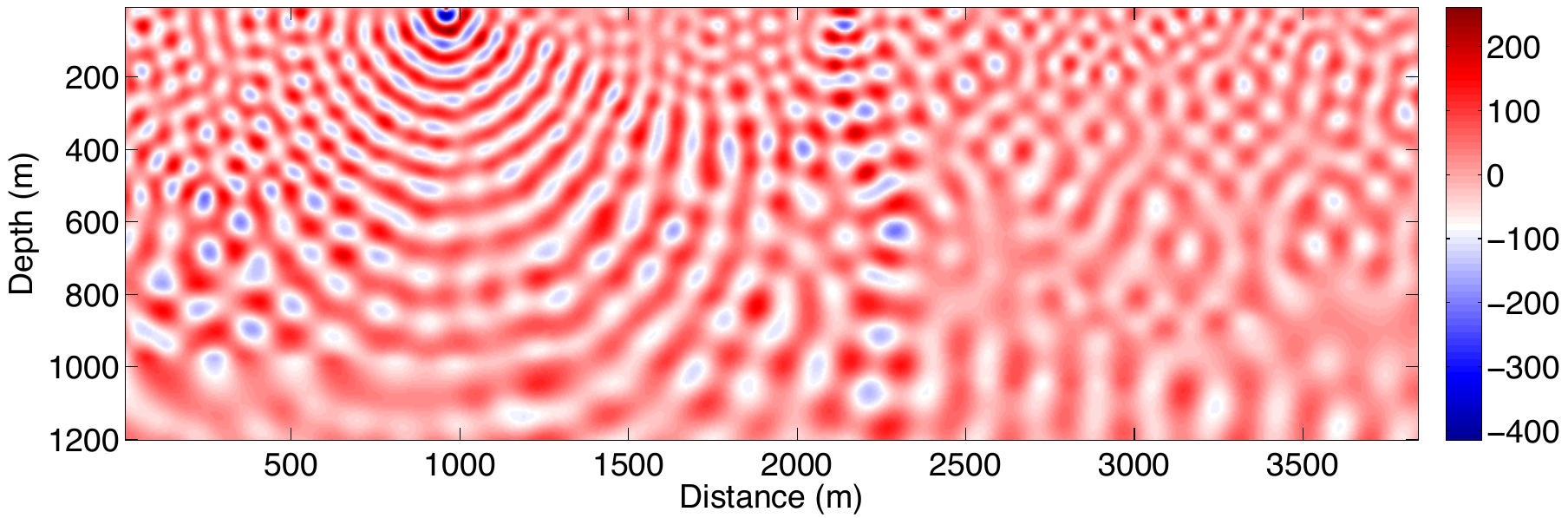}
    \end{minipage}
    \label{fig:snapshotTrueFreq_xs96_marmousi}
  }\\
  \subfloat[]
  {
    \begin{minipage}{0.9\linewidth}
      \centering
      \includegraphics[width=\textwidth]{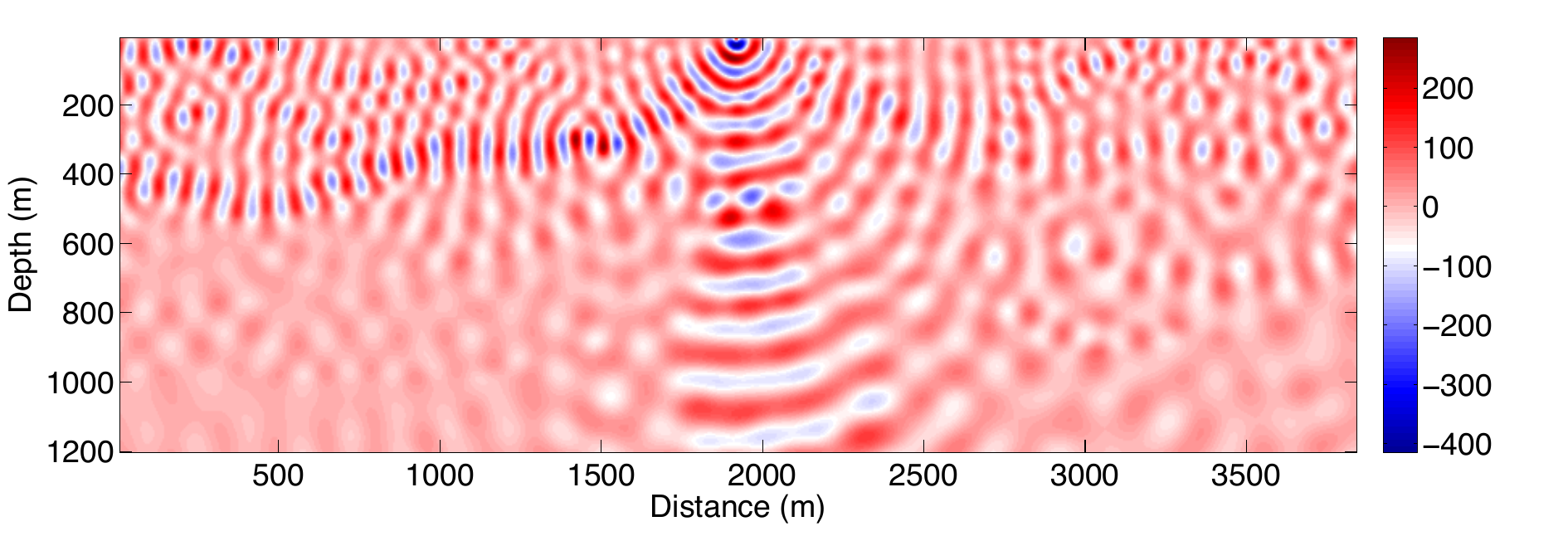}
    \end{minipage}
    \label{fig:snapshotTrueFreq_xs192_marmousi}
  }\\
  \subfloat[]
  {
    \begin{minipage}{0.9\linewidth}
      \centering
      \includegraphics[width=\textwidth]{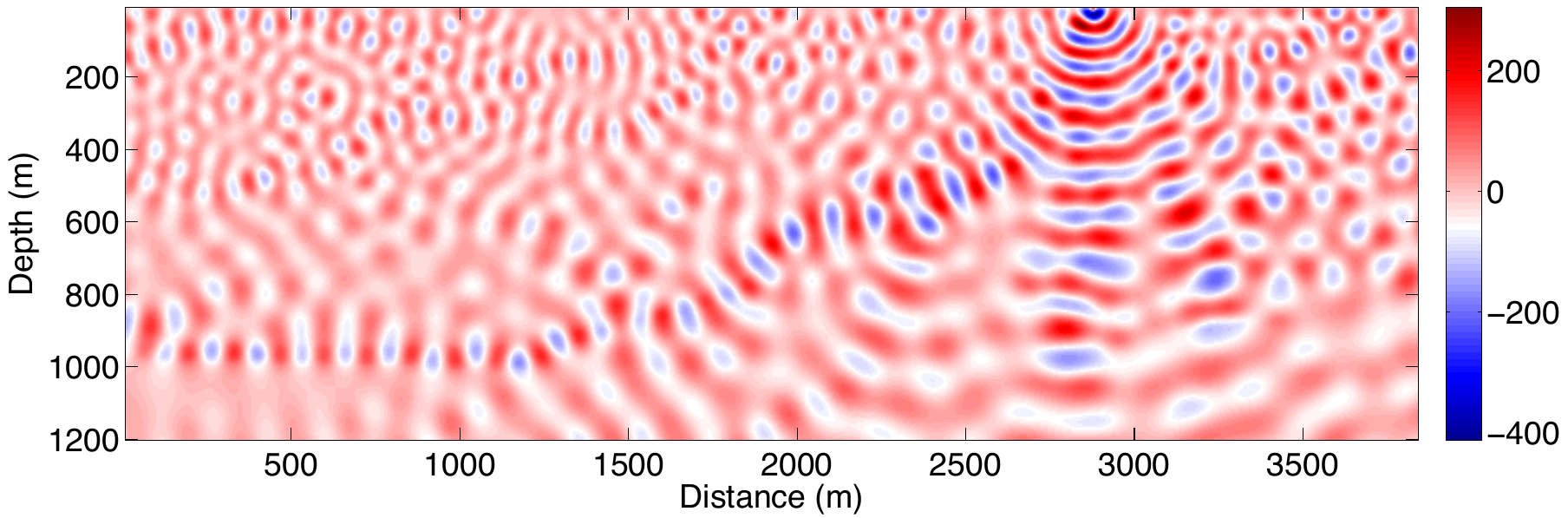}
    \end{minipage}
    \label{fig:snapshotTrueFreq_xs288_marmousi}
  }\\
  \subfloat[]
  {
    \begin{minipage}{0.9\linewidth}
      \centering
      \includegraphics[width=\textwidth]{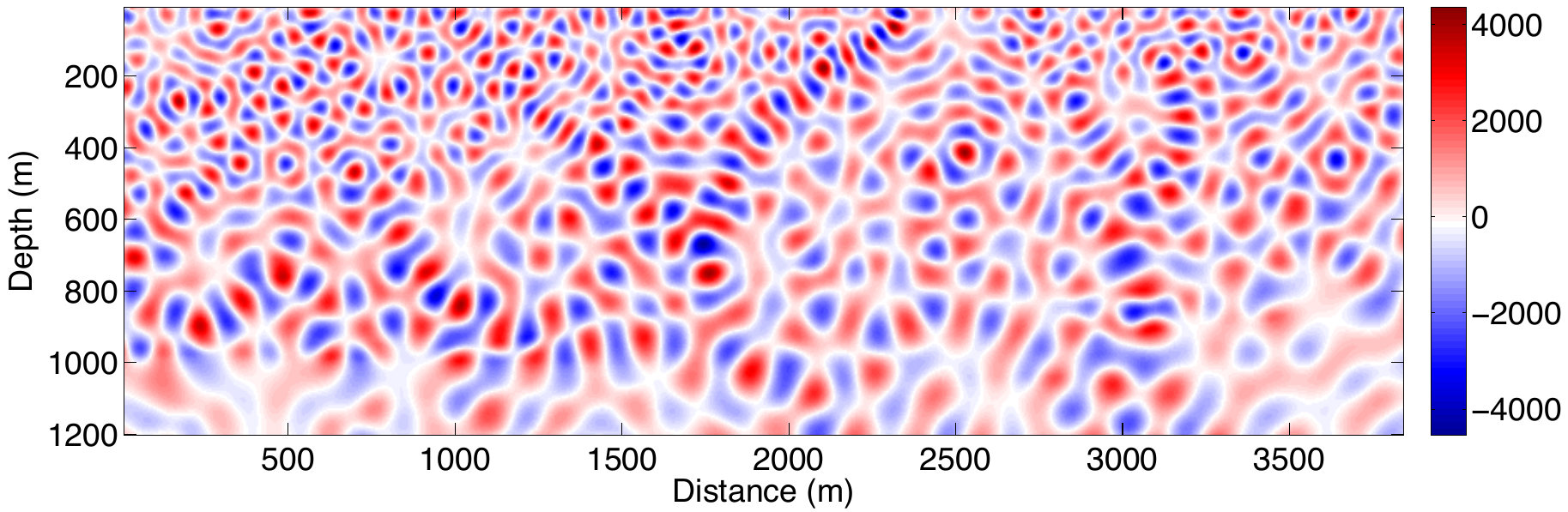}
    \end{minipage}
    \label{fig:snapshotTrueFreqSupershot_marmousi}
  }
  \caption{Wavefield examples generated by a single shot in (a) at position $\mathbf{x}_s = 960$\,m, (b) at position $\mathbf{x}_s = 1920$\,m, (c) at position $\mathbf{x}_s = 2880$\,m and (d) a supershot encoding $N_s = 384$ shots with random Gaussian weights with frequency $22.8$\,Hz}
  \label{fig:singleshot-supershot-wavefield}
\end{figure}

Figure \ref{fig:singleshot-supershot-wavefield} illustrates several frequency-domain wavefield examples at the frequency $\omega / (2\pi) = 22.8$\,Hz, in which Figures \ref{fig:snapshotTrueFreq_xs96_marmousi}, \ref{fig:snapshotTrueFreq_xs192_marmousi} and \ref{fig:snapshotTrueFreq_xs288_marmousi} show three regular wavefields $p(\mathbf{x}; \omega, \mathbf{x}_s)$ generated by three single shots at positions $\mathbf{x}_s$ = 960\,m, 1920\,m and 2880\,m, respectively, and Figure \ref{fig:snapshotTrueFreqSupershot_marmousi} shows a supershot wavefield $p_i^{(\text{s})}(\mathbf{x}; \omega)$ which encodes $N_s = 384$ shots on the surface with random Gaussian weights.

FWI uses the wavefield sample set $\mathbf{d}^{(\text{s})} \triangleq \left\{p_i^{(\text{s})}(\mathbf{x}_r; \omega)\right\}$ collected at all receiver locations $\mathbf{x}_r$ for all supershots with different frequencies $\omega$. Since each frequency is processed independently in frequency-domain modeling, the number of frequencies used in FWI can also be reduced to $N_{\omega}' < N_{\omega}$, and this set of frequencies can be randomly selected among all $N_{\omega}$ frequencies, which then reduces the dimension of $\mathbf{d}^{(\text{s})}$ to $N_{\omega}'N_s'N_r$. According to \eqref{eq:supershot-wavefield-freq}, the relationship between $\mathbf{d}^{(\text{s})}$ and the full-dimension data $\mathbf{d}$ for all receivers, single shots and frequencies can be written in a compact matrix form as
\begin{equation}
  \label{eq:supershot-data-relationship}
  \mathbf{d}^{(\text{s})} \triangleq \bm{\mathcal{F}}^{(\text{s})}(\mathbf{m}) = \mathbf{W}\mathbf{d} = \mathbf{W}\bm{\mathcal{F}}(\mathbf{m}).
\end{equation}
The subsampling matrix $\mathbf{W}$ of size $N_{\omega}'N_s'N_r \times N_{\omega}N_sN_r$ is structured as
\begin{equation}
  \label{eq:supershot-data-matrix-structure}
  \mathbf{W} \triangleq
  \left( \left( \mathbf{R}_{N_{\omega}' \times N_{\omega}} \otimes \mathbf{I}_{N_s'} \right)
  \begin{bmatrix}
  	\mathbf{w}(\omega_1) & &\\
  	& \ddots &\\
  	& & \mathbf{w}(\omega_{N_{\omega}})
  \end{bmatrix} \right)
  \otimes \mathbf{I}_{N_r}
\end{equation}
where the restriction matrix $\mathbf{R}_{N_{\omega}' \times N_{\omega}}$ of size $N_{\omega}' \times N_{\omega}$ randomly chooses $N_{\omega}'$ rows from an identity matrix of size $N_{\omega} \times N_{\omega}$ for random frequency selection.
Each $\mathbf{w}(\omega)$ is a random encoding matrix of size $N_s' \times N_s$ whose $(i, j)$-th entry is $w_{ij}(\omega)$ so that the block diagonal matrix in the center is of size $N_s'N_{\omega} \times N_sN_{\omega}$, and $\mathbf{I}_{N_s'}$, $\mathbf{I}_{N_r}$ are identity matrices of size $N_s' \times N_s'$, $N_r \times N_r$, respectively.

According to the perturbation analysis based on Born approximation theory, the relationship between the supershot Jacobian matrix $\mathbf{J}^{(\text{s})}$ and the sequential shot Jacobian matrix $\mathbf{J}$ can be written as
\begin{equation}
  \label{eq:supershot-jacobian-relationship}
  \mathbf{J}^{(\text{s})} = \mathbf{W}\mathbf{J}.
\end{equation}

For each FWI iteration $k$, different $w_{ij}(\omega)$ encoding supershots can be generated so that the random subsampling matrix, denoted as $\mathbf{W}_k$, is not static. This approach suppresses crosstalk artifacts into incoherent Gaussian noise and yields better reconstruction results. Meanwhile, no artificial bias towards a specific random source encoding pattern would be introduced into the solution by redrawing $\mathbf{W}_k$. Such an approach has been recommended in previous research on FWI that utilizes CS \cite{Herrmann:2011aa,Herrmann:2012ab,Li:2012aa,Warner:2013aa,Li:2016aa}.

Therefore, in \eqref{eq:fwi-gauss-newton-random}, $\mathbf{W}_k\delta \mathbf{d}_k$ can be obtained as a whole by calculating the difference between the recorded receiver data $\mathbf{d}_{\text{obs}}^{(\text{s})} \triangleq \mathbf{W}_k\mathbf{d}_{\text{obs}}$ encoded by $\mathbf{W}_k$ and the calculated receiver data $\mathbf{d}_k^{(\text{s})}$ generated by supershots. $\mathbf{W}_k\mathbf{J}_k$ can be regarded as the compressive Jacobian whose components includes non-altered Green's functions for receivers and random encoded Green's functions for sources.

The solution $\bm{\alpha}_k$ of the LASSO problem \eqref{eq:fwi-gauss-newton-random} relies on the choice of the sparsity constraint $\tau_k$. As suggested by \cite{Berg:2009aa}, every LASSO problem implies a convex and non-increasing function $\phi(\tau)$ that associates the least-squares residual to the sparsity level $\tau$. In this problem, each FWI iteration $k = 0, 1, 2, \dots$ needs to solve \eqref{eq:fwi-gauss-newton-random} and, therefore, has an implicit $\phi_k(\tau)$. Following the same idea used by \cite{Herrmann:2011aa,Li:2012aa} and \cite{Li:2016aa}, one can estimate $\tau_k$ by using a linear approximation of $\phi_k'(\tau)$ at $\tau = 0$, given in Theorem 2.1 of \cite{Berg:2009aa}
\begin{equation}
  \label{eq:tau-k-selection}
  \tau_k \approx -\frac{\phi_k(0)}{\phi_k'(0)} = \frac{\left\|\mathbf{W}_k\delta \mathbf{d}_k\right\|_2^2}{\left\| \bm{\mathcal{D}}_k^{\dag} \left( \left[\mathbf{W}_k\mathbf{J}_k\right]^{\dag} \left(\mathbf{W}_k\delta \mathbf{d}_k\right) \right) \right\|_{\infty}}
\end{equation}
where $\| \cdot \|_{\infty}$ is the maximum norm, $\left[\mathbf{W}_k\mathbf{J}_k\right]^{\dag}$ is the adjoint of the compressive Jacobian $\mathbf{W}_k\mathbf{J}_k$ and performs the following calculation over the vector $\mathbf{W}_k\delta \mathbf{d}_k$
\begin{equation}
  \label{eq:supershot-adjoint-jacobian}
  \begin{aligned}
  &\left[\mathbf{W}_k\mathbf{J}_k\right]^{\dag} \left(\mathbf{W}_k\delta \mathbf{d}_k\right)\\
  = &\sum\limits_{m=1}^{N_{\omega}'} \omega_m^2 f(\omega_m) \sum\limits_{n=1}^{N_r} G_k(\mathbf{x}_{r_n}; \omega_m, \mathbf{x}) \sum\limits_{i=1}^{N_s'} \sum\limits_{j=1}^{N_s} \left[\mathbf{w}_k(\omega_m)\right]_{ij} G_k(\mathbf{x}; \omega_m, \mathbf{x}_{s_j}) \left[\delta p_k^{(\text{s})}\right]_i(\mathbf{x}_{r_n}; \omega_m).
  \end{aligned}
\end{equation}
The above expression is a function with respect to the medium grid points $\mathbf{x}$, so it can be interpreted as a wavefield or image and hence can also be decomposed into global SOT coefficient by $\bm{\mathcal{D}}_k^{\dag}$.

\subsection{Projected Quasi-Newton Method for solving the LASSO problems}
The computational complexity of the FWI problem is reduced considerably after reducing the data dimensionality from $N_{\omega}N_sN_r$ to $N_{\omega}'N_s'N_r$. However, in order to minimize the objective function $J_k^{(\text{W})}(\bm{\alpha})$ in \eqref{eq:fwi-gauss-newton-random} with sparsity promotion on the global SOT coefficient $\bm{\alpha}$, the descent direction of $\bm{\alpha}$ must be projected into an $\ell_1$-norm ball with radius $\tau_k$.
\begin{algorithm}
  \caption{Spectral Projected Gradient (SPG) Algorithm}
  \label{alg:spg}
  \SetKwInOut{KwInit}{Initialization}
  \KwIn{Step length bounds $0 < a_{\min} < a_{\max}$, $\bm{\gamma}^{(l)}$, $\mathbf{B}^{(l)}$, sufficient decrease parameter $\nu$, $\ell_1$-norm bound $\tau_k$}
  \KwInit{$\bm{\alpha}_{\text{new}} \gets \bm{\alpha}^{(l)}$, initial step length $a = \dfrac{\left[\mathbf{p}^{(l-1)}\right]^T\mathbf{p}^{(l-1)}}{\left[\mathbf{p}^{(l-1)}\right]^T\mathbf{q}^{(l-1)}}$, function maximum $f_{\max} \gets -\infty$}
  \While{not converged}
  {
  	$\bm{\alpha}_{\text{old}} \gets \bm{\alpha}_{\text{new}}$\;
  	$a \gets \min\{a_{\max}, \max\{a_{\min}, a\}\}$\;
  	$\mathbf{d} \gets \bm{\mathcal{P}}_{\tau_k}^{(\ell_1)}\left( \bm{\alpha}_{\text{old}} - a\nabla Q^{(l)}(\bm{\alpha}_{\text{old}}) \right) - \bm{\alpha}_{\text{old}}$ \tcp*[l]{$\nabla Q^{(l)}(\bm{\alpha}) = \mathbf{B}^{(l)}\bm{\alpha} + \bm{\gamma}^{(l)}$}
  	$a \gets 1$\;
  	$f_{\max} \gets \max\left\{f_{\max}, J_k^{(W)}(\bm{\alpha}_{\text{old}})\right\}$\;
  	\While{$Q^{(l)}(\bm{\alpha}_{\text{old}} + a\mathbf{d}) > f_{\max} + \nu a \nabla Q^{(l)}(\bm{\alpha}_{\text{old}})^T\mathbf{d}$}
  	{
  	  Choose $a \in (0, a)$ by backtracking\;
  	}
  	$\bm{\alpha}_{\text{new}} \gets \bm{\alpha}_{\text{old}} + a\mathbf{d}$\;
  	$a \gets \dfrac{\left[\bm{\alpha}_{\text{new}} - \bm{\alpha}_{\text{old}}\right]^T\left[\bm{\alpha}_{\text{new}} - \bm{\alpha}_{\text{old}}\right]}{\left[\bm{\alpha}_{\text{new}} - \bm{\alpha}_{\text{old}}\right]^T\left[\nabla Q^{(l)}(\bm{\alpha}_{\text{new}}) - \nabla Q^{(l)}(\bm{\alpha}_{\text{old}})\right]}$
  }
  \KwOut{$\widehat{\bm{\alpha}} \gets \bm{\alpha}_{\text{new}}$}
\end{algorithm}

A limited-memory projected quasi-Newton method (l-PQN) proposed by \cite{Schmidt:2009aa} can solve the LASSO problem \eqref{eq:fwi-gauss-newton-random} iteratively, based on a two-layer strategy. In each iteration $l = 0, 1, 2, \dots$, the outer layer formulates a quadratic approximation function $Q^{(l)}(\bm{\alpha})$ of the objective function $J_k^{(\text{W})}(\bm{\alpha})$ around the current iterate $\bm{\alpha}^{(l)}$
\begin{equation}
  \label{eq:gnobj-quasi-newton}
  Q^{(l)}(\bm{\alpha}) \triangleq J_k^{(\text{W})}\left(\bm{\alpha}^{(l)}\right) + \left( \bm{\alpha} - \bm{\alpha}^{(l)} \right)^T\bm{\gamma}^{(l)} + \frac{1}{2}\left( \bm{\alpha} - \bm{\alpha}^{(l)} \right)^T \mathbf{B}^{(l)} \left( \bm{\alpha} - \bm{\alpha}^{(l)} \right)
\end{equation}
where $\bm{\gamma}^{(l)}$ is the gradient of $J_k^{(\text{W})}(\bm{\alpha})$ evaluated for $\bm{\alpha}^{(l)}$
\begin{equation}
  \label{eq:gradient-jw-alpha}
  \bm{\gamma}^{(l)} \triangleq \frac{\partial J_k^{(\text{W})}}{\partial \bm{\alpha}}\left( \bm{\alpha}^{(l)} \right) = -\Re \left\{ \bm{\mathcal{D}}_k^{\dag} \left( \left[\mathbf{W}_k\mathbf{J}_k\right]^{\dag} \left( \mathbf{W}_k\delta \mathbf{d}_k - \mathbf{W}_k\mathbf{J}_k \bm{\mathcal{D}}_k\left( \bm{\alpha}^{(l)} \right) \right) \right) \right\}
\end{equation}
and $\mathbf{B}^{(l)}$ denotes a positive-definite approximation matrix of $\left[\mathbf{W}_k\mathbf{J}_k\bm{\mathcal{D}}_k\right]^{\dag}\mathbf{W}_k\mathbf{J}_k\bm{\mathcal{D}}_k$, the Hessian matrix of $J_k^{(\text{W})}(\bm{\alpha})$, at the $l$-th iteration of l-PQN. The inner layer iteratively searches for a feasible descent direction by minimizing $Q^{(l)}(\bm{\alpha})$ subject to the $\ell_1$-norm constraints
\begin{equation}
  \label{eq:quadratic-minimization}
  \widehat{\bm{\alpha}} = \argmin\limits_{\bm{\alpha}} Q^{(l)}(\bm{\alpha}) \quad \text{subject to} \quad \|\bm{\alpha}\|_1 \leq \tau_k.
\end{equation}

This problem can be solved via the spectral projected gradient (SPG) algorithm \cite{Birgin:1999aa,Birgin:2014aa} shown in Algorithm \ref{alg:spg}, where the following variables are used
\begin{equation}
  \label{eq:lbfgs-difference-definition}
  \begin{aligned}
  	\mathbf{p}^{(l)} &\triangleq \bm{\alpha}^{(l+1)} - \bm{\alpha}^{(l)}\\
  	\mathbf{q}^{(l)} &\triangleq \bm{\gamma}^{(l+1)} - \bm{\gamma}^{(l)}.
  \end{aligned}
\end{equation}

In Algorithm \ref{alg:spg}, the Euclidean projection operator $\bm{\mathcal{P}}_{\tau}^{(\ell_1)}(\bm{\alpha})$ that projects the vector $\bm{\alpha}$ onto the $\ell_1$-norm ball with radius $\tau$ is defined as
\begin{equation}
  \label{eq:l1-projection}
  \bm{\mathcal{P}}_{\tau}^{(\ell_1)}(\bm{\alpha}) \triangleq \argmin\limits_{\bm{\beta}} \|\bm{\alpha} - \bm{\beta}\|_2^2 \quad \text{subject to} \quad \|\bm{\beta}\|_1 = \tau.
\end{equation}
A randomized algorithm that efficiently solves this projection problem is shown in Algorithm \ref{alg:l1-projection} \cite{Duchi:2008aa}.
\begin{algorithm}
  \caption{Projection onto an $\ell_1$-norm ball}
  \label{alg:l1-projection}
  \KwIn{$\bm{\alpha} \in \mathbb{R}^N$, $\tau > 0$}
  Sort $\bm{\alpha}$ s.t. $|\alpha_1| \geq |\alpha_2| \geq \dots \geq |\alpha_N|$\;
  Find $\rho = \argmax\limits_j \left( |\alpha_j| - \frac{1}{j} \left( \sum\limits_{r=1}^j |\alpha_r| - \tau \right) \right)$\;
  Define $\theta = \frac{1}{\rho} \left( \sum\limits_{i=1}^{\rho} |\alpha_i| - \tau \right)$\;
  \KwOut{$\bm{\beta} \in \mathbb{R}^N$ such that $\beta_i = \text{sign}(\alpha_i) \cdot \max\{|\alpha_i| - \theta, 0\}$}
\end{algorithm}

After solving the inner-layer problem \eqref{eq:quadratic-minimization}, the direction $\mathbf{d}^{(l)} \triangleq \widehat{\bm{\alpha}} - \bm{\alpha}^{(l)}$ is guaranteed to be a feasible descent direction since $\mathbf{B}^{(l)}$ is a positive-definite matrix. In order to obtain the next iterate $\bm{\alpha}^{(l+1)}$ for the outer layer, a backtracking line search method along the search direction $\mathbf{d}^{(l)}$ can be applied to find a step length $a \in (0, 1]$ that obeys the Armijo condition \cite{Armijo:1966aa}
\begin{equation}
  \label{eq:armijo}
  J_k^{(\text{W})}\left( \bm{\alpha}^{(l)} + a\mathbf{d}^{(l)} \right) \leq J_k^{(\text{W})}\left( \bm{\alpha}^{(l)} \right) + \nu a \left[ \bm{\gamma}^{(l)} \right]^T\mathbf{d}^{(l)}
\end{equation}
which then ensures a sufficient decrease of the objective function. In \eqref{eq:armijo} the sufficient decrease parameter $\nu$ is set to $10^{-4}$ as suggested by \cite{Nocedal:2006aa}. Because $\mathbf{d}^{(l)}$ takes the $\ell_1$-norm constraint into account, the next iterate $\bm{\alpha}^{(l+1)}$ also satisfies the constraint for the selected value of $a$.

The positive-definite matrix $\mathbf{B}^{(l)}$ that approximates the Hessian matrix of $J_k^{(\text{W})}(\bm{\alpha})$ can be built with the quasi-Newton methods such as the l-BFGS algorithm \cite{Nocedal:1980aa} which is one of the most popular choices. The l-BFGS algorithm maintains at most $m$ past $\mathbf{p}^{(l)}$ and $\mathbf{q}^{(l)}$ vectors for the Hessian approximation. It initializes $\mathbf{B}^{(0)} = \sigma^{(0)}\mathbf{I}$, and for $l > 0$, updates $\mathbf{B}^{(l)}$ by the following formula
\begin{equation}
  \label{eq:lbfgs}
  \mathbf{B}^{(l)} = \sigma^{(l)}\mathbf{I}
  -
  \begin{bmatrix}
  	\sigma^{(l)}\mathbf{P}^{(l)} & \mathbf{Q}^{(l)}
  \end{bmatrix}
  \begin{bmatrix}
  	\sigma^{(l)} \left[\mathbf{P}^{(l)}\right]^T \mathbf{P}^{(l)} & \mathbf{L}^{(l)}\\
  	\left[\mathbf{L}^{(l)}\right]^T & -\mathbf{X}^{(l)}
  \end{bmatrix}^{-1}
  \begin{bmatrix}
  	\sigma^{(l)} \left[\mathbf{P}^{(l)}\right]^T\\
  	\left[\mathbf{Q}^{(l)}\right]^T
  \end{bmatrix}
\end{equation}
where the scalar $\sigma^{(l)} \triangleq \dfrac{\left[\mathbf{q}^{(l)}\right]^T\mathbf{p}^{(l)}}{\left[\mathbf{q}^{(l)}\right]^T\mathbf{q}^{(l)}}$, the matrices $\mathbf{P}^{(l)} \triangleq \left[ \mathbf{p}^{(l-m)}, \dots, \mathbf{p}^{(l-1)} \right]$, $\mathbf{Q}^{(l)} \triangleq \left[ \mathbf{q}^{(l-m)}, \dots, \mathbf{q}^{(l-1)} \right]$, $\mathbf{X}^{(l)} \triangleq \text{diag}\left\{ \left[\mathbf{p}^{(l-m)}\right]^T\mathbf{q}^{(l-m)}, \dots, \left[\mathbf{p}^{(l-1)}\right]^T\mathbf{q}^{(l-1)} \right\}$ and $\mathbf{L}^{(l)}$ is defined by
\begin{equation*}
  \left[\mathbf{L}^{(l)}\right]_{ij} = \left\{
  \begin{aligned}
  	&\left[\mathbf{p}^{(l-m-1+i)}\right]^T\mathbf{q}^{(l-m-1+j)}, \quad &&\text{if} \quad i > j\\
  	&0, \quad &&\text{otherwise}.
  \end{aligned}
  \right.
\end{equation*}

\begin{figure}[htbp]
\centering
  \begin{minipage}{\linewidth}
    \centering
    \includegraphics[width=\textwidth]{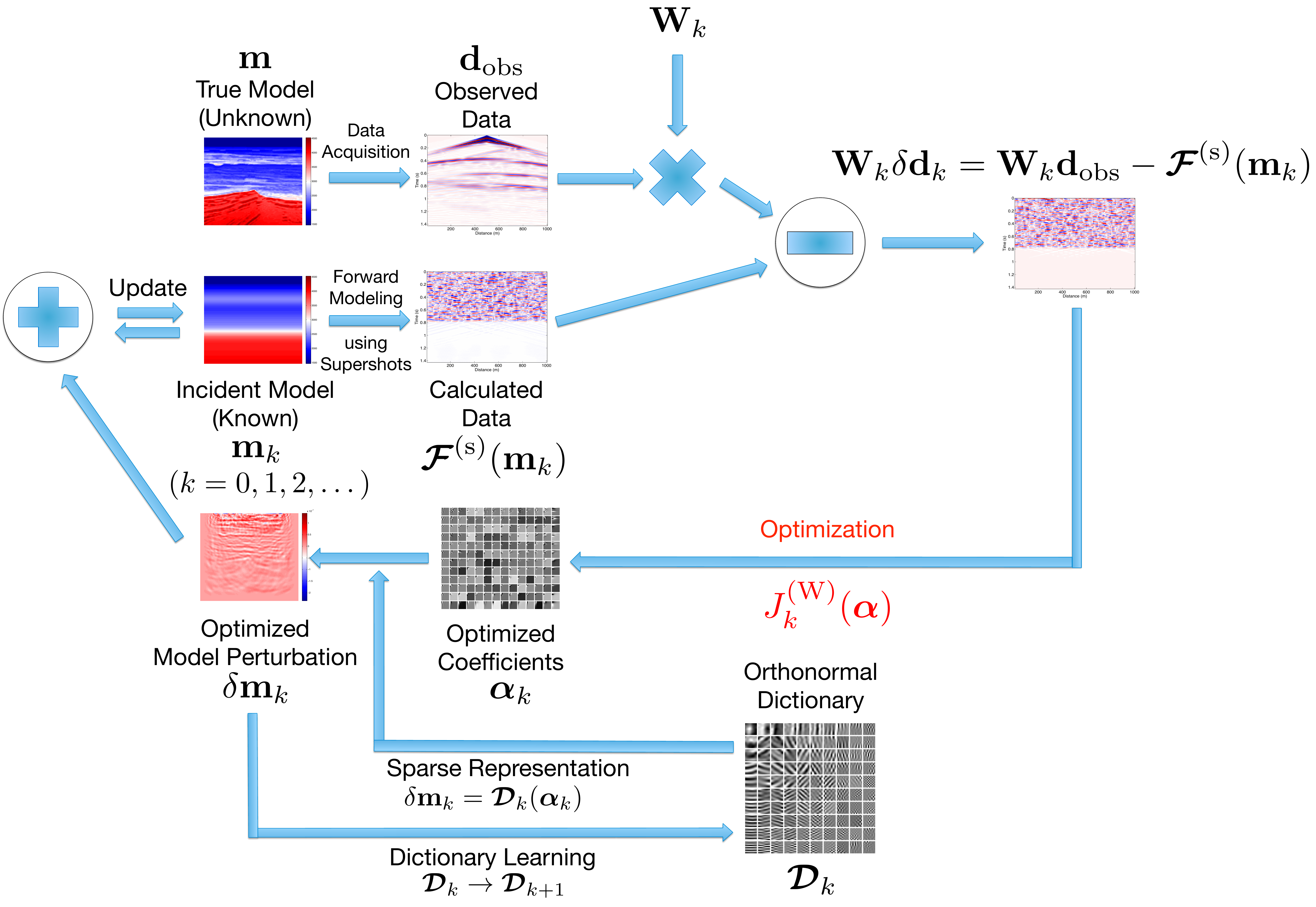}
  \end{minipage}
  \caption{FWI workflow using SOT-domain sparsity promotion with adaptive transform $\bm{\mathcal{D}}_k$ based on online orthonormal dictionary learning}
\label{fig:sotfwiWorkflow}
\end{figure}

Finally, Algorithm \ref{alg:dl-sparse-fwi} summarizes the overall SOT-based sparse-promoting FWI optimization procedure which is initialized by a smooth model $\mathbf{m}_{\text{smth}}$. The accuracy of $\mathbf{m}_{\text{smth}}$ directly affects the performance of FWI. To avoid FWI becoming trapped in local minima, a good initial model can be found using other inversion methods such as traveltime tomography \cite{Brenders:2007aa}, or migration velocity analysis \cite{Symes:2008aa}. Each newly optimized $\delta \mathbf{m}_k$ becomes the source for $R$ new patches for online dictionary learning in order to update the dictionary to $\mathbf{D}_{k+1}$, which will then be used in the corresponding SOT operator $\bm{\mathcal{D}}_{k+1}$ for the sparse representation of $\delta \mathbf{m}_{k+1}$ in the next FWI iteration. The entire workflow of the compressive FWI using the SOT is depicted in Figure \ref{fig:sotfwiWorkflow}.
\begin{algorithm}
  \caption{Sparsity-Promoting FWI based on the SOT}
  \label{alg:dl-sparse-fwi}
  \SetAlgoLined
  \SetAlFnt{\footnotesize}
  \SetKwInOut{KwInit}{Initialization}
  \AlFnt
  \KwIn{Recorded seismic data $\mathbf{d}_{\text{obs}} \triangleq \left\{p_{\text{obs}}(\mathbf{x}_r; \omega, \mathbf{x}_s)\right\}$, initial smooth model $\mathbf{m}_{\text{smth}}$, number of FWI iterations $K$, receiver locations $\mathbf{x}_r \in \mathcal{S}$, number of receivers $N_r$, sequential shot locations $\mathbf{x}_s \in \mathcal{S}$, number of sequential shots $N_s$, number of supershots $N_s'$, number of frequencies $N_{\omega}$, reduced number of frequencies $N_{\omega}'$, patch height $n_z$, patch width $n_x$, atom size $N = n_zn_x$, convergence error bound $\epsilon$}
  \KwOut{FWI result $\mathbf{m}_K$}
  \KwInit{$k \gets 0$, $\mathbf{m}_0 \gets \mathbf{m}_{\text{smth}}$, relative model change $\Delta_0 \gets \infty$}
  \While{$\Delta_k > \epsilon$ and $k < K$}
  {
  	Randomly draw $N_{\omega}'$ out of $N_{\omega}$ frequencies to form a set $\Omega'$\;
  	Generate $N_{\omega}'$ random Gaussian matrices $\mathbf{w}_k(\omega) \triangleq \{w_{ij}(\omega)\} \in \mathbb{R}^{N_s' \times N_s}$ for all frequencies $\omega \in \Omega'$ to produce $\mathbf{W}_k \triangleq \text{diag}\left\{ \mathbf{w}_k(\omega_1), \dots, \mathbf{w}_k(\omega_{N_{\omega}'}) \right\} \otimes \mathbf{I}$\;
  	Generate supershots $f_i^{(\text{s})}(\mathbf{x}; \omega) = \sum\limits_{j=1}^{N_s} w_{ij}(\omega)f(\omega)\delta(\mathbf{x} - \mathbf{x}_{s_j}),\ \forall i = 1, \dots, N_s'$\;
  	Encode the recorded seismic data $\mathbf{d}_{\text{obs}}^{(\text{s})} \triangleq \mathbf{W}_k\mathbf{d}_{\text{obs}}$\;
  	Solve \eqref{eq:acoustic-wave-equation-supershot} to get $p_i^{(\text{s})}(\mathbf{x}; \omega)$ for all supershots $\forall i = 1, \dots, N_s'$, and frequencies $\omega \in \Omega'$\;
  	Collect $\mathbf{d}_k^{(\text{s})} \triangleq \left\{p_i^{(\text{s})}(\mathbf{x}_r; \omega)\right\}$ for all receivers $\mathbf{x}_r \in \mathcal{S}$, supershots $\forall i = 1, \dots, N_s'$, and frequencies $\omega \in \Omega'$\;
  	$\mathbf{W}_k\delta\mathbf{d}_k = \mathbf{d}_{\text{obs}}^{(\text{s})} - \mathbf{d}_k^{(\text{s})}$\;
  	Collect Green's functions $G_i^{(\text{s})}(\mathbf{x}; \omega) \triangleq \sum\limits_{j=1}^{N_s} w_{ij}(\omega) G(\mathbf{x}; \omega, \mathbf{x}_{s_j})$ for all supershots $\forall i = 1, \dots, N_s'$, and frequencies $\omega \in \Omega'$\;
  	Collect Green's functions $G(\mathbf{x}_r; \omega, \mathbf{x}_j)$ for all receivers $\mathbf{x}_r \in \mathcal{S}$\;
  	$\text{Solve } \left\{
  	\begin{aligned}
  	&\bm{\alpha}_k = \argmin\limits_{\bm{\alpha}}\frac{1}{2} \left\| \mathbf{W}_k\delta \mathbf{d}_k - \mathbf{W}_k\mathbf{J}_k \bm{\mathcal{D}}_k(\bm{\alpha}) \right\|_2^2 \\
  	&\text{s.t.}\ \|\bm{\alpha}\|_1 \leq \tau_k \approx \frac{\left\|\mathbf{W}_k\delta \mathbf{d}_k\right\|_2^2}{\left\| \bm{\mathcal{D}}_k^{\dag} \left( \left[\mathbf{W}_k\mathbf{J}_k\right]^{\dag} \left(\mathbf{W}_k\delta \mathbf{d}_k\right) \right) \right\|_{\infty}}
  	\end{aligned}
  	\right\}
  	$ with l-PQN\;
  	$\delta \mathbf{m}_k = \bm{\mathcal{D}}_k(\bm{\alpha}_k)$\;
  	Learn $\mathbf{D}_{k+1}$ from $R$ patches of $\delta \mathbf{m}_k$ using Algorithm \ref{alg:online-ortho-dl} inside the outer \textbf{for} loop\;
  	$\mathbf{m}_{k+1} = \mathbf{m}_k + \delta \mathbf{m}_k$\;
  	$\Delta_k = \|\mathbf{m}_{k+1} - \mathbf{m}_k\|_2 / \|\mathbf{m}_k\|_2$\;
  	$k \gets k + 1$\;
  }
\end{algorithm}

\section{Results}
\begin{figure}[htbp]
  \centering
  \subfloat[]
  {
    \begin{minipage}{0.8\linewidth}
      \centering
      \includegraphics[width=\textwidth]{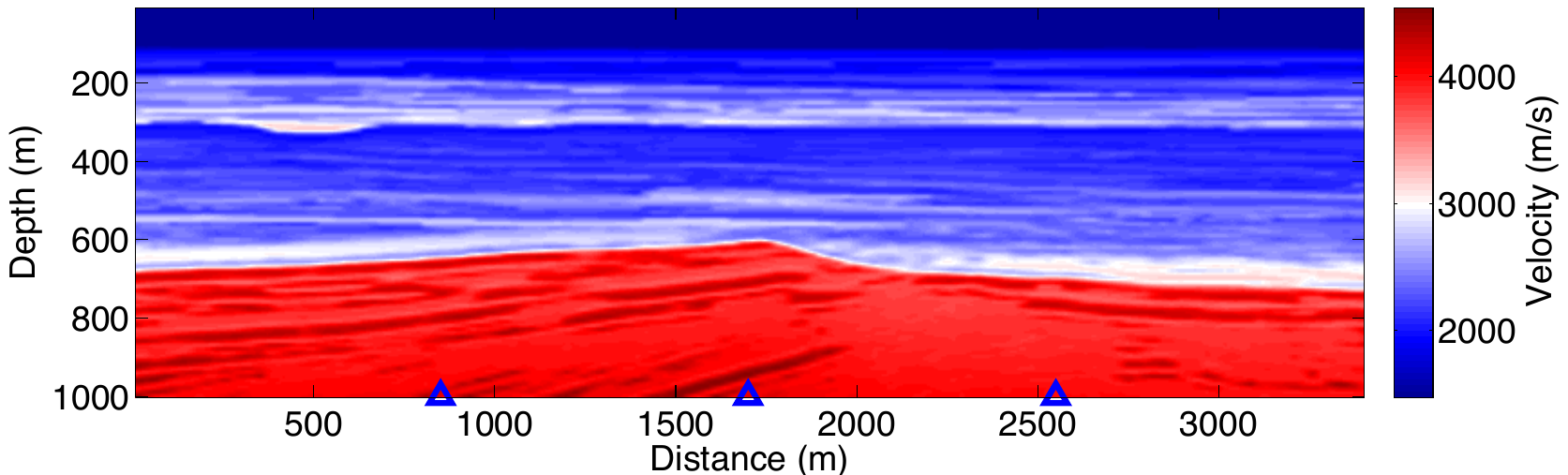}
    \end{minipage}
    \label{fig:vmTrue-bgcompass}
  }\\
  \subfloat[]
  {
    \begin{minipage}{0.8\linewidth}
      \centering
      \includegraphics[width=\textwidth]{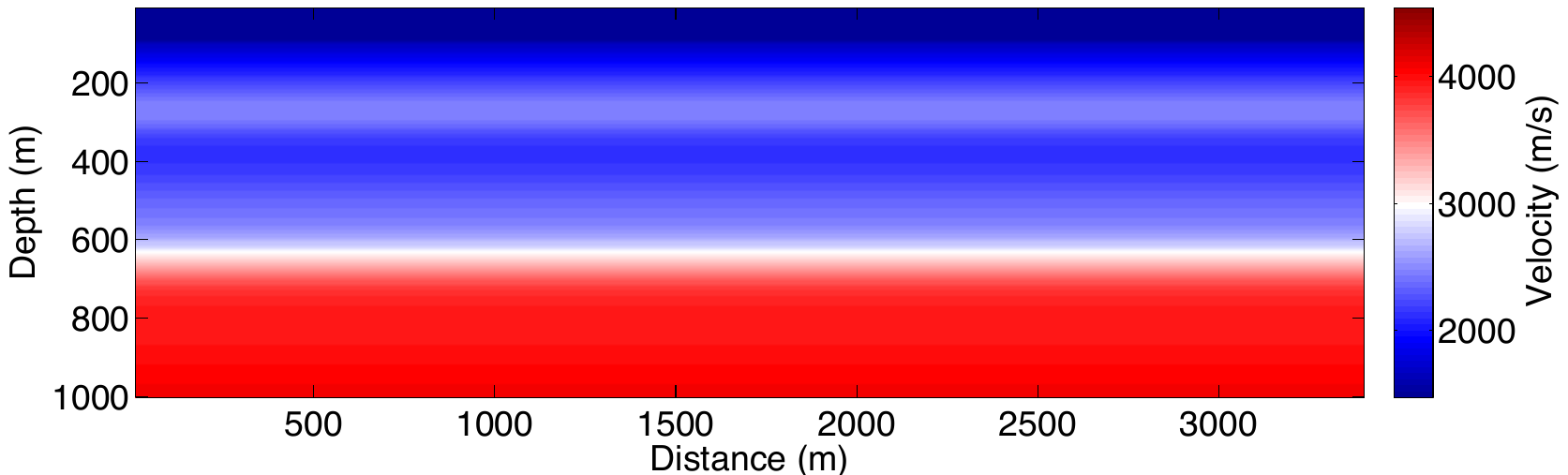}
    \end{minipage}
    \label{fig:vmSmooth-bgcompass}
  }\\
  \subfloat[]
  {
    \begin{minipage}{0.8\linewidth}
      \centering
      \includegraphics[width=\textwidth]{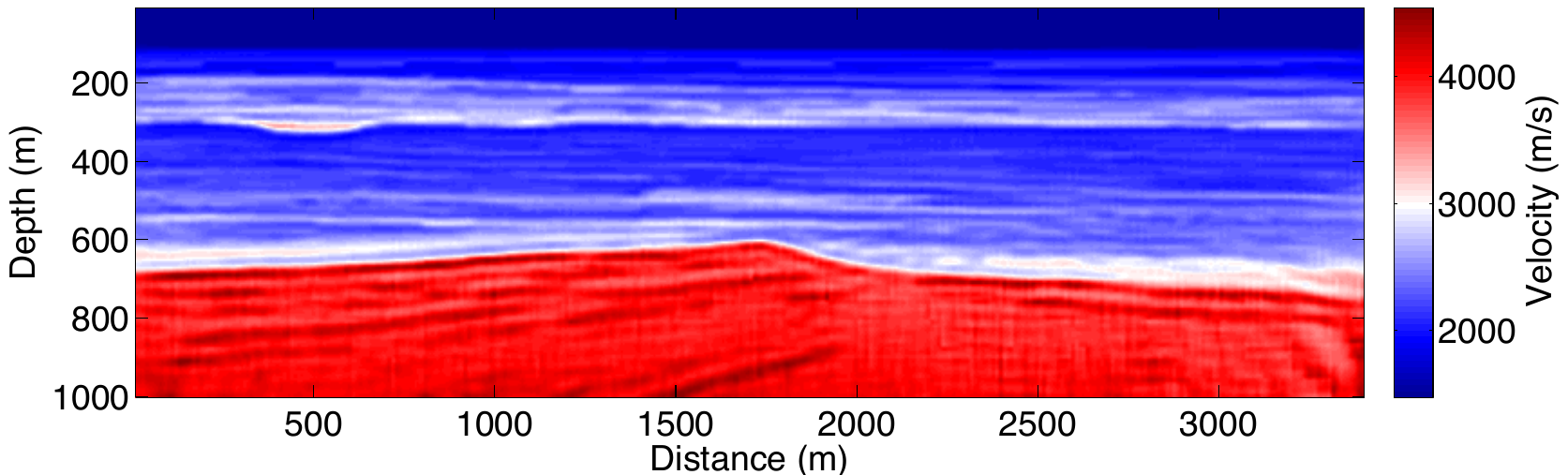}
    \end{minipage}
    \label{fig:vmNew5x20-bgcompass-sotcs}
  }
  \\
  \subfloat[]
  {
    \begin{minipage}{0.8\linewidth}
      \centering
      \includegraphics[width=\textwidth]{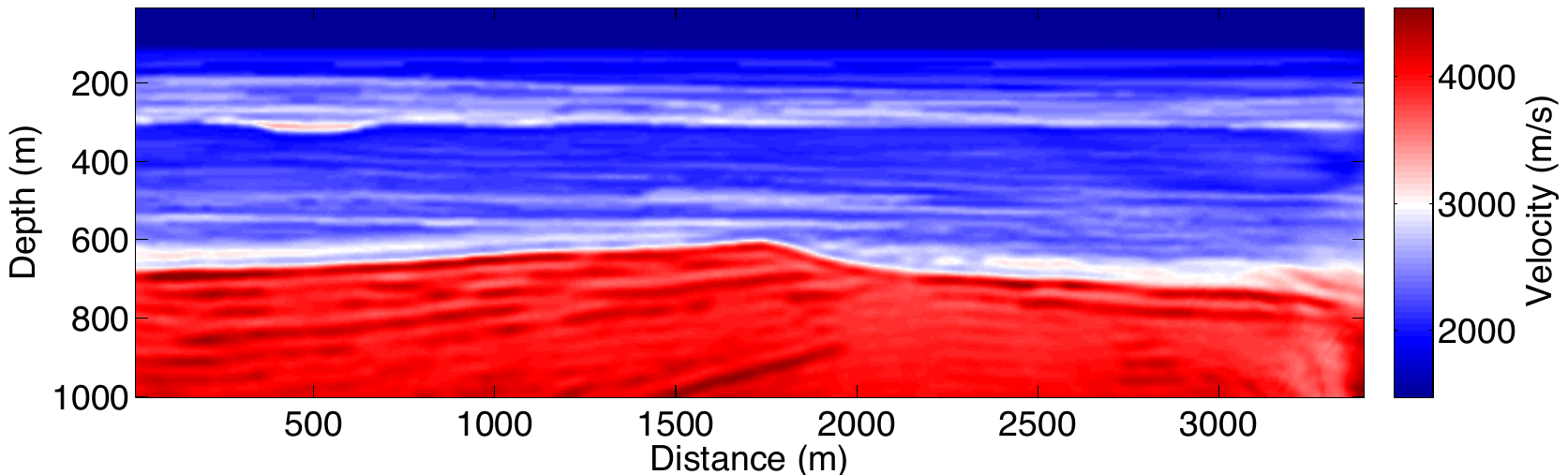}
    \end{minipage}
    \label{fig:vmNew5x20-bgcompass}
  }
  \caption{FWI results for the BG-Compass model with velocity range of 1500 to 4500\,m/s. (a) Original model $\mathbf{v}_{\text{true}} = 1/\sqrt{\mathbf{m}_{\text{true}}}$; blue triangles mark horizontal positions for vertical velocity logs shown in Figure \ref{fig:vertical-vlogs-bgcompass}, (b) Initial smooth model $\mathbf{v}_{\text{smth}} = 1/\sqrt{\mathbf{m}_{\text{smth}}}$, (c) Result from sparsity-promoting FWI based on the SOT with 3 supershots after all 100 iterations, (d) Result on the full data with all shots and frequencies after all 100 iterations.}
  \label{fig:results-bgcompass}
\end{figure}

\begin{figure}[htbp]
  \centering
  \subfloat[]
  {
    \begin{minipage}{0.75\linewidth}
      \centering
      \includegraphics[width=\textwidth]{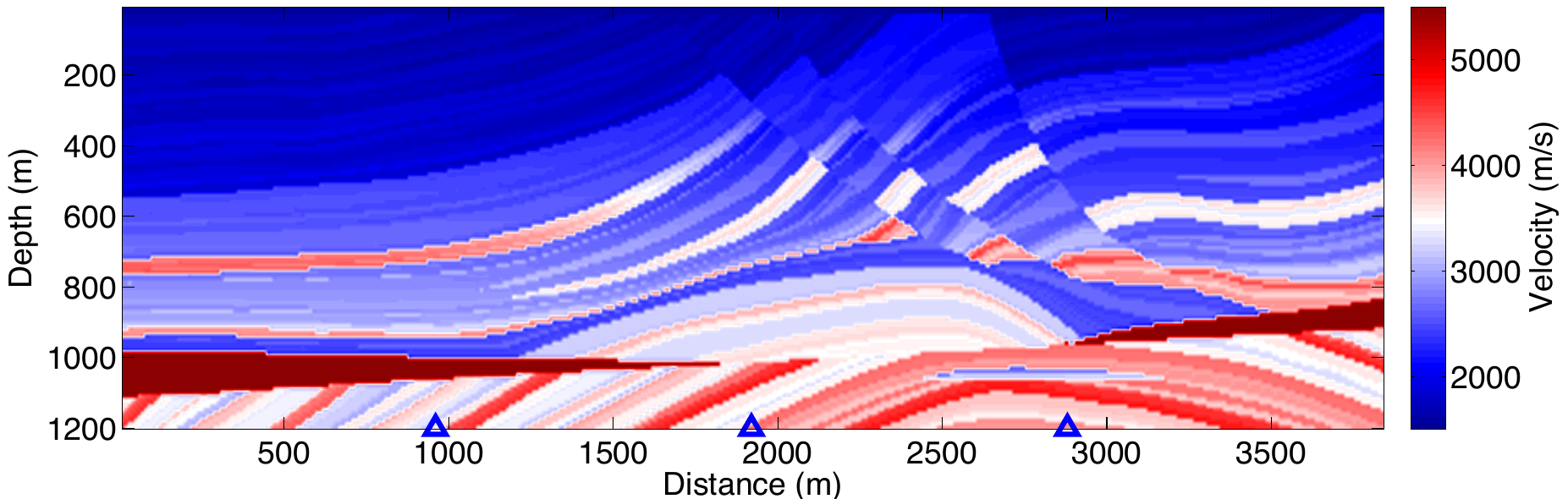}
    \end{minipage}
    \label{fig:vmTrue-marmousi}
  }\\
  \subfloat[]
  {
    \begin{minipage}{0.75\linewidth}
      \centering
      \includegraphics[width=\textwidth]{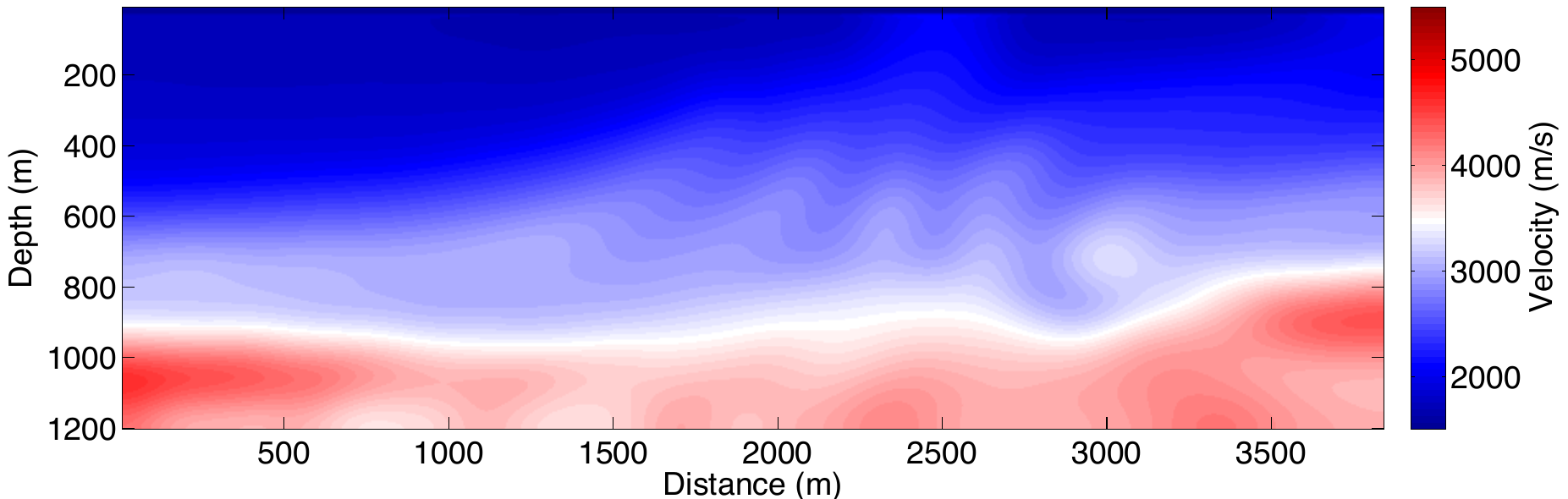}
    \end{minipage}
    \label{fig:vmSmooth-marmousi}
  }\\
  \subfloat[]
  {
    \begin{minipage}{0.75\linewidth}
      \centering
      \includegraphics[width=\textwidth]{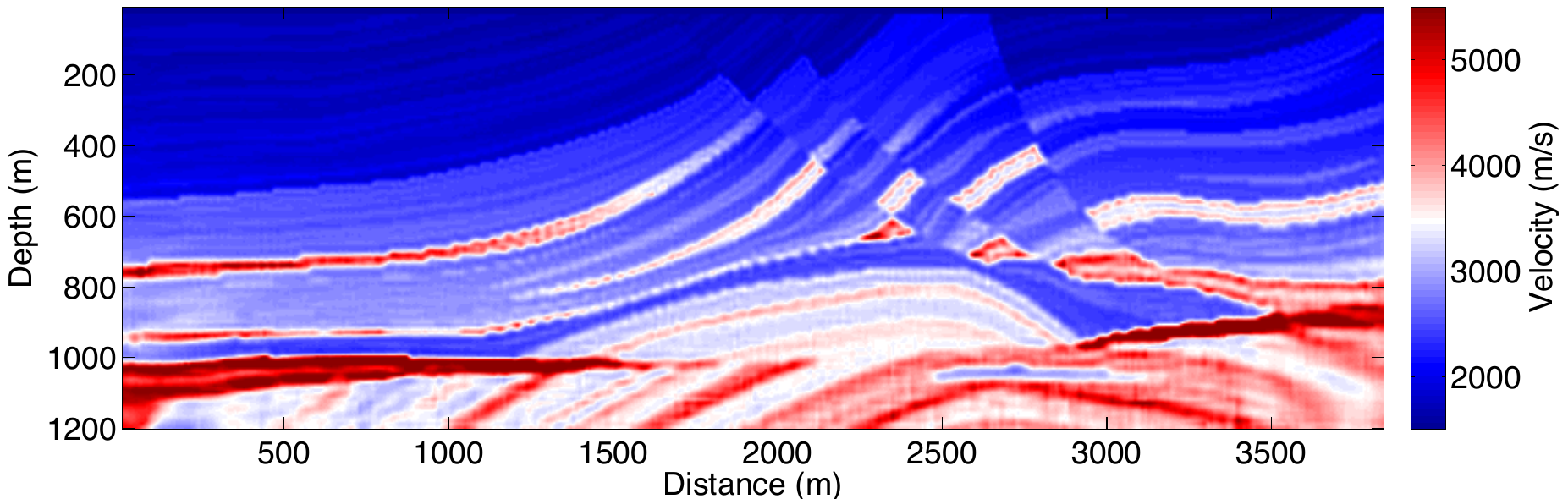}
    \end{minipage}
    \label{fig:vmNew5x20-marmousi-sotcs}
  }
  \\
  \subfloat[]
  {
    \begin{minipage}{0.75\linewidth}
      \centering
      \includegraphics[width=\textwidth]{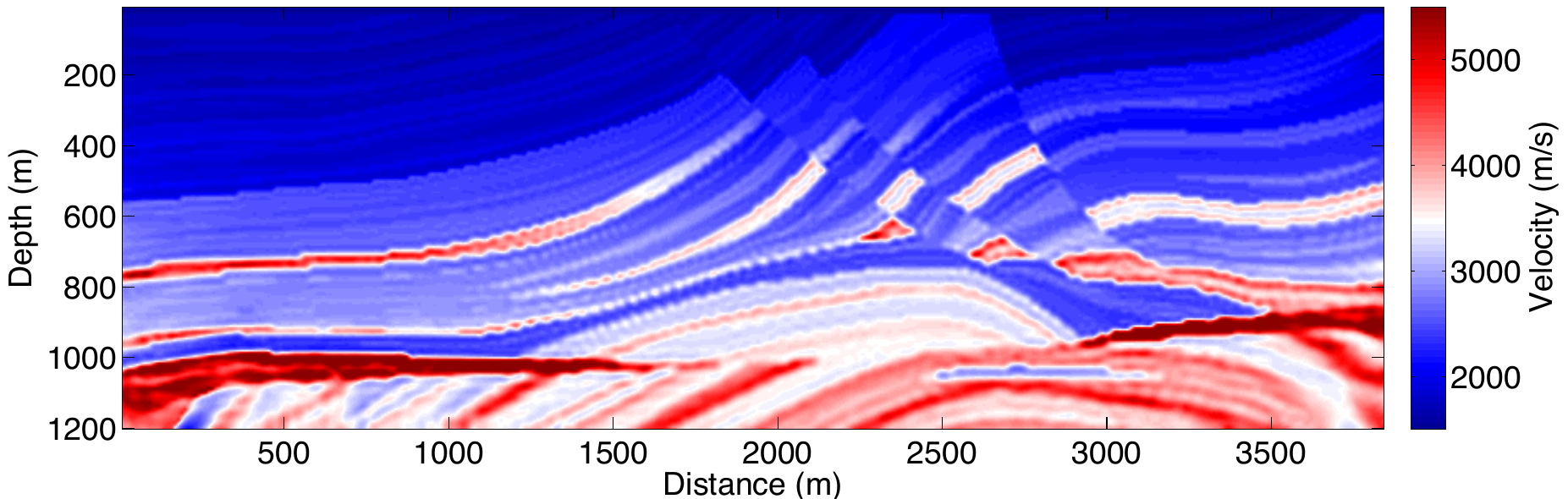}
    \end{minipage}
    \label{fig:vmNew5x20-marmousi}
  }
  \caption{FWI results for the Marmousi model with velocity range of 1500 to 5800\,m/s. (a) Original model $\mathbf{v}_{\text{true}} = 1/\sqrt{\mathbf{m}_{\text{true}}}$; blue triangles mark horizontal positions for vertical velocity logs shown in Figure \ref{fig:vertical-vlogs-bgcompass}, (b) Initial smooth model $\mathbf{v}_{\text{smth}} = 1/\sqrt{\mathbf{m}_{\text{smth}}}$, (c) Result from sparsity-promoting FWI based on the SOT with 4 supershots and 16 random frequencies after all 100 iterations, (d) Result on the full data with all shots and frequencies after all 100 iterations.}
  \label{fig:results-marmousi}
\end{figure}

In the following experiments, the proposed FWI method is tested on two synthetic velocity models that are often used to verify inversion algorithms in realistic settings. One velocity model is from the BG-Compass benchmark model whose exact form is shown in Figure \subref*{fig:vmTrue-bgcompass}. This model is rescaled to $N_z \times N_x = 100 \times 350$ grid points and covers a width of 3.5\,km and a depth of 1\,km. Full data generated by $N_s = 350$ sequential shots are recorded by $N_r = 350$ receivers equispaced along the surface of the model. Another benchmark velocity model is the well-known Marmousi model shown in Figure \subref*{fig:vmTrue-marmousi}. This model is rescaled to $N_z \times N_x = 120 \times 384$ grid points and covers a width of 3.84\,km and a depth of 1.2\,km. Full data from $N_s = 384$ sequential shots on the model surface are recorded over $N_r = 384$ equispaced receivers. The grid spacing of $10\,$m guarantees that a sufficient number of grid points are used to represent the expected wavelengths and no grid dispersion happens.
We simulate wavefields by discretizing the PDE \eqref{eq:acoustic-wave-equation-freq} as a Helmholtz system and solving with a direct solver based on the 8\textsuperscript{th}-order staggered-grid finite difference frequency-domain method \cite{Ajo-Franklin:2005aa} in which the left, right and bottom boundary reflections are absorbed by perfectly matched layers \cite{Komatitsch:2007aa}.

The shot source is a Ricker wavelet centered at 20\,Hz with 256 frequency components spanning 3.0 to 48.1\,Hz, and its spectrum $f(\omega)$ is assumed known and fixed. FWI starts from an initial smooth model shown in Figure \subref*{fig:vmSmooth-bgcompass} for BG-Compass, or Figure \subref*{fig:vmSmooth-marmousi} for Marmousi. In practical implementations, FWI is carried out sequentially in several consecutive frequency bands to avoid local minima caused by cycle skipping \cite{Bunks:1995aa,Sirgue:2004aa}. Here we use five frequency bands within 3.0 to 48.1\,Hz, so the average number of frequencies per band is $N_{\omega} = 256/5 \approx 52$.
In each frequency band, $K = 20$ FWI iterations are executed. After 20 FWI iterations are completed on one frequency band, the resulting more accurate model serves as the initial model for another 20 FWI iterations on the next higher frequency band.

For every data-reduced FWI iteration in our experiments based on CS, we only use $N_s' = (1/100)N_s$ supershots (3 for BG-Compass and 4 for Marmousi), and $N_{\omega}' = 16$ random frequencies from one of five frequency bands. Thus, the problem dimensionality of the compressed objective function $J_k^{(\text{W})}(\bm{\alpha})$ in Equation \eqref{eq:fwi-gauss-newton-random} is $(N_{\omega}N_s)/(N_{\omega}'N_s') \approx 320$ times smaller than that of the full-data Gauss-Newton objective function $J_k(\delta \mathbf{m})$ in \eqref{eq:fwi-gauss-newton-linear}. This does not necessarily mean that a data-reduced FWI iteration runs 320 times faster than a full-data FWI iteration as actual implementations may vary, but the reduced time on both forward modeling and objective function minimization, as well as the reduced RAM costs, are still considerable in our case.
The data-reduced FWI results are shown in Figures \subref*{fig:vmNew5x20-bgcompass-sotcs} and \subref*{fig:vmNew5x20-marmousi-sotcs} for BG-Compass and Marmousi, respectively. As a comparison, the FWI results on the full data with all shots and frequencies are shown in Figures \subref*{fig:vmNew5x20-bgcompass} and \subref*{fig:vmNew5x20-marmousi} for both models, respectively. The sparsity promotion introduced by the adaptive SOT moves the problem to a much lower dimensional space, so we are able to achieve comparable inversion results with much lower computational cost.

\begin{figure*}[htbp]
  \centering
  \subfloat[]
  {
    \begin{minipage}{0.33\linewidth}
      \centering
      \includegraphics[width=\textwidth]{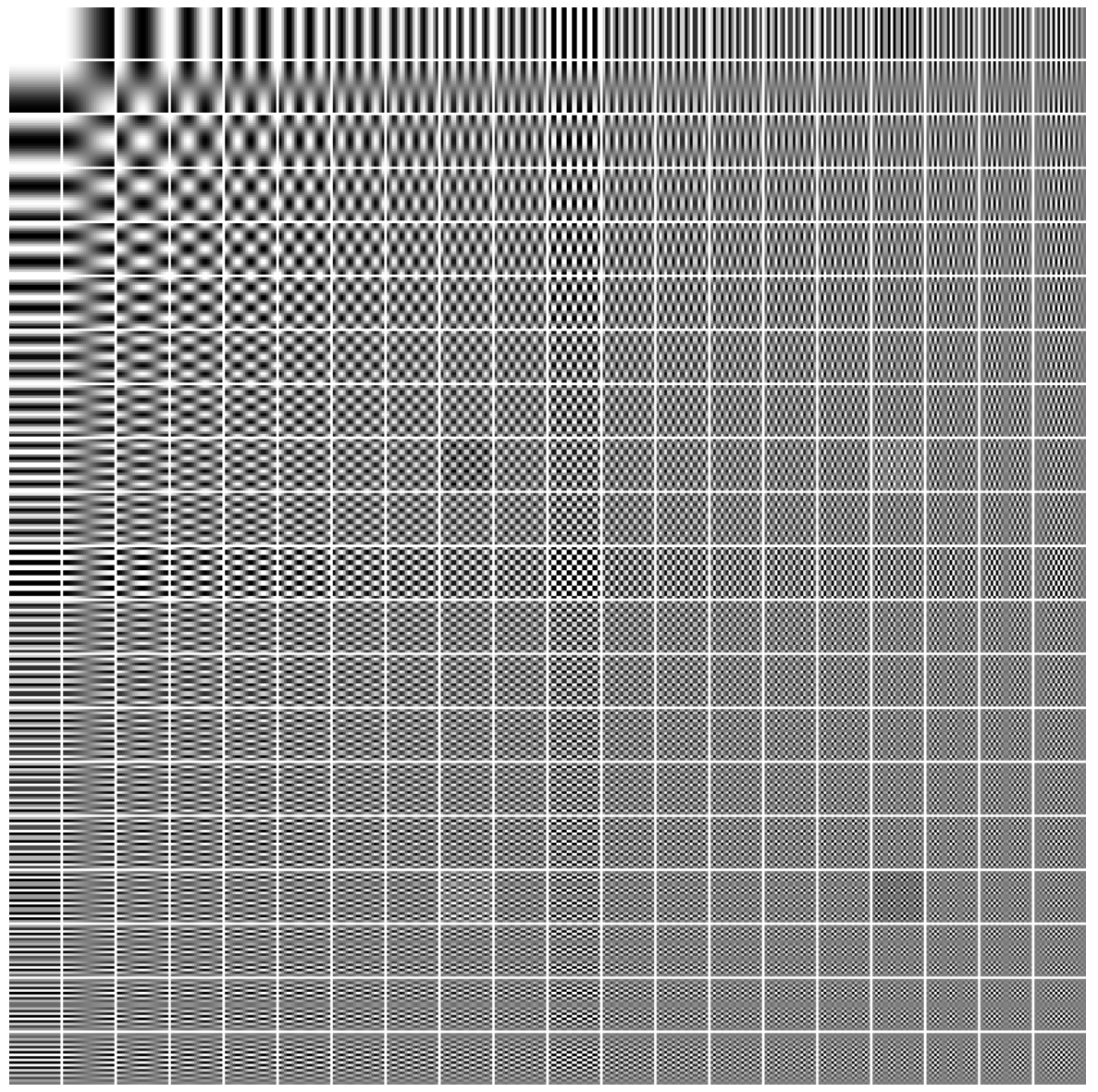}
    \end{minipage}
    \label{fig:dictDCT_20x20}
  }
  \subfloat[]
  {
    \begin{minipage}{0.33\linewidth}
      \centering
      \includegraphics[width=\textwidth]{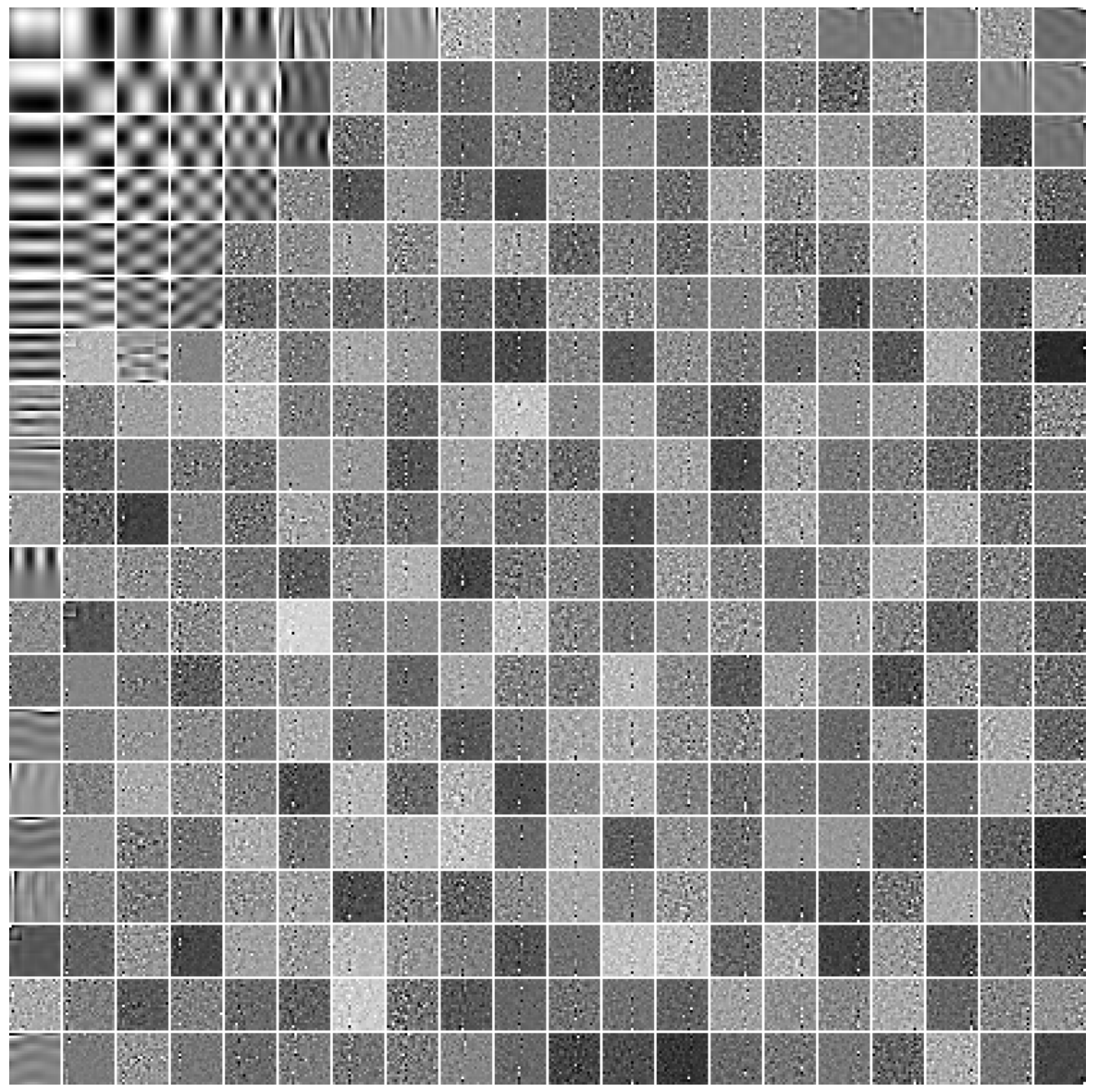}
    \end{minipage}
    \label{fig:dict1x20_bgcompass}
  }
  \subfloat[]
  {
    \begin{minipage}{0.33\linewidth}
      \centering
      \includegraphics[width=\textwidth]{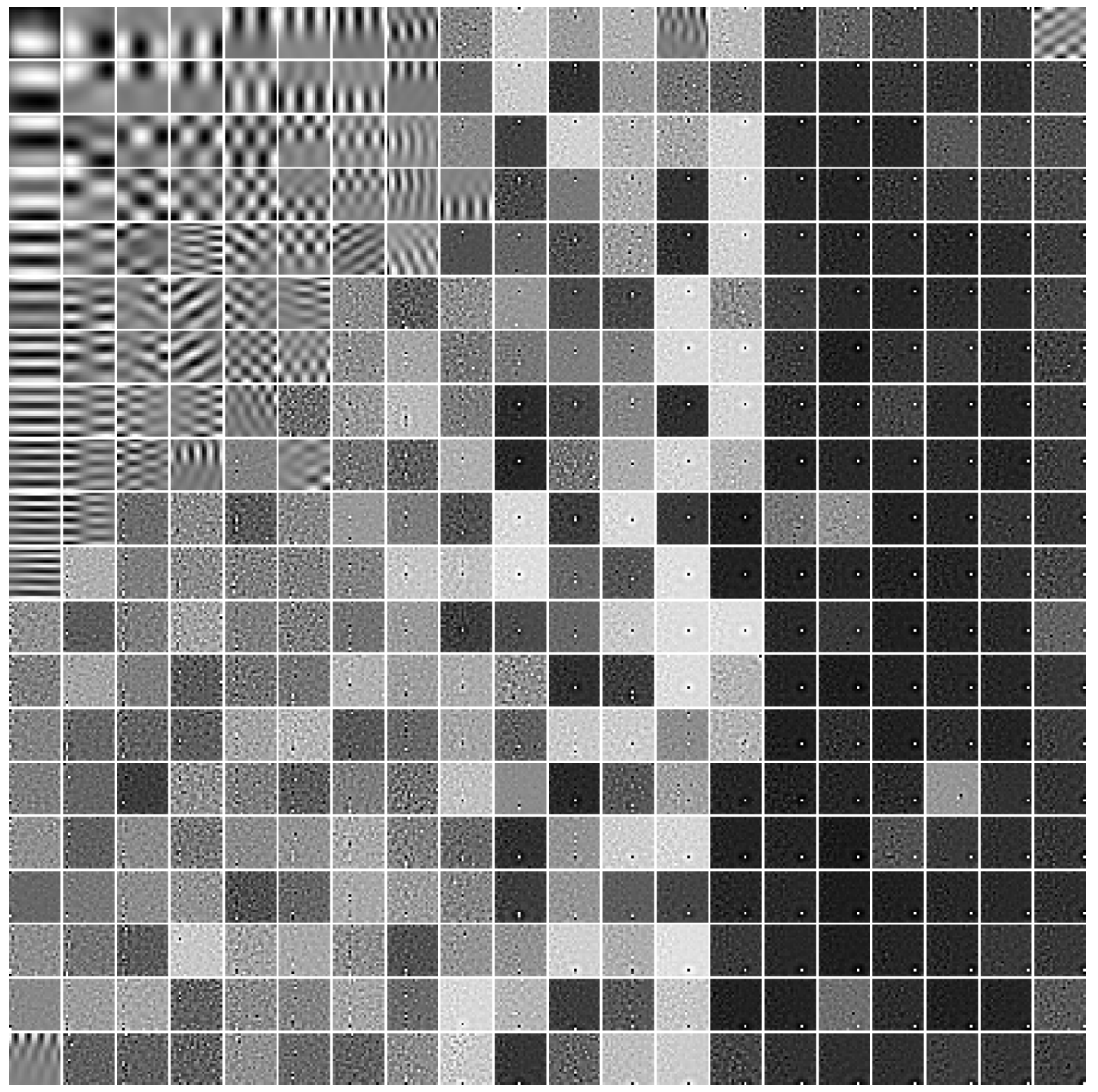}
    \end{minipage}
    \label{fig:dict2x20_bgcompass}
  }\\
  \subfloat[]
  {
    \begin{minipage}{0.33\linewidth}
      \centering
      \includegraphics[width=\textwidth]{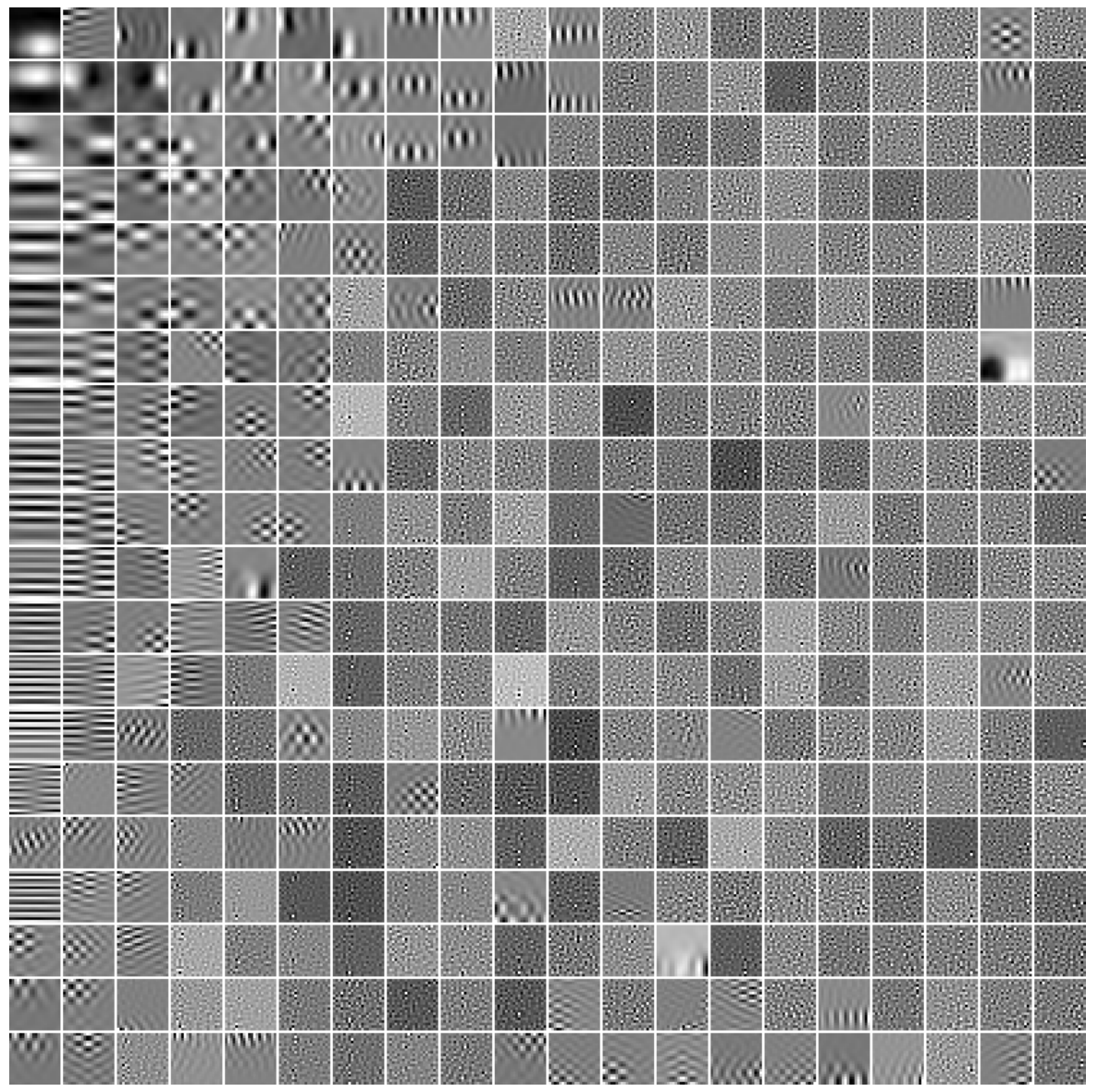}
    \end{minipage}
    \label{fig:dict3x20_bgcompass}
  }
  \subfloat[]
  {
    \begin{minipage}{0.33\linewidth}
      \centering
      \includegraphics[width=\textwidth]{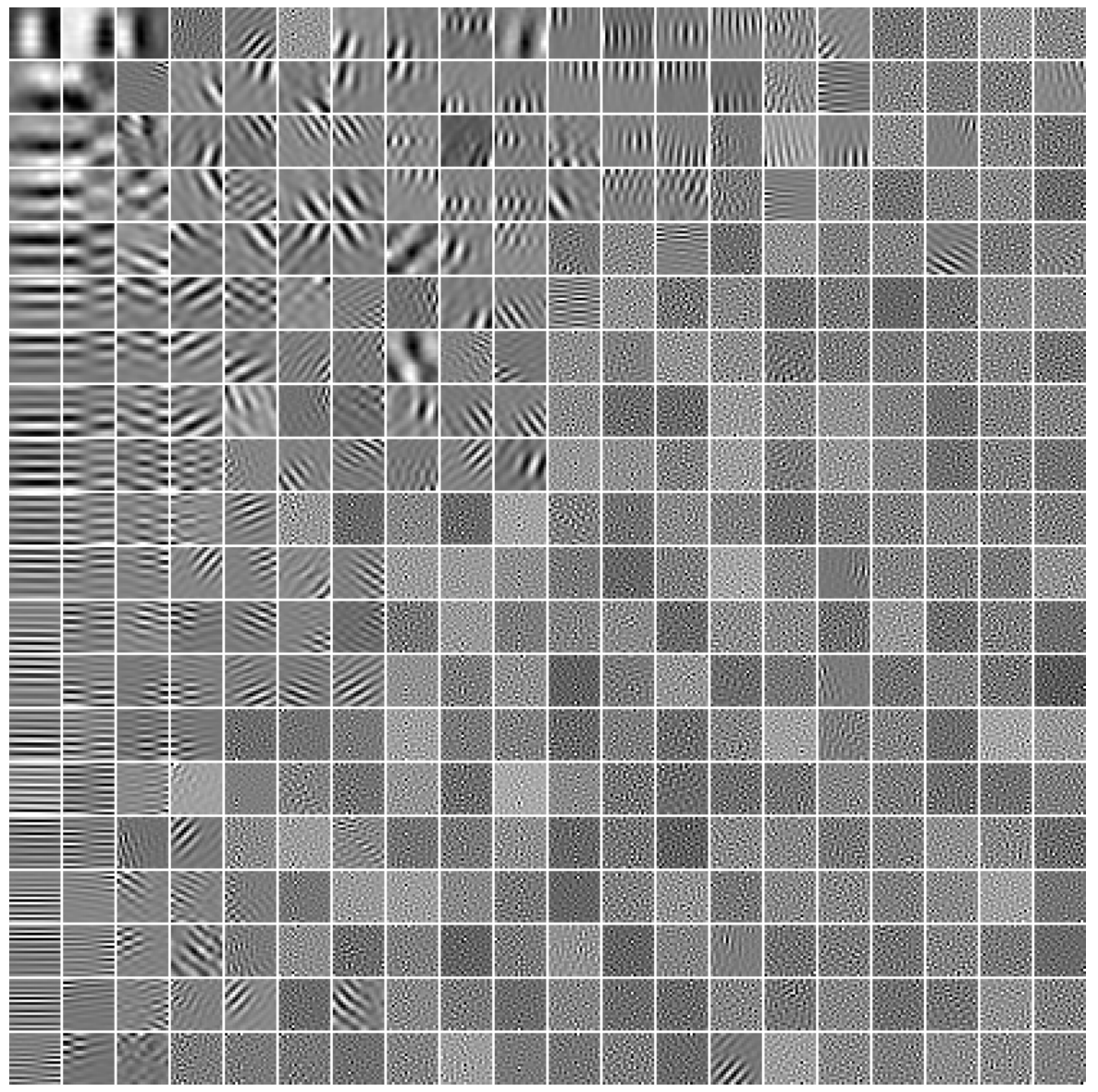}
    \end{minipage}
    \label{fig:dict4x20_bgcompass}
  }
  \subfloat[]
  {
    \begin{minipage}{0.33\linewidth}
      \centering
      \includegraphics[width=\textwidth]{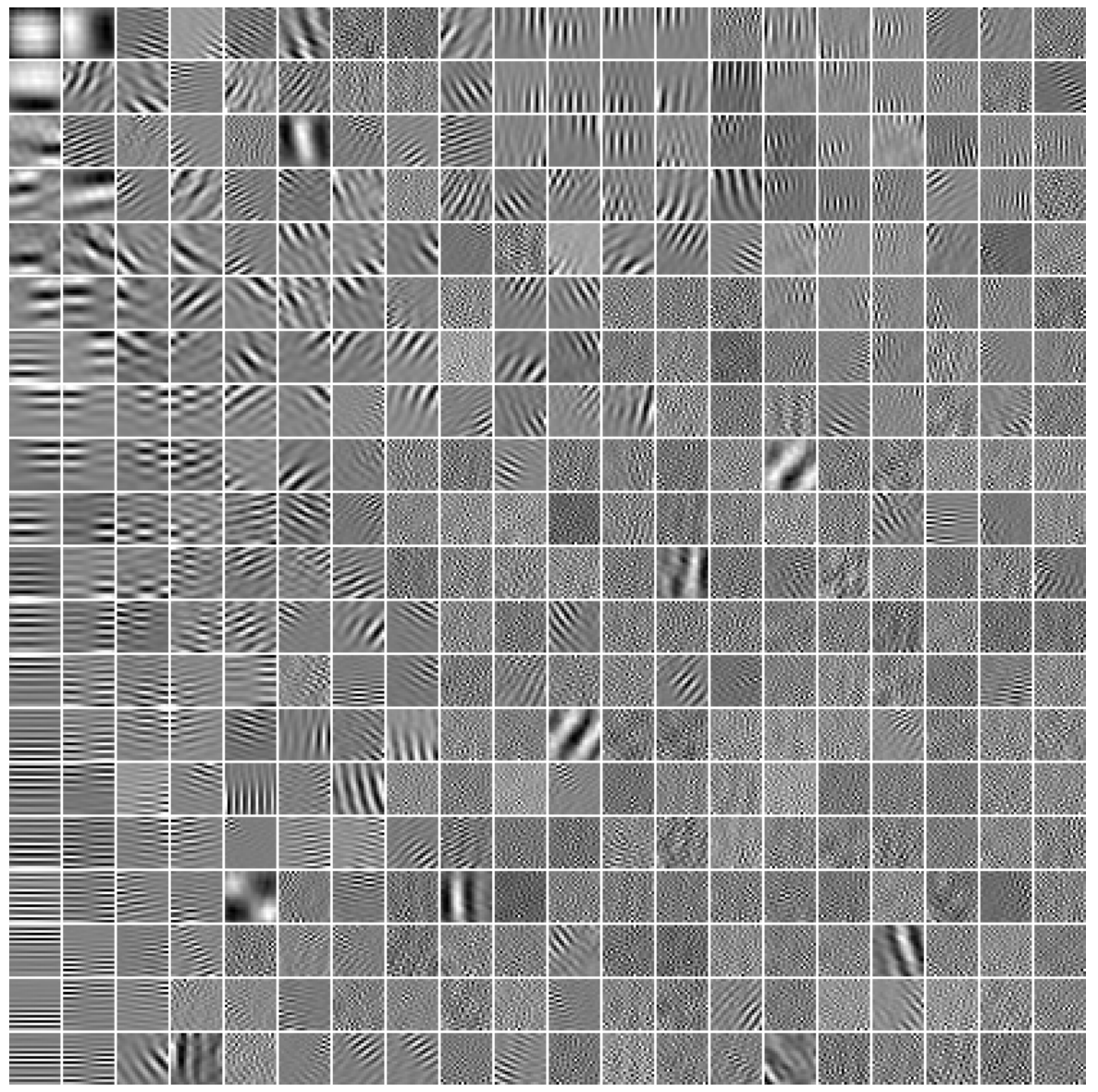}
    \end{minipage}
    \label{fig:dict5x20_bgcompass}
  }
  \caption{Initial dictionary $\mathbf{D}_0$ and the learned dictionaries $\mathbf{D}_{20}$ by Algorithm \ref{alg:online-ortho-dl} after $K = 20$ FWI iterations in each frequency band on the BG-Compass model. Dictionary size is $400 \times 400$; each column is displayed as a $20 \times 20$ block in the images.
  (a) Initial DCT matrix $\mathbf{D}_0$,
  (b) Trained dictionary $\mathbf{D}_{20}$ for the first frequency band, 3.0--11.6\,Hz,
  (c) second band, 12.1--20.8\,Hz,
  (d) third band, 21.3--29.9\,Hz,
  (e) fourth band, 30.4--39.0\,Hz, and (f) fifth band, 39.5--48.1\,Hz.}
  \label{fig:dict-bgcompass}
\end{figure*}

\begin{figure*}[htbp]
  \centering
  \subfloat[]
  {
    \begin{minipage}{0.5\linewidth}
      \centering
      \includegraphics[width=\textwidth]{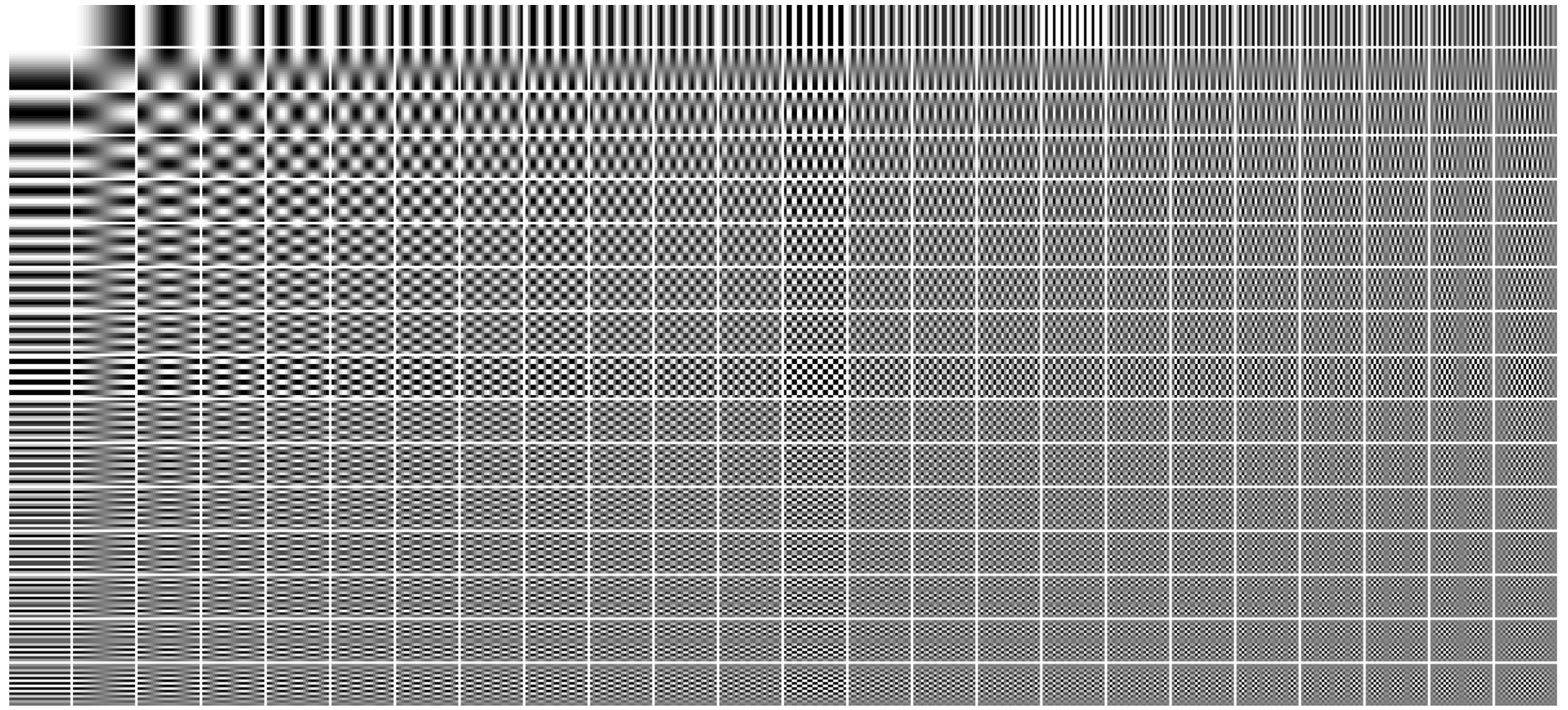}
    \end{minipage}
    \label{fig:dictDCT_16x24}
  }
  \subfloat[]
  {
    \begin{minipage}{0.5\linewidth}
      \centering
      \includegraphics[width=\textwidth]{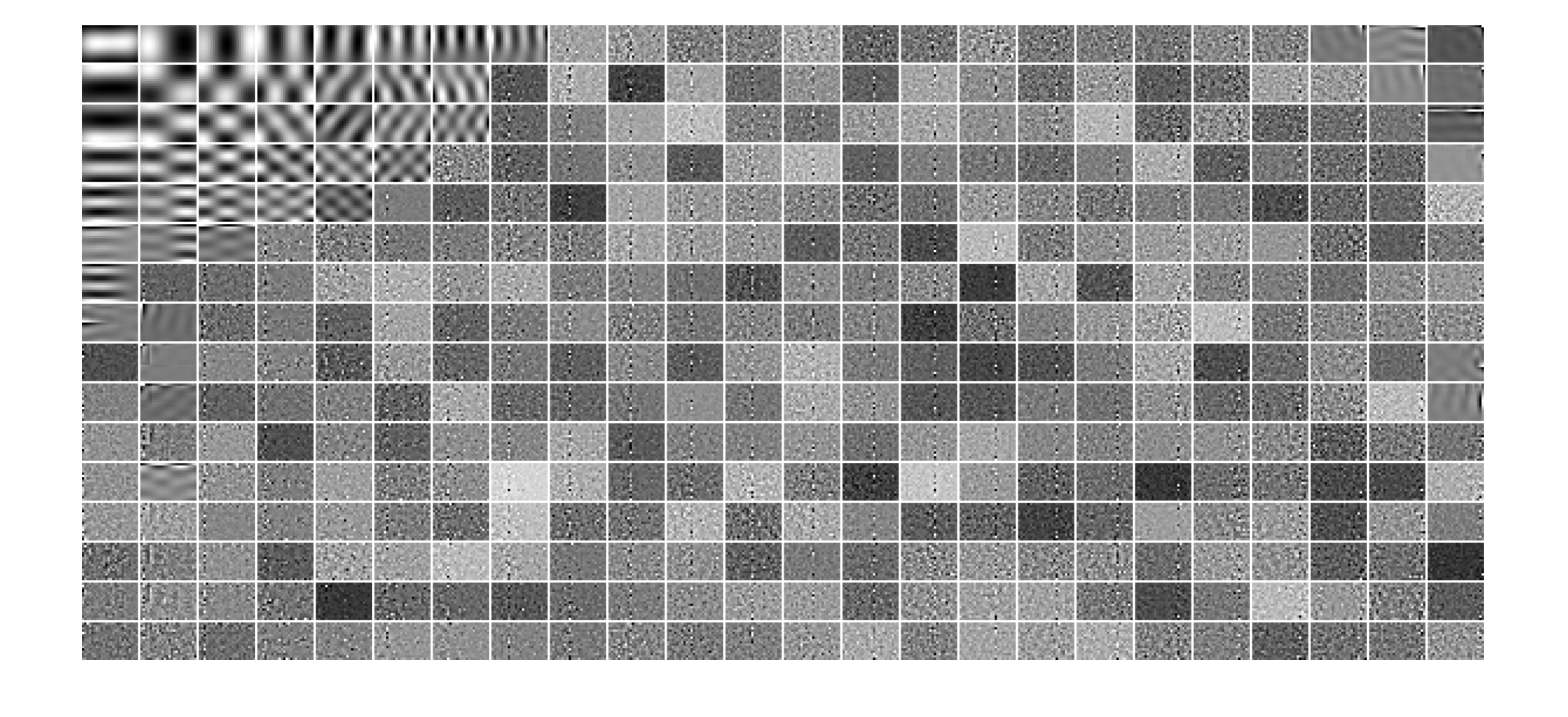}
    \end{minipage}
    \label{fig:dict1x20_marmousi}
  }\\
  \subfloat[]
  {
    \begin{minipage}{0.5\linewidth}
      \centering
      \includegraphics[width=\textwidth]{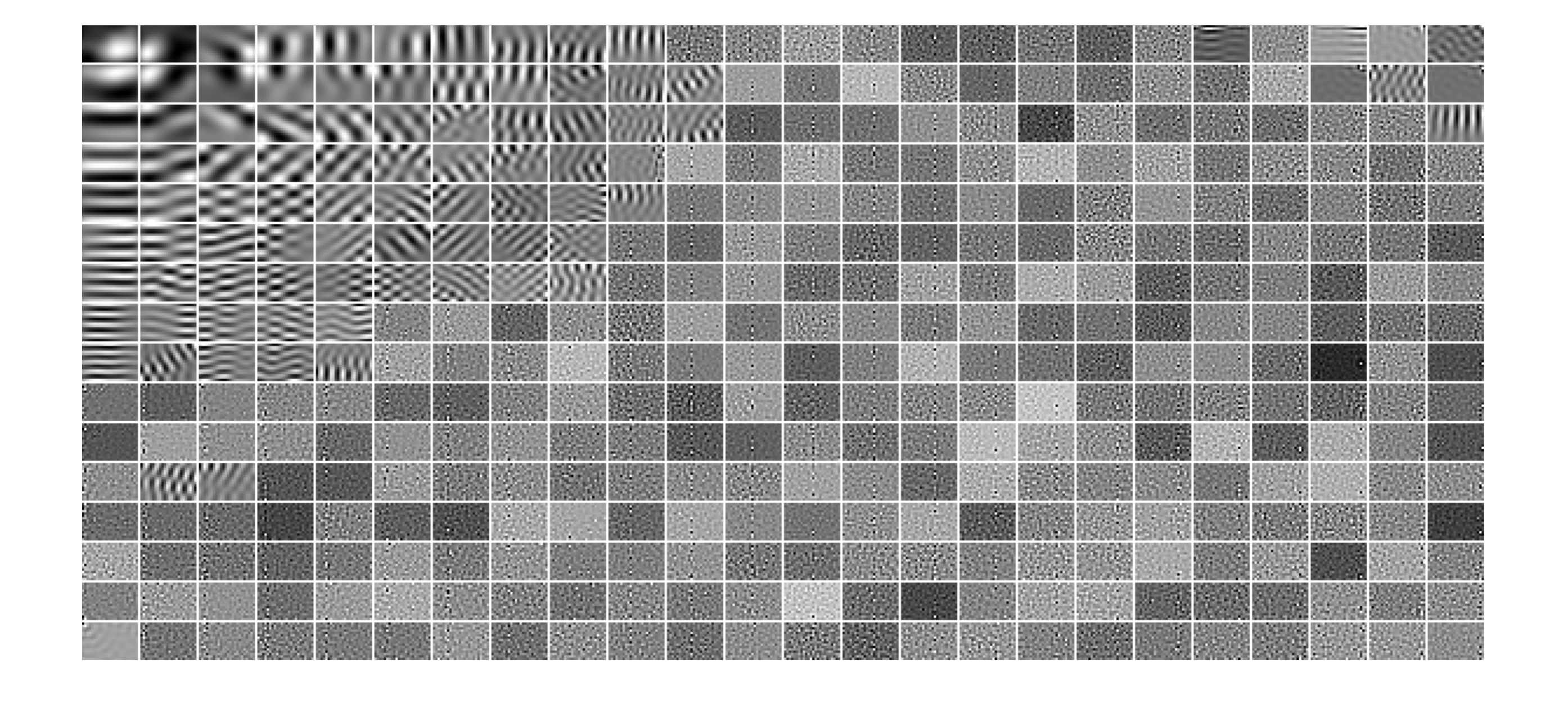}
    \end{minipage}
    \label{fig:dict2x20_marmousi}
  }
  \subfloat[]
  {
    \begin{minipage}{0.5\linewidth}
      \centering
      \includegraphics[width=\textwidth]{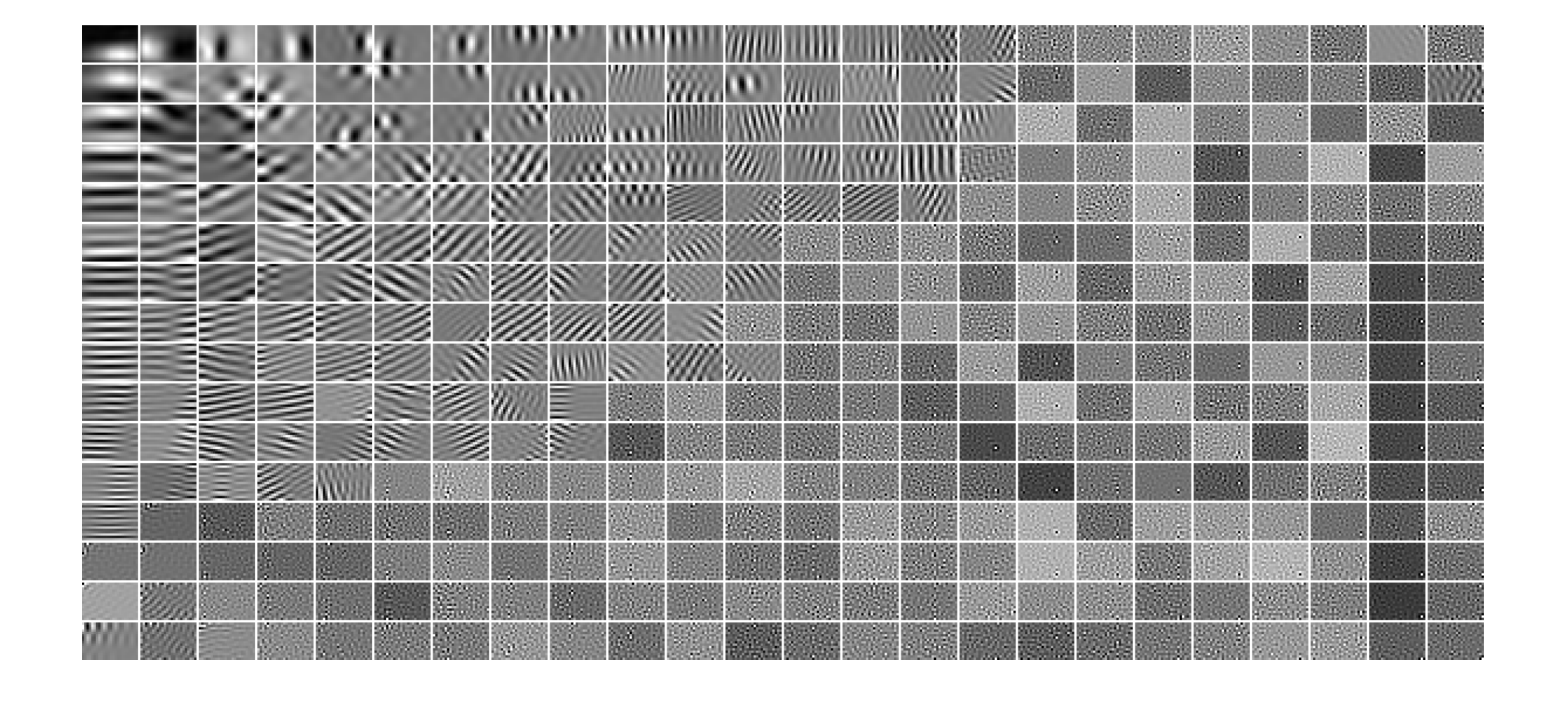}
    \end{minipage}
    \label{fig:dict3x20_marmousi}
  }\\
  \subfloat[]
  {
    \begin{minipage}{0.5\linewidth}
      \centering
      \includegraphics[width=\textwidth]{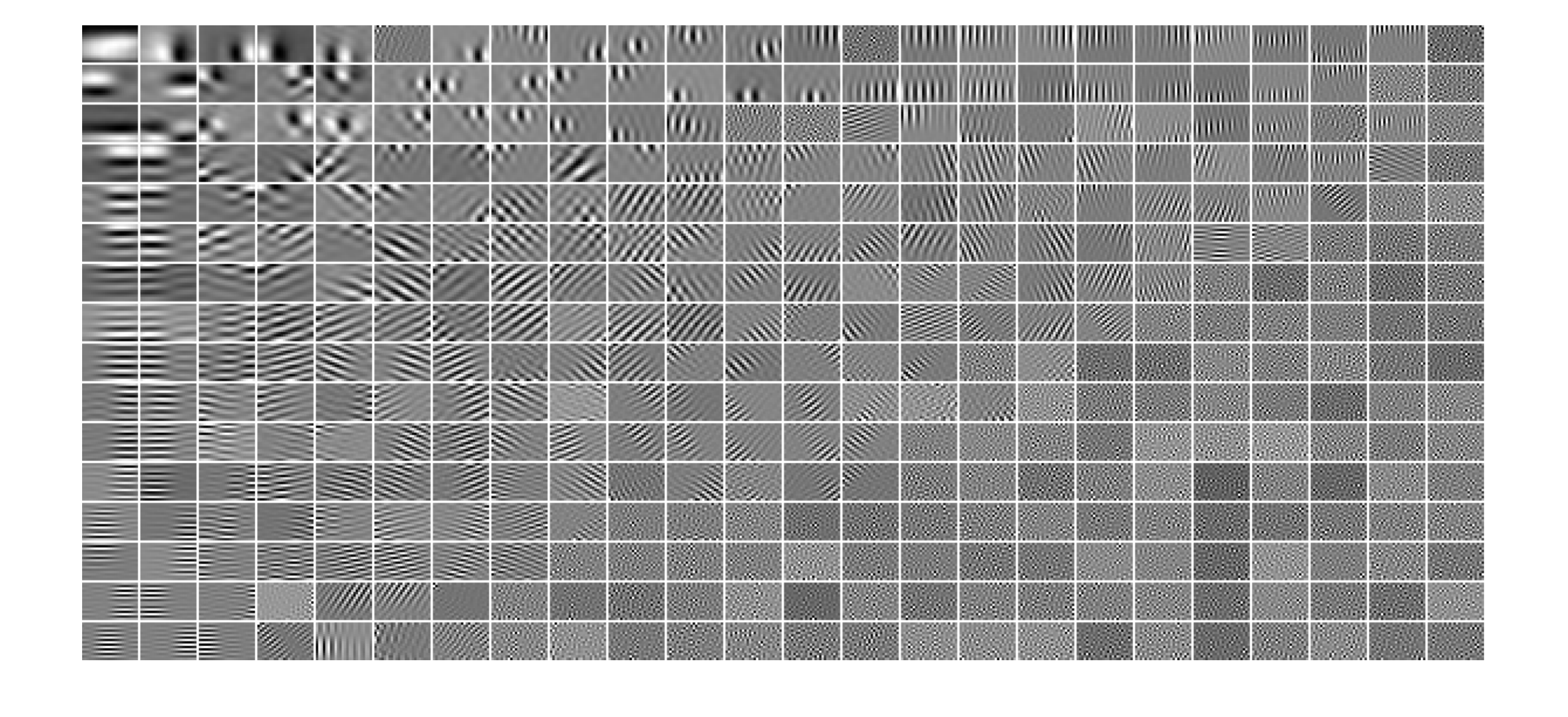}
    \end{minipage}
    \label{fig:dict4x20_marmousi}
  }
  \subfloat[]
  {
    \begin{minipage}{0.5\linewidth}
      \centering
      \includegraphics[width=\textwidth]{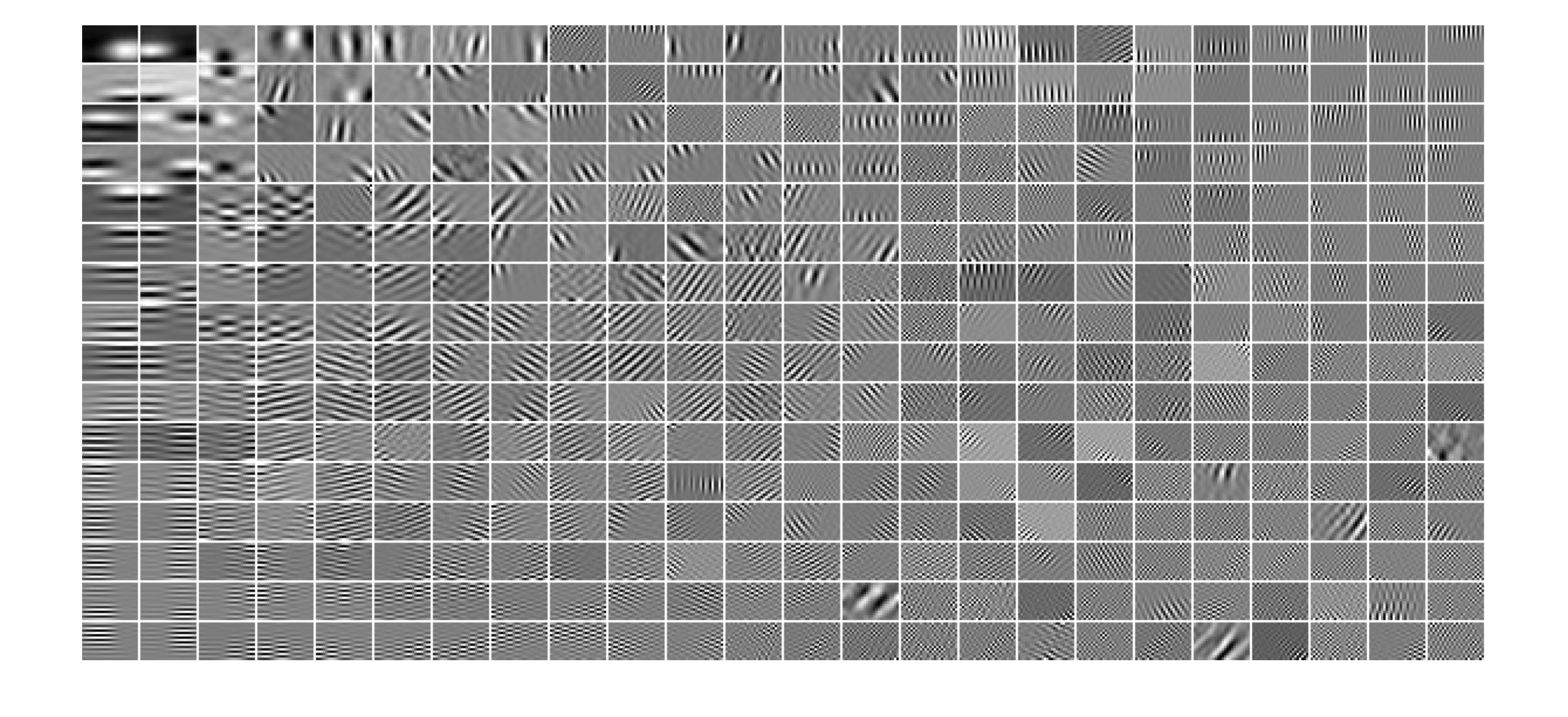}
    \end{minipage}
    \label{fig:dict5x20_marmousi}
  }
  \caption{Initial dictionary $\mathbf{D}_0$ and the learned dictionaries $\mathbf{D}_{20}$ by Algorithm \ref{alg:online-ortho-dl} after $K = 20$ FWI iterations in each frequency band on the Marmousi model. Dictionary size is $384 \times 384$; each column is displayed as a $16 \times 24$ block in the images.
  (a) Initial DCT matrix $\mathbf{D}_0$,
  (b) Trained dictionary $\mathbf{D}_{20}$ for the first frequency band, 3.0--11.6\,Hz,
  (c) second band, 12.1--20.8\,Hz,
  (d) third band, 21.3--29.9\,Hz,
  (e) fourth band, 30.4--39.0\,Hz, and (f) fifth band, 39.5--48.1\,Hz.}
  \label{fig:dict-marmousi}
\end{figure*}

\begin{figure}[htbp]
  \centering
  \subfloat[]
  {
    \begin{minipage}{0.33\linewidth}
      \centering
      \includegraphics[width=\textwidth]{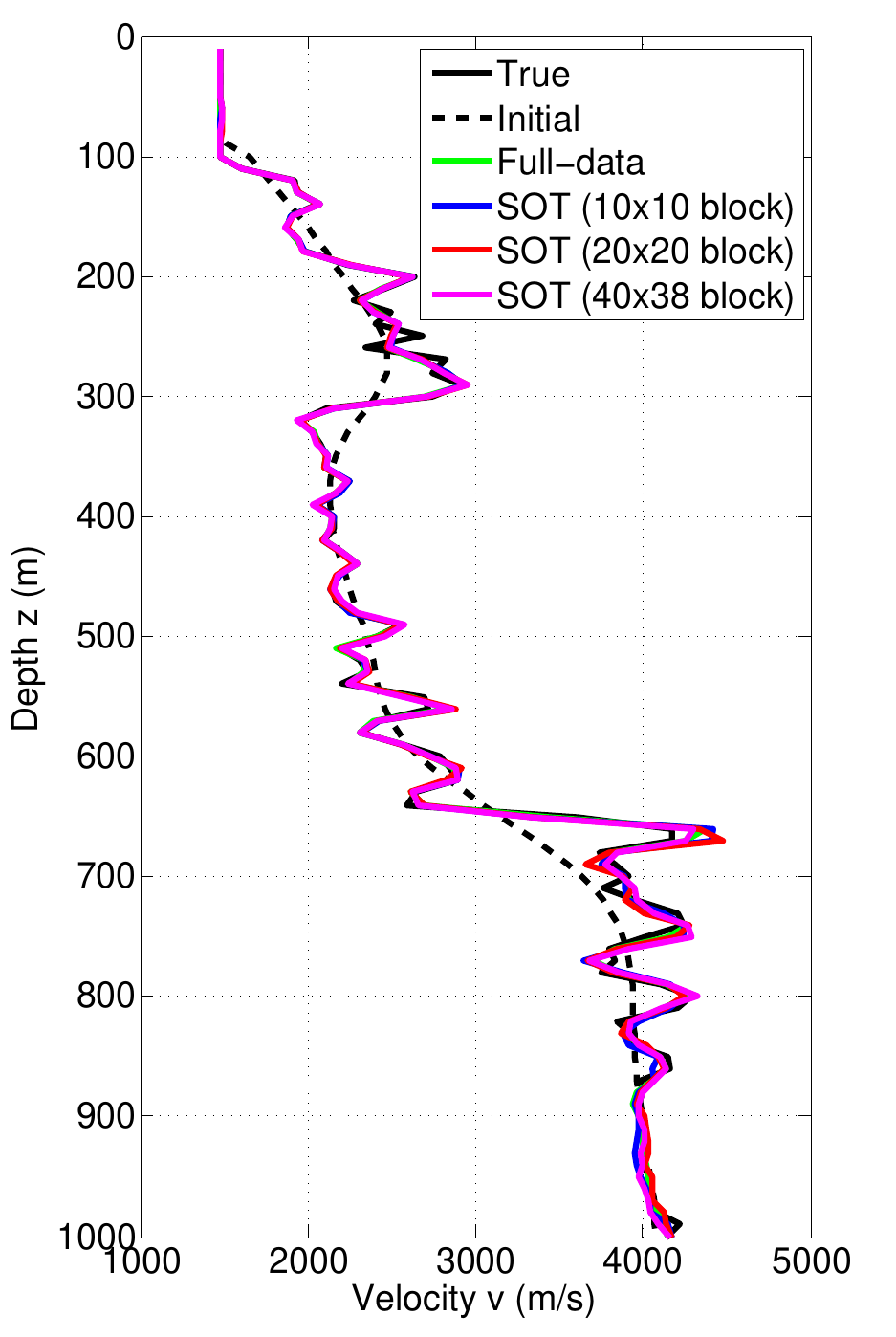}
    \end{minipage}
    \label{fig:vertical-log-x85-bgcompass}
  }
  \subfloat[]
  {
    \begin{minipage}{0.33\linewidth}
      \centering
      \includegraphics[width=\textwidth]{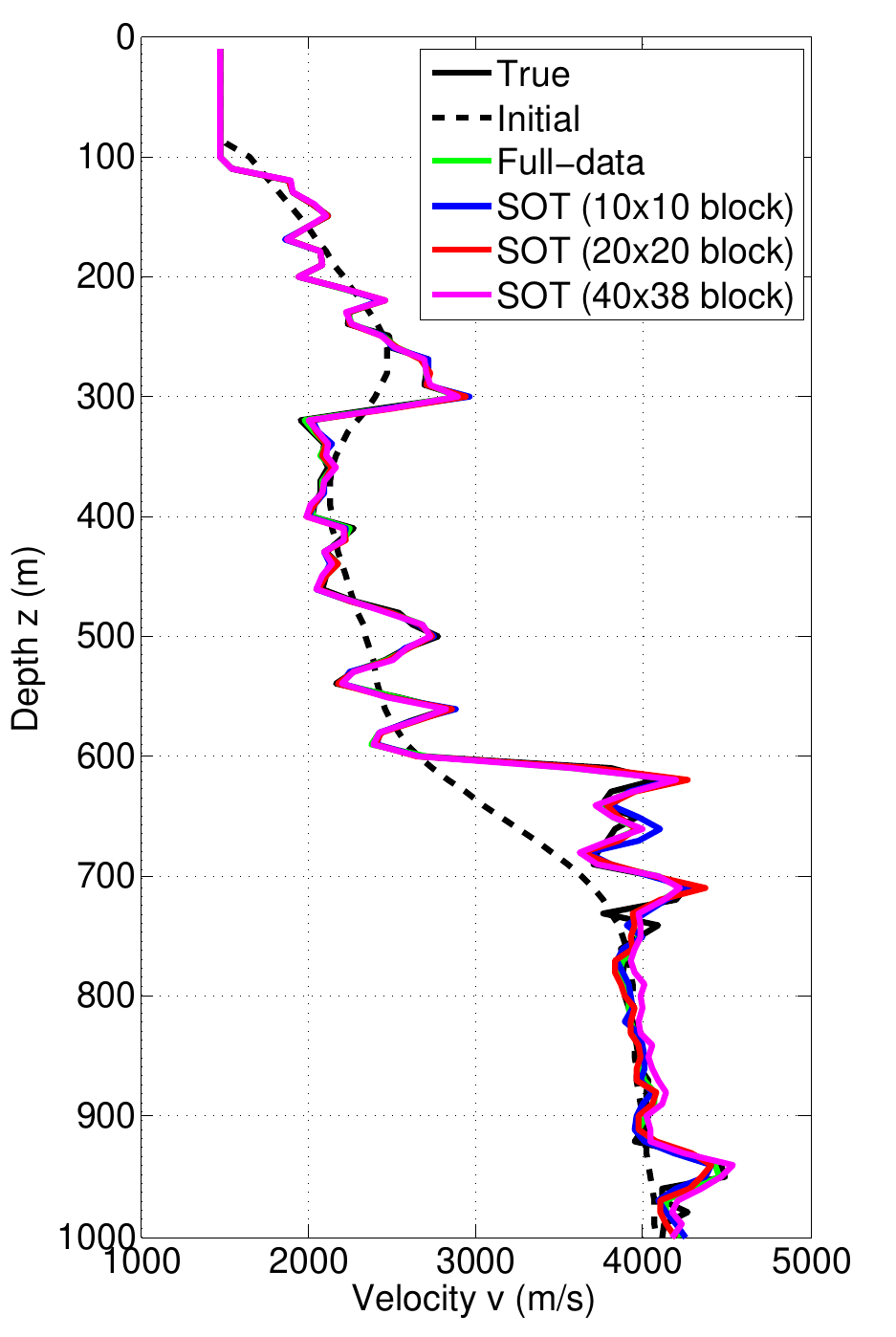}
    \end{minipage}
    \label{fig:vertical-log-x170-bgcompass}
  }
  \subfloat[]
  {
    \begin{minipage}{0.33\linewidth}
      \centering
      \includegraphics[width=\textwidth]{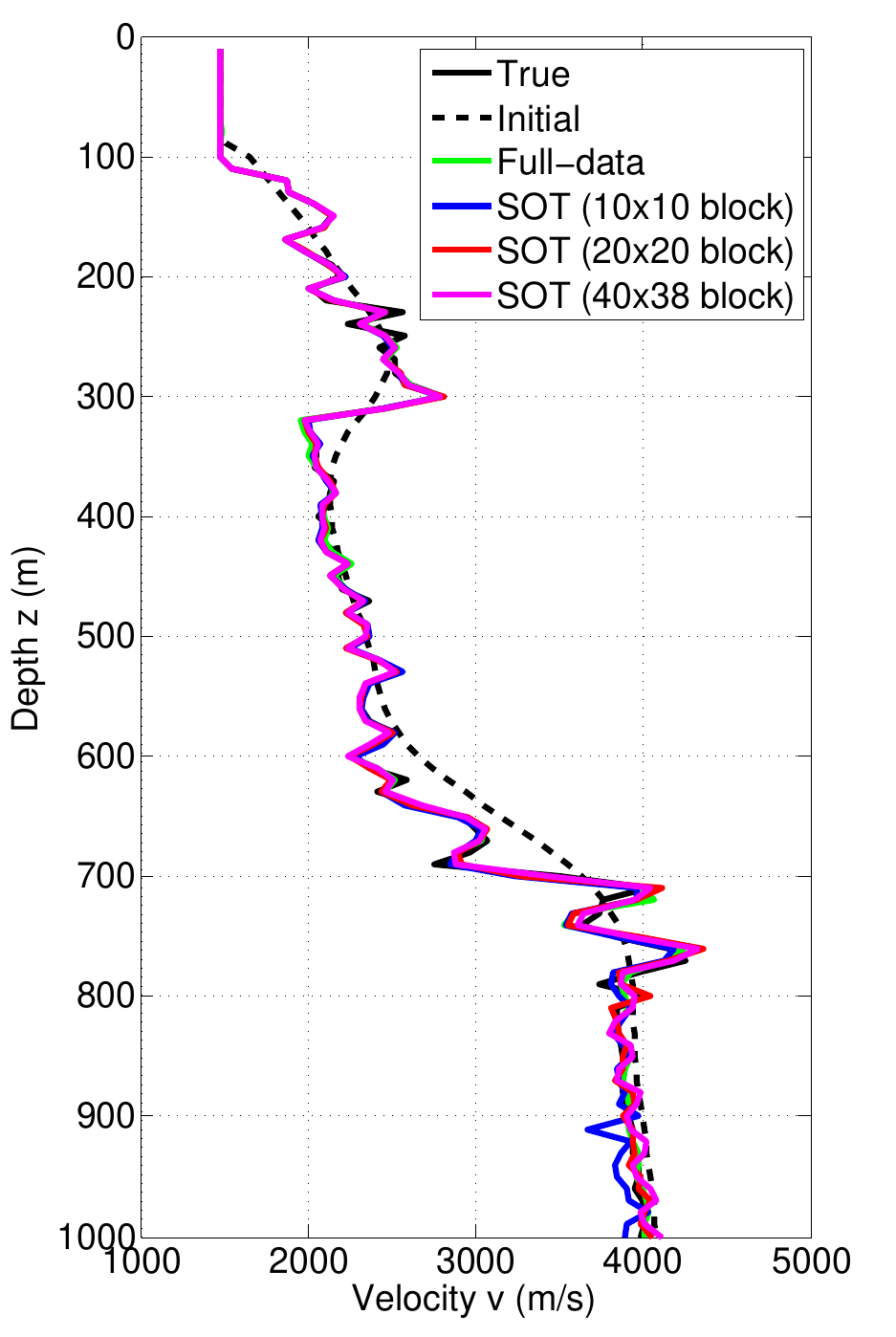}
    \end{minipage}
    \label{fig:vertical-log-x255-bgcompass}
  }
  \caption{Vertical velocity logs for the BG-Compass model. (a) $x = 875$\,m, (b) $x = 1750$\,m, (c) $x = 2625$\,m. Three different patch sizes are tested.}
  \label{fig:vertical-vlogs-bgcompass}
\end{figure}

\begin{figure}[htbp]
  \centering
  \subfloat[]
  {
    \begin{minipage}{0.33\linewidth}
      \centering
      \includegraphics[width=\textwidth]{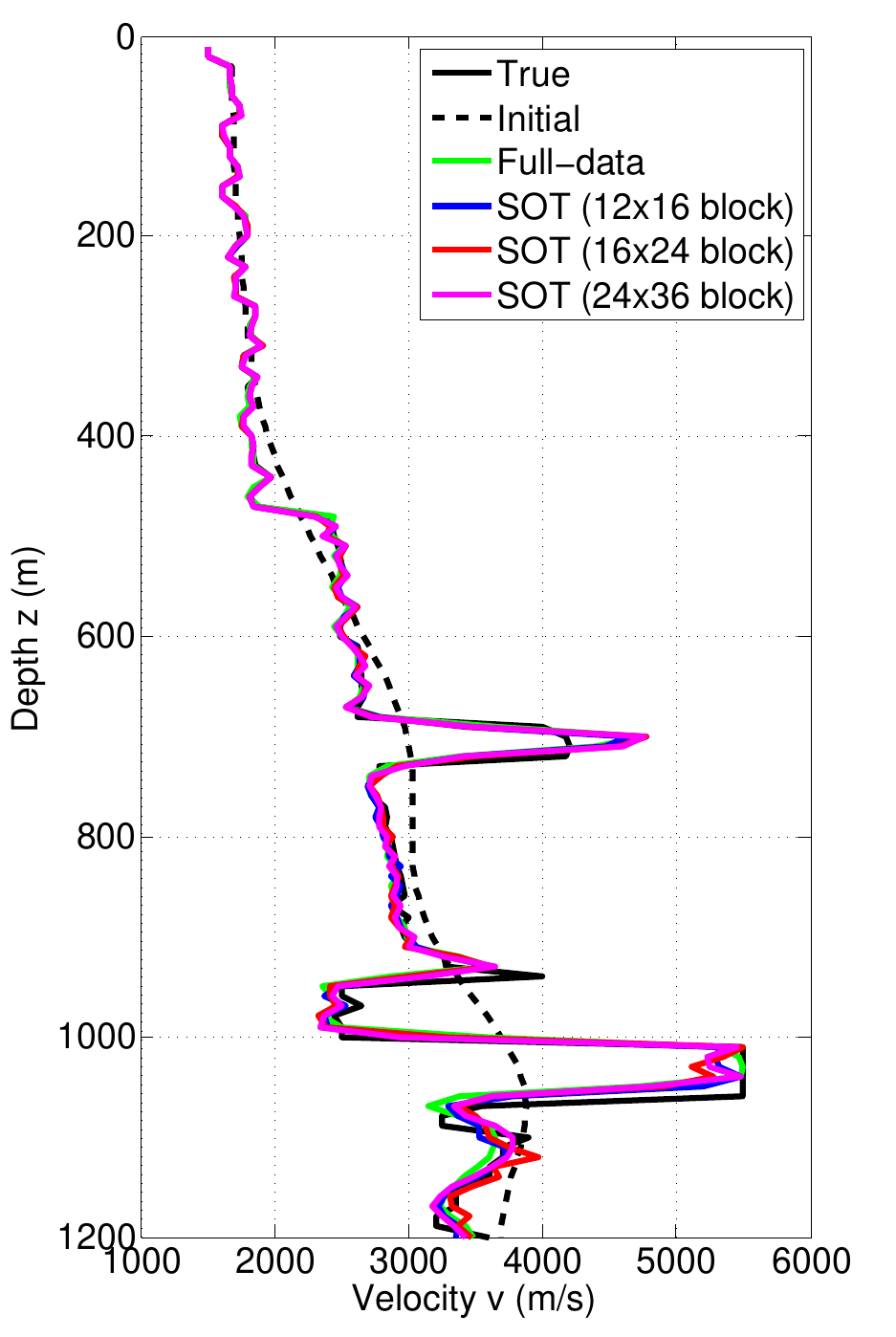}
    \end{minipage}
    \label{fig:vertical-log-x96-marmousi}
  }
  \subfloat[]
  {
    \begin{minipage}{0.33\linewidth}
      \centering
      \includegraphics[width=\textwidth]{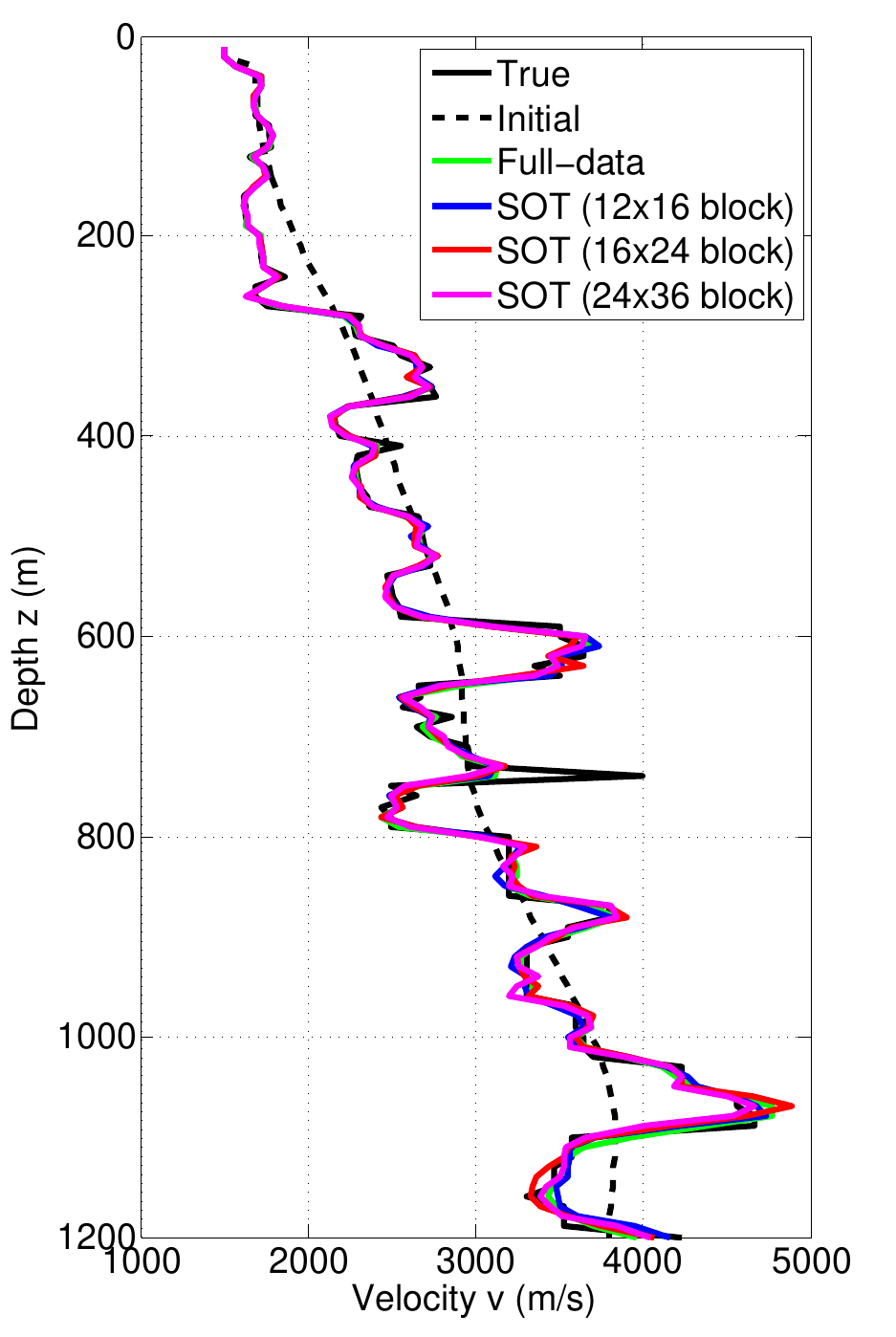}
    \end{minipage}
    \label{fig:vertical-log-x192-marmousi}
  }
  \subfloat[]
  {
    \begin{minipage}{0.33\linewidth}
      \centering
      \includegraphics[width=\textwidth]{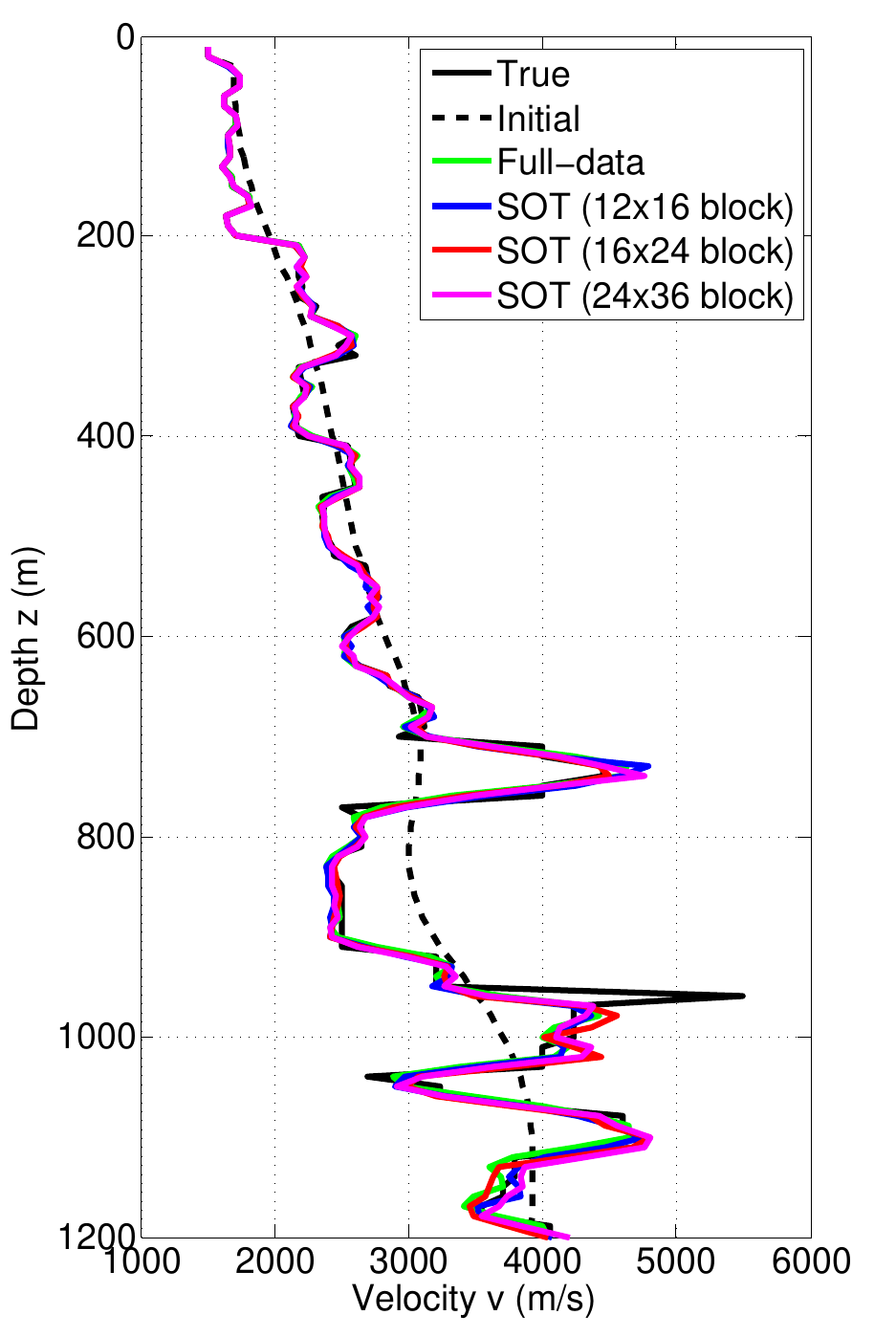}
    \end{minipage}
    \label{fig:vertical-log-x288-marmousi}
  }
  \caption{Vertical velocity logs for the Marmousi model. (a) $x = 960$\,m, (b) $x = 1920$\,m, (c) $x = 2880$\,m. Three different patch sizes are tested.}
  \label{fig:vertical-vlogs-marmousi}
\end{figure}

\begin{figure}[htbp]
  \centering
  \subfloat[]
  {
    \begin{minipage}{0.5\linewidth}
      \centering
      \includegraphics[width=\textwidth]{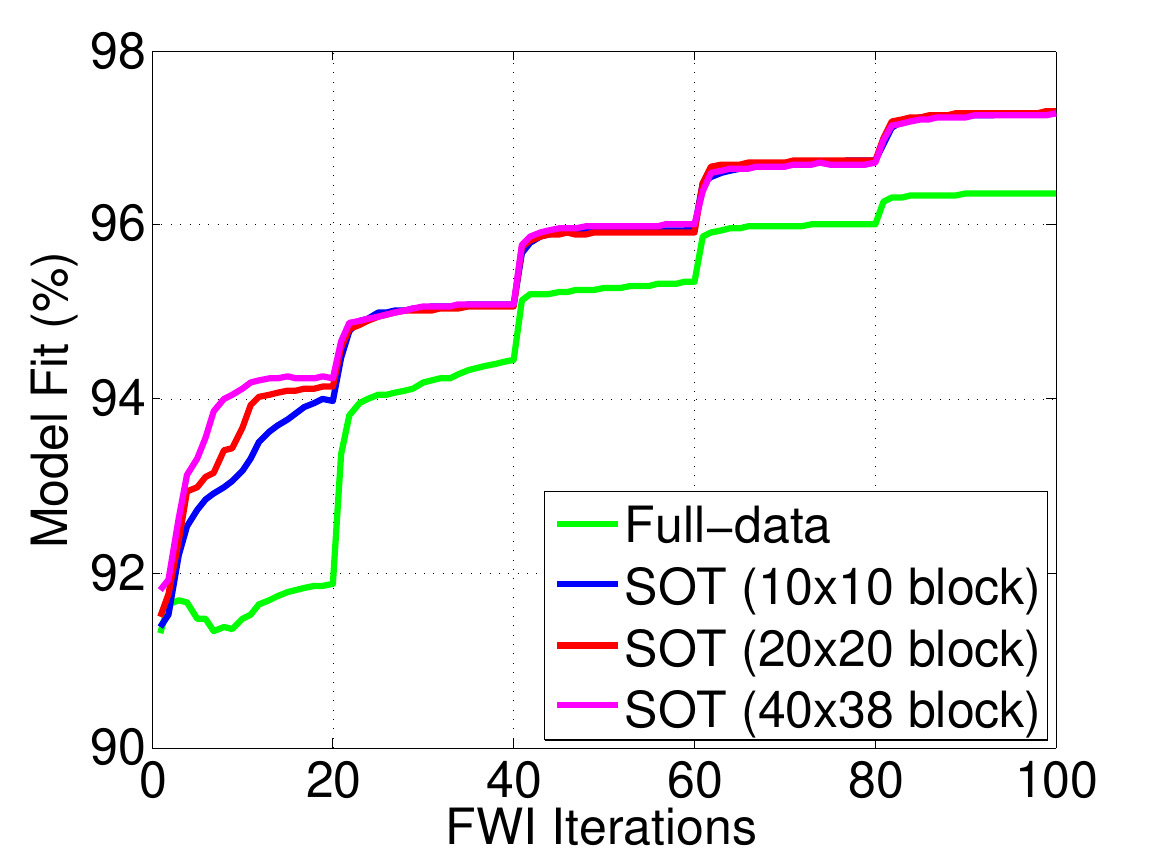}
    \end{minipage}
    \label{fig:modelfit-bgcompass}
  }
  \subfloat[]
  {
    \begin{minipage}{0.5\linewidth}
      \centering
      \includegraphics[width=\textwidth]{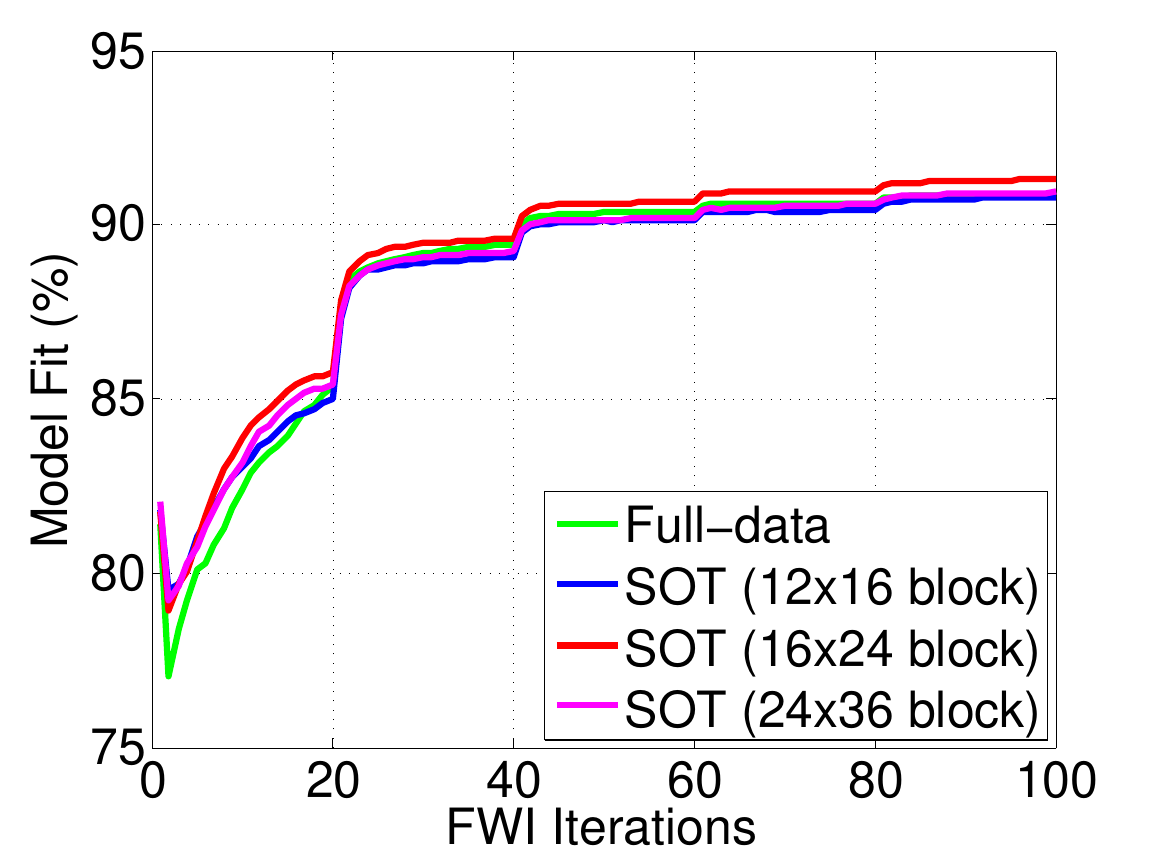}
    \end{minipage}
    \label{fig:modelfit-marmousi}
  }
  \caption{Model fit versus FWI iteration number for SOT-domain sparsity regularization for (a) BG-Compass model and (b) Marmousi model. Three different patch sizes are tested.}
  \label{fig:modelfit}
\end{figure}

In our experiments, the details of the orthonormal dictionaries $\mathbf{D}_k$ learned for SOT within the five frequency bands are as follows. Since the $\delta \mathbf{m}$ inverted in different frequency bands contain different wavenumber components and their features have different scales, we reinitialize the online orthonormal dictionary learning algorithm from $k = 0$ with a Discrete Cosine Transform (DCT) orthonormal dictionary $\mathbf{D}_0$ every time we move forward to a new frequency band.
For the case of the BG-Compass model, the default size of the training patches from $\delta \mathbf{m}$ is $n_z \times n_x = 20 \times 20$, $N = n_zn_x = 400$, so the dictionaries are $\mathbf{D}_k \in \mathbb{R}^{400 \times 400}$. Similarly, for the Marmousi model, the default size of training patches from $\delta \mathbf{m}$ is $n_z \times n_x = 16 \times 24$, $N = n_zn_x = 384$, so $\mathbf{D}_k \in \mathbb{R}^{384 \times 384}$. Figures \ref{fig:dict-bgcompass} and \ref{fig:dict-marmousi} show how the dictionaries evolve by Algorithm \ref{alg:online-ortho-dl} during FWI iterations on different frequency bands, in which each $n_z \times n_x$ patch of $\delta \mathbf{m}$ is a linear combination of the atoms visualized as blocks. Figures \subref*{fig:dictDCT_20x20} and \subref*{fig:dictDCT_16x24} are the DCT dictionaries $\mathbf{D}_0$ that initialize Algorithm \ref{alg:online-ortho-dl} when FWI starts processing a new frequency band. After completing $K = 20$ iterations of FWI as well as online dictionary learning, Figures \subref*{fig:dict1x20_bgcompass} -- \subref*{fig:dict5x20_bgcompass} and \subref*{fig:dict1x20_marmousi} -- \subref*{fig:dict5x20_marmousi} show the updated orthonormal dictionaries $\mathbf{D}_K$ in each frequency band. The Lagrange multiplier is empirically set as $\lambda = 0.2^2$ for both models so that an appropriate trade-off can be maintained between speed of convergence and capability of sparse representation.
Different patch sizes $N = n_z \times n_x$ are also tested with different sized dictionaries $\mathbf{D}_k \in \mathbb{R}^{N \times N}$  in order to study the robustness of the method (see Figures \ref{fig:vertical-vlogs-bgcompass}, \ref{fig:vertical-vlogs-marmousi}, and \ref{fig:modelfit}).
%

Figures \ref{fig:vertical-vlogs-bgcompass} and \ref{fig:vertical-vlogs-marmousi} show vertical velocity logs for several lateral positions $x$ on the recovered images of both models, marked by blue triangles underneath Figures \ref{fig:vmTrue-bgcompass} and \ref{fig:vmTrue-marmousi}, respectively. Besides traditional vertical velocity logs, the quality of FWI can also be measured by the following model fit metric proposed by \cite{Guitton:2012aa}
\begin{equation}
  \label{eq:model-similarity}
  M(k) \triangleq \left( 1 - \frac{\|\mathbf{v}_{\text{true}} - \mathbf{v}_k\|_2}{\|\mathbf{v}_{\text{true}}\|_2} \right) \times 100\%
\end{equation}
where $\mathbf{v}_{\text{true}} = 1/\sqrt{\mathbf{m}_{\text{true}}}$ is the exact velocity model and $\mathbf{v}_k = 1/\sqrt{\mathbf{m}_k}$ is the intermediate velocity model obtained at the $k$-th FWI iteration.
The curves in Figure \ref{fig:modelfit} compare the model fit metric $M(k)$ versus FWI iteration number for both velocity models. These results indicate that different patch sizes yield very similar curves and we should choose a moderate patch size $N = n_z \times n_x$ that it is neither too large to train nor too small to represent.
\pagebreak[4]  
%

\begin{figure}[htbp]
	\centering
	\begin{minipage}{0.9\linewidth}
	\centering
	\includegraphics[width=\textwidth]{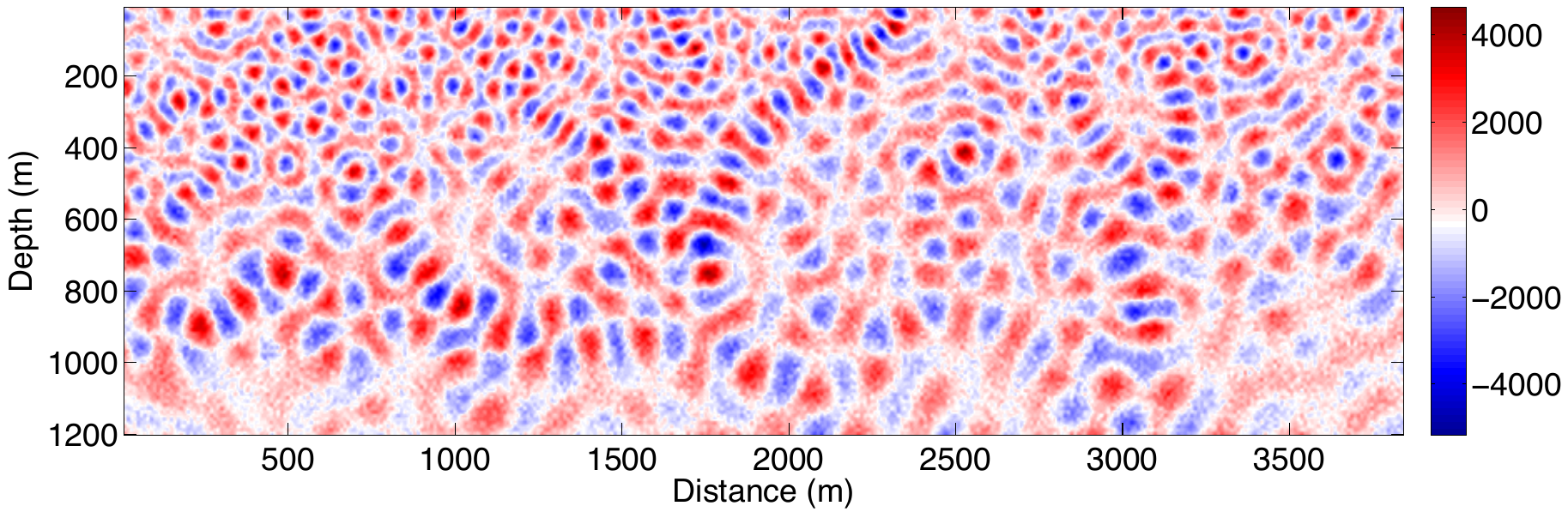}
	\end{minipage}
	\caption{Noisy wavefield examples generated by a supershot with frequency $22.8$\,Hz, SNR = 10\,dB}
\label{fig:snapshotTrueFreqSupershot_marmousi_noise}
\end{figure}

\begin{figure}[htbp]
	\centering
	\begin{minipage}{\linewidth}
	\centering
	\includegraphics[width=0.9\textwidth]{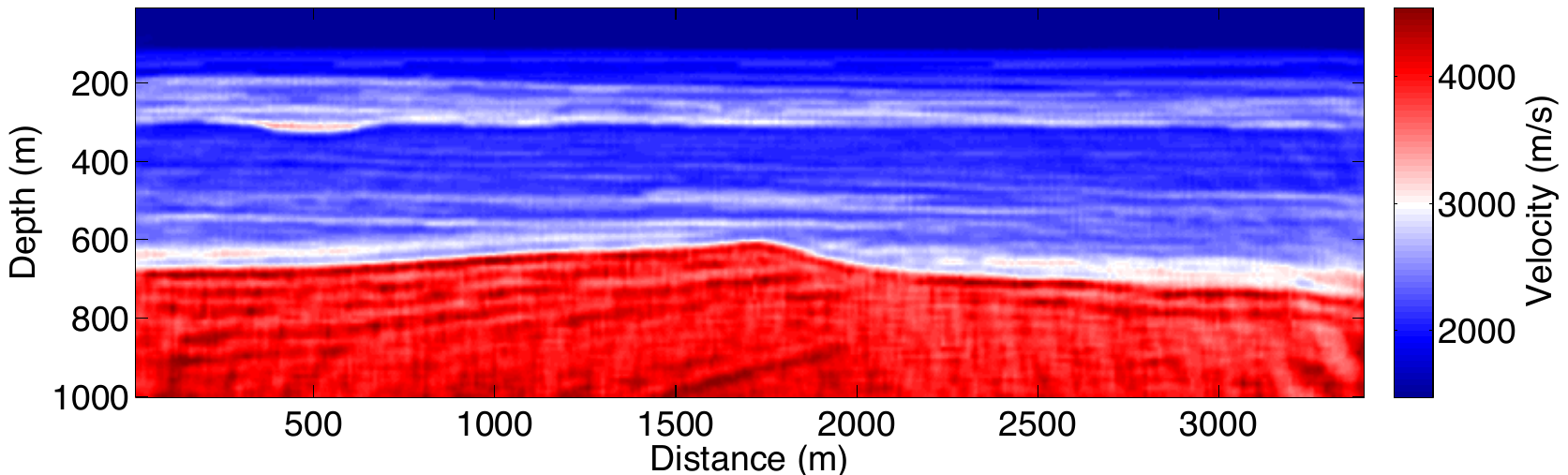}
	\end{minipage}
	\caption{FWI results for the BG-Compass model with noisy dataset; average SNR = 10\,dB.}
\label{fig:vmNew5x20-bgcompass-sotcs-noise}
\end{figure}

\begin{figure}[htbp]
	\centering
	\begin{minipage}{\linewidth}
	\centering
	\includegraphics[width=0.9\textwidth]{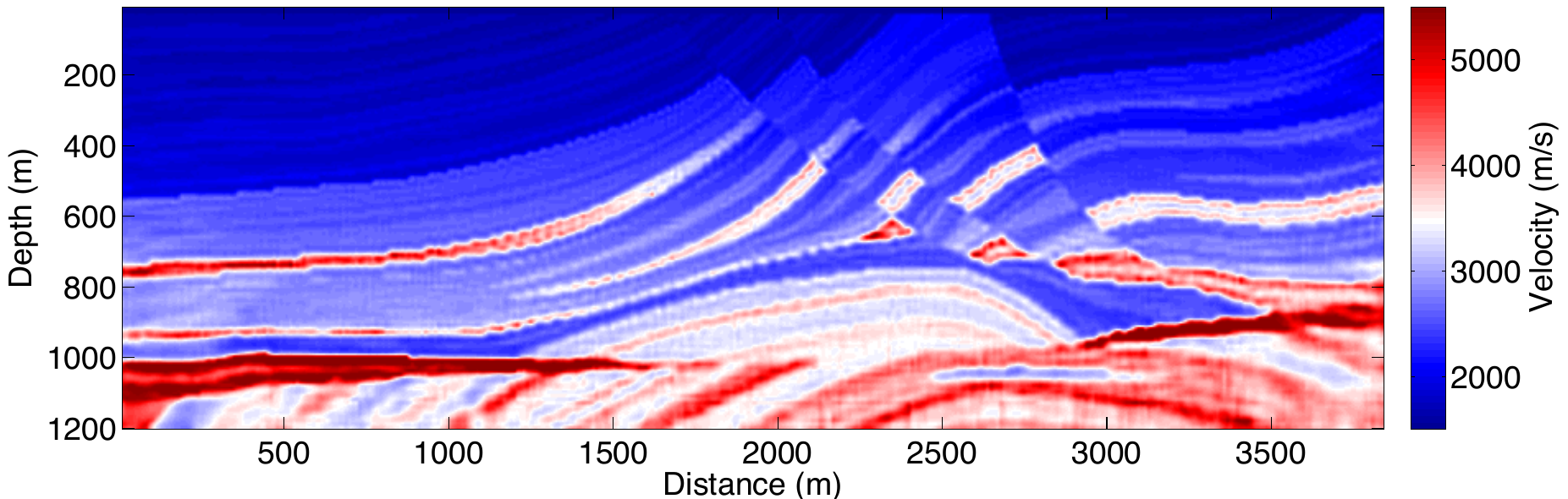}
	\end{minipage}
	\caption{FWI results for the Marmousi model with noisy dataset; average SNR = 10\,dB.}
\label{fig:vmNew5x20-marmousi-sotcs-noise}
\end{figure}

\begin{figure}[htbp]
  \centering
  \subfloat[]
  {
    \begin{minipage}{0.5\linewidth}
      \centering
      \includegraphics[width=\textwidth]{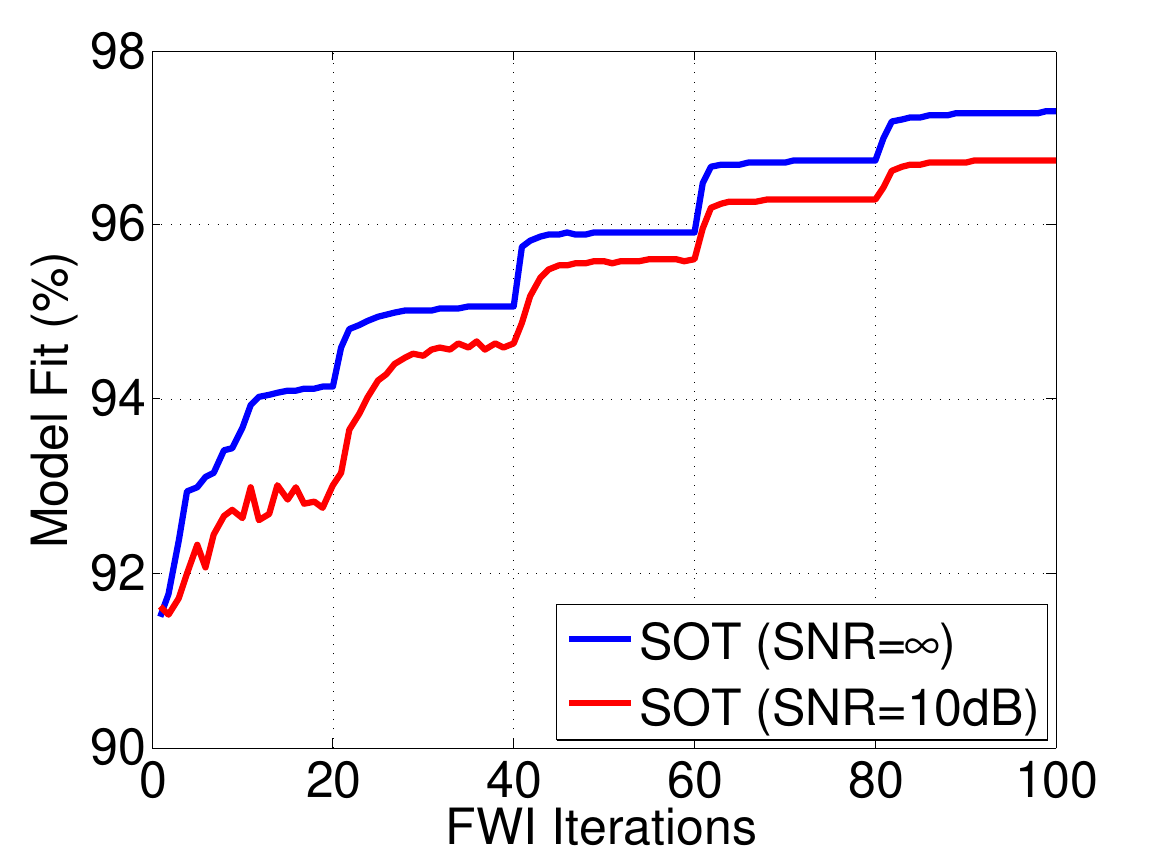}
    \end{minipage}
    \label{fig:modelfit-bgcompass-noise}
  }
  \subfloat[]
  {
    \begin{minipage}{0.5\linewidth}
      \centering
      \includegraphics[width=\textwidth]{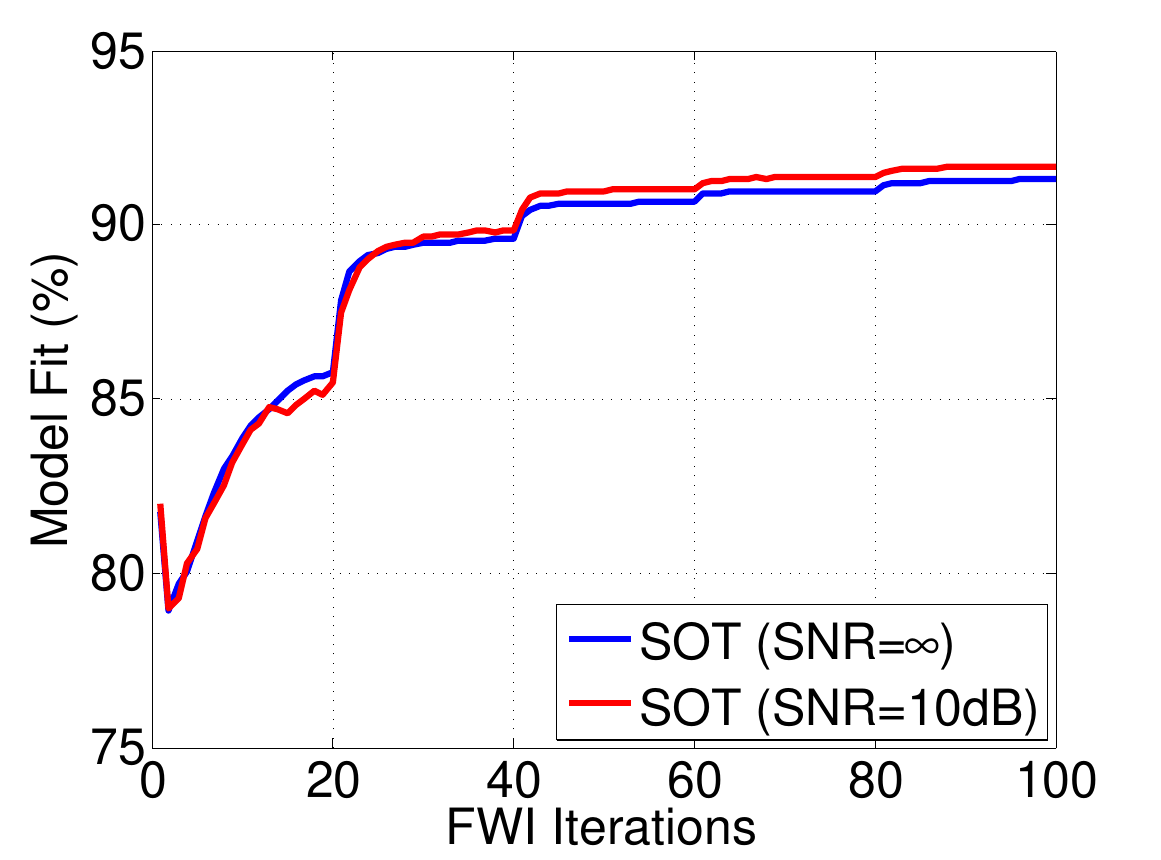}
    \end{minipage}
    \label{fig:modelfit-marmousi-noise}
  }
  \caption{Model fit versus FWI iteration number of SOT-domain sparsity regularization with noiseless data (blue line) and noisy data at average SNR = 10\,dB. (red line) for (a) BG-Compass model, (b) Marmousi model.}
  \label{fig:modelfit-noise}
\end{figure}

To further test the robustness of our method, we create a noisy seismic dataset by adding white Gaussian noise (WGN), and then perform FWI without any prior denoising process.
Figure \ref{fig:snapshotTrueFreqSupershot_marmousi_noise} illustrates the wavefield generated by a supershot with the frequency $\omega / (2\pi) = 22.8$\,Hz, where WGN is added such that the average signal to noise ratio (SNR) equals to 10\,dB. As before, $N_s' = 3$ supershots for the BG-Compass model, $N_s' = 4$ supershots for the Marmousi model, and $N_{\omega}' = 16$ random frequencies are used for each frequency band.
Figures \ref{fig:vmNew5x20-bgcompass-sotcs-noise} and \ref{fig:vmNew5x20-marmousi-sotcs-noise} show the FWI results based on the noisy data for both velocity models, followed by curves of model fit metric $M(k)$ versus FWI iterations in Figure \ref{fig:modelfit-noise}. Comparing Figures \ref{fig:vmNew5x20-bgcompass-sotcs-noise} and \ref{fig:vmNew5x20-marmousi-sotcs-noise} with Figures \ref{fig:vmNew5x20-bgcompass-sotcs} and \ref{fig:vmNew5x20-marmousi-sotcs} respectively, we see that our method still obtains a good FWI result with noisy data, which is a result of the SOT-domain sparsity regularization.

\section{Discussion}
FWI is a computationally expensive problem because it is iterative and must repeatedly solve wave equations on a large-scale heterogeneous medium for many sources. In order to reduce the computational cost without compromising the results, CS techniques suggest using only a small fraction of the data volume after some randomness has been injected as long as the model perturbation $\delta \mathbf{m}$ can be transformed into sparse coefficients.
Then it is still possible to invert the velocity model by iteratively minimizing a modified Gauss-Newton objective function with an $\ell_1$-norm constraint. Different from using predefined transforms with implicit dictionaries such as wavelets or curvelets, this paper introduces a sparse transform called the SOT based on learning patch-sized adaptive dictionaries from the inverted model perturbations. Predefined transforms are designed to optimally represent a specific class of signals, so they cannot guarantee sparse representations of all signals with different features. Recent results in signal processing have shown that learning adaptive dictionaries by training on the signals themselves leads to sparse signal representation and state-of-the-art performance in various applications. 

To introduce dictionary learning into FWI, two features were emphasized. First, the learned dictionaries are made orthonormal, yielding a fast and straightforward alternating optimization scheme for SOT dictionary learning. An orthonormal dictionary is a Parseval frame, therefore finding sparse coefficient vectors only requires simple matrix multiplication. On the other hand, an overcomplete dictionary loses this simplicity and makes sparse representation a basis pursuit problem. Second, SOT dictionary learning is configured to work online, taking the iterative Gauss-Newton optimization process of FWI into account. After each iteration, the optimized model perturbation not only serves to update the velocity model, but also provides a resource to get many training patches for learning a new dictionary that adapts to the specific feature variations of different model perturbations.

\begin{figure}[htbp]
\centering
  \begin{minipage}{0.7\linewidth}
    \centering
    \includegraphics[width=\textwidth]{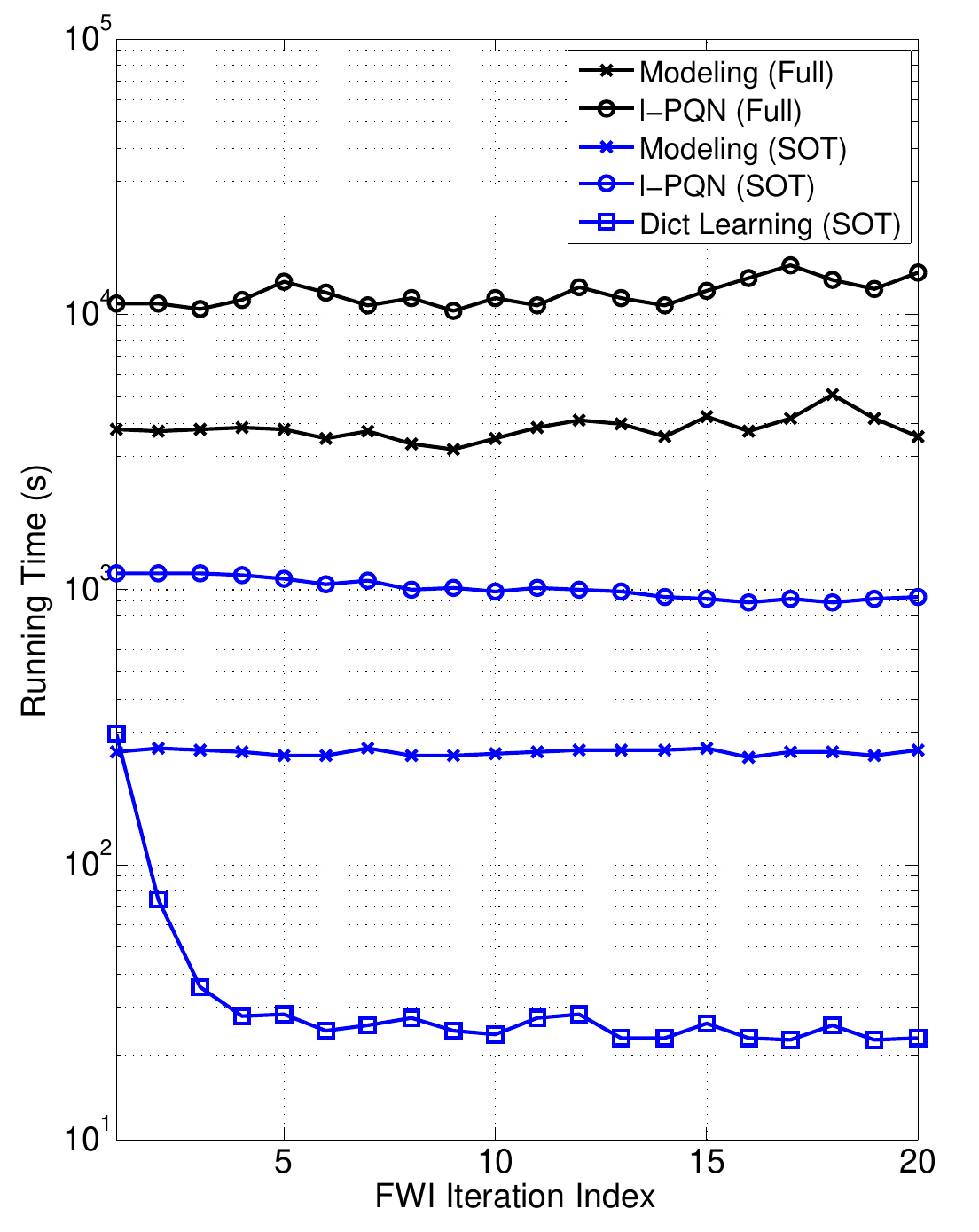}
  \end{minipage}
  \caption{Running time profile of forward modeling, l-PQN optimization and dictionary learning versus FWI iterations.}
\label{fig:runtime}
\end{figure}

The extra computational overhead involved in learning the orthonormal dictionary for SOT has been analyzed in the previous section ``Sparse Orthonormal Transform''. In addition to the theoretical complexity analysis, we now report one instance of the actual running time of forward modeling, l-PQN optimization and orthonormal dictionary learning (with $16 \times 24$ patches) for 20 FWI iterations (in one frequency band) on the Marmousi model with 4 supershots and 16 random frequencies.
As a comparison, the running times of forward modeling and l-PQN optimization for the full-data FWI are also provided (see Figure \ref{fig:runtime}). The computing cluster is based on 12-core Intel\textsuperscript{\textregistered} Xeon\textsuperscript{\textregistered} CPU with 64GB RAM and we have accelerated the forward modeling by parallel computing.

Figure \ref{fig:runtime} shows that the running time of forward modeling and l-PQN optimization for the full-data FWI is over 10 times of that for the data-reduced FWI.
It is also noticeable that, after the first FWI iteration, the running time for orthonormal dictionary learning falls rapidly to a negligible level compared to the cost of the other two phases. The online learning approach exhibits this behavior because it always updates the latest and best dictionary for the incoming model perturbation at each FWI iteration. Once a good dictionary has been obtained, many fewer training iterations are required for the updates. Therefore, it is safe to say that the actual overhead from orthonormal dictionary learning is not significant.

The extension of orthonormal dictionary learning, SOT, as well as the FWI using SOT for sparse representation to 3D models should be feasible. Though solving 3D forward modeling problems is much more expensive than 2D problems, recent development on the direct solvers eventually relieves the difficulty \cite{Operto:2015aa}. Besides, compressive sensing schemes using SOT could offer more flexibility and efficiency to 3D problems as 3D grid geometry provides more freedom for subsampling.
Since dictionary learning reshapes model patches and atoms into vectors no matter how many dimensions in space they occupy, we can reasonably infer their computational complexity according to the learning steps in Algorithm \ref{alg:online-ortho-dl}. The 3D model size is $N_z \times N_x \times N_y$ and the 3D patch size is $n_z \times n_x \times n_y$, where $n_z \ll N_z$, $n_x \ll N_x$, $n_y \ll N_y$. If all possible overlapping patches are used for training, then each orthonormal dictionary learning iteration costs $\mathcal{O}((n_zn_xn_y)^2(N_z-n_z+1)(N_x-n_x+1)(N_y-n_y+1)+(n_zn_xn_y)^3)$, and applying the SOT with the 3D dictionary to a 3D model costs $\mathcal{O}(n_zn_xn_yN_zN_xN_y)$.
The training patch size is the one parameter that needs to be chosen in order to get sparse representation of the model perturbations. Small patches are desirable for training the orthonormal dictionary.
The other two parameters in the algorithm, $\tau_k$ and $\lambda$ are chosen based on the data or via empirical testing.

\section{Conclusion}
In this paper, a novel and efficient compressive sensing scheme that significantly reduces the computational complexity of the FWI problem has been presented. The new method exploits the sparsity representation of model perturbations with a sparse orthonormal transform (SOT) such that each block of model perturbation can be represented with sparse coefficients over adaptive data-driven dictionaries trained from previous results. Compared to traditional model-based transforms that are only optimal for objects with piecewise smoothness, the SOT is better able to adapt to nonintuitive signal regularities such as complex geophysical features. Compared to the traditional overcomplete dictionary learning methods, the orthonormal dictionary learning method is much more efficient and can adapt in an online manner. Therefore, the SOT enables us to significantly reduce the amount of data used in FWI by invoking the strategy of compressive sampling. We randomly select sources and frequencies and encode the data with random phases so that this scheme works for both land and marine scenarios. After that, the original Gauss-Newton problem becomes a LASSO problem and can be effectively solved using a projected quasi-Newton algorithm.

The experiments clearly indicate that we can obtain high-quality inverted velocity models with both simple and complex geophysical features by working with a small subset of the full seismic dataset, even in the presence of noise.

\bibliographystyle{seg}
\bibliography{dictLearningFWI}

\end{document}